\documentclass[iop]{emulateapj-rtx4}
\shortauthors{Sekanina \& Kracht}
\shorttitle{Comet Pairs and Groups, and the 1921 Lick Object}
\slugcomment{Version \today }
\newcommand{\Rsun}{$R_{\mbox{\scriptsize \boldmath $\odot$}}\!$}

\newcommand{\lapeq}{$\;$\raisebox{0.3ex}{$<$}\hspace{-0.28cm}\raisebox{-0.75ex}{$\sim$}$\;$}
\newcommand{\gapeq}{$\;$\raisebox{0.3ex}{$>$}\hspace{-0.28cm}\raisebox{-0.75ex}{$\sim$}$\;$}

\begin{document}
\title{Pairs and Groups of Genetically Related Long-Period Comets and\\Probable
Identity of the Mysterious Lick Object of 1921}
%
\author{Zdenek Sekanina$^1$ \& Rainer Kracht$^2$}
\affil{$^1$Jet Propulsion Laboratory, California Institute of Technology,
  4800 Oak Grove Drive, Pasadena, CA 91109, U.S.A.\\
$^2$Ostlandring 53, D-25335 Elmshorn, Schleswig-Holstein, Germany}
\email{Zdenek.Sekanina@jpl.nasa.gov\\
{\hspace*{2.59cm}}R.Kracht@t-online.de{\vspace{-0.2cm}}}

\begin{abstract}
We present the history of investigation of the dynamical properties of pairs
and groups of genetically related long-period comets (other than the Kreutz
sungrazing system).  Members of a comet pair or group move in nearly
identical orbits and their origin as fragments of a common parent comet is
unquestionable.  The only variable is the time of perihelion passage, which
differs from member to member considerably due primarily to an orbital-momentum
increment acquired during breakup.  Meter-per-second separation velocities
account for gaps of years or tens of years, thanks to the orbital periods of
many millennia. The physical properties of individual members may not
at all be alike, as illustrated by the trio of C/1988~A1, C/1996~Q1, and
C/2015~F3.  We exploit orbital similarity to examine whether the celebrated
and as yet unidentified object, discovered from the Lick Observatory near the
Sun at sunset on 1921 August~7, happened to be a member of such a pair and to
track down the long-period comet to which it could be genetically related.
Our search shows that the Lick object, which could not be a Kreutz sungrazer,
was most probably a companion to comet C/1847~C1 (Hind), whose perihelion
distance was $\sim$9~{\Rsun} and true orbital period approximately 8300~yr.
{\vspace{-0.03cm}}The gap of 74.4~yr between their perihelion times is
consistent with a separation velocity of $\sim$1~m~s$^{-1}$ that~set~the
fragments apart following the parent's breakup in a general proximity of
perihelion during the previous return to the Sun in the 7th millennium
BCE.{\vspace{0.05cm}}
\end{abstract}

\section{Introduction}
Unverified sightings of bright objects near the Sun, reported from time
to time, are usually explained as incidental observations of known cosmic
bodies, such as the planets Venus, Jupiter, or Mercury, when stationary
over a period of seconds or minutes; or of various man-made items in the
atmosphere, such as weather or other baloons, when seen to float.

It happens much less often that an unidentified object near the~Sun is
reported by an expert.  The proper procedure is to promptly alert the
scientific community, so that the body could be detected by other observers,
its existence confirmed, and its nature established.  If the object orbits
the Sun, then depending on the quality of gathered observations, which in
each case rests on the conditions and circumstances at the times of discovery
and follow-up observations, information may sometimes be gained to secure the
motion of the object accurately enough to eventually allow an orbit to be
computed.

If the observations do not allow an orbit determination, the lost object
almost always remains unidentified.  Rare exceptions are lucky situations,
when the object happens to be a long-period comet (defined in this study
by an orbital period between $\sim$100 and $\sim$100\,000 yr; Section~4.3)
that is genetically related to another, previously observed comet and is, with
reasonable confidence, recognized as such.  As fragments of a common parent,
the two make up a {\it comet pair\/}.  If there are more than two~fragments,
they all belong to the same {\it comet group\/} or {\it family\/}.  Since the
members of a comet pair or group move about the Sun in virtually identical
orbits, the motion of any one of them can readily be {\it emulated\/} to a
fairly high degree of accuracy by equating it with the motion of any other
one, the perihelion time being the only variable.

A well-known case in point is a comet discovered in the Sun's outer corona
during the total solar eclipse on 1882 May 17.  Scientific expeditions,
dispatched to \mbox{Sohag},~on the Nile in Upper Egypt, to observe the eclipse,
reported visual and photographic detections of a bright streak not aligned with
the radial coronal ejections but nearly tangential to the occulted limb --- the
comet's narrow tail (e.g., Abney \& Schuster 1884).  The official designation
of the comet is now X/1882~K1, but informally it has been known as {\it
Tewfik\/}, named in honor~of~\mbox{Tewfik} Pasha, then reigning Khedive of
Egypt.  Observed for only 72 seconds (the duration of totality at Sohag;
e.g., Tacchini 1883, Abney \& Schuster 1884), the comet has never again been
seen since.  It turned out that the astrometric position of the sharply
defined nucleus of X/1882~K1, published by Abney~\& Schuster (1884), was
within $\sim$0$^\prime\!$.5 of the orbit of comet C/1843~D1~(\mbox{(Marsden}
1967), a major member of the Kreutz sungrazer system; and that the position
of X/1882~K1 reported~by~\mbox{Tacchini} (1882, 1883)\footnote{Kreutz
(1901) mistakenly attributed Tacchini's (1882, 1883) position of X/1882~K1
to C.\,Tr\'epied, and this error seems to have been propagating in the
literature throughout the 20th century.  The idea that the prominent
feature was a comet did not occur to Tr\'epied (1882) until after he saw
Schuster's photographs.} was similarly close to the orbit of C/1880~C1
(Mars\-den 1989a), another Kreutz member, which reached perihelion only
2.3~years before the eclipse.  This evidence is the basis for a general
consensus that X/1882~K1, too, was a member of the Kreutz system.

\begin{table*}[t]
\begin{center}
{\footnotesize {\bf Table 1}\\[0.08cm]
{\sc Orbital Elements of Comets C/1988 F1 (Levy) and C/1988 J1 (Shoemaker-Holt)
(Equinox J2000.0).}\\[0.1cm]
\begin{tabular}{l@{\hspace{1.25cm}}c@{\hspace{1.25cm}}c}
\hline\hline\\[-0.22cm]
Orbital element/Quantity & Comet C/1988 F1 & Comet C/1988 J1 \\[0.08cm]
\hline\\[-0.2cm]
Osculation epoch (TT) & 1987 Nov.\,21.0 & 1988 Feb.\,9.0 \\
Time of perihelion passage $t_\pi$ (TT) & 1987 Nov.\,29.94718
 & 1988 Feb.\,14.22162 \\
Argument of perihelion $\omega$ & 326$^\circ\!$.51491 & 326$^\circ\!$.51498 \\
Longitude of ascending node $\Omega$ & 288$^\circ\!$.76505
 & 288$^\circ\!$.76487 \\
Orbit inclination $i$ & \,\,62$^\circ\!$.80744 & \,\,62$^\circ\!$.80658 \\
Perihelion distance $q$ & 1.1741762 & 1.1744657 \\
Orbital eccentricity $e$ & 0.9978157 & 0.9978301 \\
\hspace*{3.09cm}osculation & 12\,460$\:\pm\:$460$\;\;$
                                & 12\,590$\:\pm\:$230$\;\;$ \\[-0.259cm]
Orbital period $P$ (yr)
 $\!\left\{ \raisebox{1.5ex}{}\raisebox{-1.5ex}{} \right.$\\[-0.259cm]
\hspace*{3.09cm}original$\:\!^{\rm a}$   & 13\,960 & 13\,840 \\[0.05cm]
\hline\\[-0.22cm]
Orbital arc covered by observations & 1988 Mar.\,22--1988 July 18
          & 1988 May 13--1988 Oct.\,20\\
Number of observations employed & 30 & 60 \\
Root-mean-squares residual & $\pm$0$^{\prime\prime}\!.7$
 & $\pm$0$^{\prime\prime}\!$.9 \\
Orbit-quality code$^{\rm b}$ & 2A & 2A \\
Reference & {\scriptsize Marsden (1989b)}
 & {\scriptsize Marsden (1989b)} \\[0.06cm]
\hline\\[-0.25cm]
\multicolumn{3}{l}{\parbox{12.9cm}{\scriptsize $^{\rm a}$\,With respect to
 the barycenter of the Solar System.}}\\[-0.01cm]
\multicolumn{3}{l}{\parbox{14.02cm}{\scriptsize $^{\rm b}$\,Following the
{\vspace{-0.03cm}}classification system introduced by Marsden et al.\
(1978); the errors of the elements other than the orbital period are
unavailable.}}\\[0.2cm]
\end{tabular}}
\end{center}
\end{table*}

\section{Pairs and Groups of Genetically Related\\Long-Period Comets}
Because of implications for the evolution of comets and a relevance to the
problem of fragmentation, comet pairs and groups have been of scientific
interest for a long time.  While the genetic makeup of the Kreutz sungrazing
system is certain (Section~1), lists of pairs and groups of other comets,
based on apparent orbital similarity, were published, for example, by
Pickering (1911) (49~pairs and 17~larger groups) and by Porter (1952, 1963)
(more than a dozen groups).  The issue of genetically associated comets got
{\vspace{-0.06cm}}to the forefront of scientific debate in the 1970s, when
a major statistical investigation by \"{O}pik (1971) led to his startling
conclusion that at least 60\% of all comets with aphelia beyond $\sim$10~AU
were members of one of a large number of comet groups.  In response, Whipple
(1977), combining a Monte Carlo approach with probability methods, countered
that a random sample of comets has{\vspace{-0.04cm}} clumping properties very
similar to those of \"{O}pik's set of observed comets and that there is no
compelling evidence for any overabundance of genetically related comets.
Supporting Whipple's conclusion, Kres\'ak (1982) found no true comet pairs
at all.

\subsection{The First Pair}
By coincidence, strong evidence for the existence of pairs of genetically
related long-period comets --- other than the Kreutz sungrazers --- became
available several years after the debate ended.  The circumstances clearly
favored Whipple's (1977) conservative approach.

The first comet pair consisted of C/1988~F1 (Levy)~and C/1988~J1
(Shoemaker-Holt).  They were discovered within two months of each other and
their perihelion times were only 76~days apart.  Their osculating orbits,
computed by Marsden (1989b) and listed in Table~1, were virtually identical,
yet their true orbital period, defined by the original barycentric semimajor
axis, was close to 14\,000~yr.  The striking orbital similarity and a very
small temporal separation both suggested that the two comets still had been
part of a single body during the previous return to perihelion around
12\,000~BCE --- and for a long time afterwards.

On the other hand, the relative motions of C/1988~F1 and C/1988~J1 are
inconsistent with those of fragments of an ordinary split comet except when
observed many years after separation.  Survival of secondary fragments (or
companions) over such long periods of time is untypical --- weeks or
months are the rule --- among the split short-period comets, the most notable
exceptions being 3D/Biela (Marsden \& Sekanina 1971; Sekanina 1977, 1982)
and {\vspace{-0.04cm}}73P/Schwassmann-Wachmann (Seka\-nina 2006).\footnote{A
record of sorts might be held by the short-period comets 42/Neujmin and
53P/Van Biesbroeck, which according to Carusi et al.'s (1985) orbital
integrations had had nearly identical orbits before 1850, implying that
they are fragments of a single comet that split at that time near Jupiter.
However, the proposed genetic relationship of the two comets appears to be
a matter of ongoing discussion (e.g., Pittichov\'a et al.\ 2003).}  And
the gap of 76~days in the perihelion time, minuscule in terms of the orbital
period of C/1988~F1, is unusually protracted when compared to an average
split comet, whose fragments are commonly separated by a fraction of a day
or a few days at~the~most.

The motion of a companion relative to the primary is known to be affected by
both an orbital-momentum change at breakup (resulting in a nonzero separation
velocity) and a continuous outgassing-driven differential nongravitational
acceleration after separation (Sekanina 1978, 1982).  As the dominant radial
component of the nongravitational acceleration causes the less massive ---
and usually the intrinsically fainter --- companion to start trailing the
more massive primary in the orbit, the brighter leading fragment represents
a signature of the role of the nongravitational forces at work.\footnote{There
are rare exceptions to this rule, but even then the more massive fragment
is intrinsically fainter for only a limited time.} We now examine the two
mechanisms as potential triggers of the 76~day gap between C/1988~F1 and
C/1988~J1.

\subsubsection{Effects of a separation velocity}
Well-determined separation velocities of split comets' fragments are known
to range from $\sim$0.1 to $\sim$2~m~s$^{-1}$ (e.g., Sekanina 1982).  Keeping
this in mind, we employ a simple model to demonstrate that the fragmentation
event that led to the birth of C/1988~F1 and C/1988~J1 could not have occurred
during the previous return to perihelion.  We begin by combining the virial
theorem with Kepler's third law,
\begin{eqnarray}
V_{\rm frg} & \,=\, & c_0 \, \sqrt{\frac{2}{r_{\rm frg}}-\frac{1}{a}}\;,
 \nonumber\\[0.1cm]
P & \,=\, & c \, a^{\frac{3}{2}},
\end{eqnarray}
where $r_{\rm frg}$ and $V_{\rm frg}$ are, respectively, the heliocentric
distance and the orbital velocity at the time of fragmentation, $a$ is
the semimajor axis of the orbit, $P$ is the orbital period, and $c_0$
and $c$ are constants.  Expressing $V_{\rm frg}$ in m~s$^{-1}$,
$r_{\rm frg}$ and $a$ in AU,{\vspace{-0.04cm}} and $P$ in yr, the values of
the constants are \mbox{$c_0 = 2.978 \times 10^4$} and \mbox{$c = 1$}.  By
differentiating both equations, requiring that \mbox{$r_{\rm frg} \ll 2 a$},
and writing an increment $\Delta (1/a)$ in terms of an increment $\Delta P$
of the orbital period and $1/a$ in terms of $P$, we obtain the following
expression for the relationship between $\Delta V$, a separation velocity
(in m~s$^{-1}$) in the direction of the orbital-velocity vector, and
$\Delta P$:
\begin{equation}
\Delta V = \frac{c_0 c^{\frac{2}{3}} \sqrt{2}}{6} \, r_{\rm frg}^{\frac{1}{2}}
\, P^{-\frac{5}{3}} \Delta P = C \, r_{\rm frg}^{\frac{1}{2}} \,
P^{-\frac{5}{3}} \Delta P,
\end{equation}
where $C$ is a constant whose value is 7019 when $\Delta P$ is in yr or
19.22 when in days.  For the pair of C/1988~F1 and C/1988~J1, we find from
Table~1 \mbox{$P \simeq 14\,000$ yr} and \mbox{$\Delta P = 76.3$ days}, so
that the inferred separation velocity (in m~s$^{-1}$) in the direction of
the orbital-velocity vector becomes
\begin{equation}
\Delta V = 0.00018 \, r_{\rm frg}^{\frac{1}{2}}.
\end{equation}
Since \mbox{$a_{\rm orig} \simeq 580$ AU}, $r_{\rm frg}$ is to be {\it
much\/} smaller than 1160~AU and the separation velocity at the time of
fragmentation {\it near the previous perihelion\/} is predicted to be
\mbox{$\Delta V < 0.006$ m s$^{-1}$}, orders of magnitude lower than the
least separation velocities for the split comets.  Only if the separation
velocity should be {\it exactly\/} normal to the orbital-velocity vector,
rather an absurd premise, might this condition be satisfied.  A much more
plausible scenario is C/1988~F1 and C/1988~J1 having separated from their
common parent fairly recently, when it already was on its way from aphelion
to the 1987 perihelion.

This argument is fully supported by a more rigorous treatment of the
effects on the companion's perihelion time that are caused by its separation
velocity (reckoned relative to the primary) acquired upon the parent comet's
fragmentation.  Let the primary's perihelion time be $t_\pi$, its perihelion
distance $q$, its eccentricity \mbox{$e < 1$}, and its parameter \mbox{$p
= q (1 \!+ e)$}.  Furthermore, let {\boldmath $P$}, {\boldmath $Q$},
{\boldmath $R$} be the unit orbit-orientation vectors with the components
\mbox{$P_{\rm x}, \ldots, R_{\rm z}$} in the right-handed ecliptical
coordinate system.  We wish to determine the temporal gap between the
perihelion passages of the companion and the primary,
\begin{equation}
\Delta t_\pi = t_{\pi}^{\prime} \!-\! t_\pi,
\end{equation}
where $t_{\pi}^{\prime}$ is the companion's perihelion time; a positive
$\Delta t_\pi$ signifies that the companion trails the primary.

With the planetary perturbations and nongravitational forces ignored, the
conditions at the time of fragmentation, $t_{\rm frg}$, are described as
follows:\ the comet's heliocentric distance is $r_{\rm frg}$ and the position
vector's coordinates in the ecliptical system are \mbox{$(x_{\rm frg}, y_{\rm
frg}, z_{\rm frg})$}; the primary fragment's true anomaly is \mbox{$v_{\rm
frg} =  \arccos [(p/r_{\rm frg} \!-\! 1)/e]$}, its orbital velocity is $V_{\rm
frg}$, and the components of its velocity vector in the ecliptical system are
\mbox{$(\dot{x}_{\rm frg}, \dot{y}_{\rm frg}, \dot{z}_{\rm frg})$}; and the
companion's separation velocity relative to the primary is $U$, its components
in the cardinal directions of the {\bf RTN} coordinate system,\footnote{This
is a right-handed orthogonal coordinate system, whose origin is the primary
fragment, with the {\bf R} axis pointing radially away from the Sun, the
%
%
%
transverse {\bf T} axis located in the primary's orbital plane, and the {\bf N}
axis normal to this plane.} $U_{\rm R}$, $U_{\rm T}$, and $U_{\rm N}$, being
related to the ecliptical components $U_{\rm x}$, $U_{\rm y}$, and $U_{\rm z}$
by
\begin{equation}
\left[\! \begin{array}{c}
U_{\rm x} \\
U_{\rm y} \\
U_{\rm z}
\end{array}
\!\right] \!\!=\!\! \left[\! \begin{array}{ccc}
P_{\rm x} & \! Q_{\rm x} & \! R_{\rm x} \\
P_{\rm y} & \! Q_{\rm y} & \! R_{\rm y} \\
P_{\rm z} & \! Q_{\rm z} & \! R_{\rm z}
\end{array}
\!\right] \!\cdot\! \left[\! \begin{array}{ccc}
\cos v_{\rm frg} \! & -\sin v_{\rm frg} & \! 0 \\
\sin v_{\rm frg} \! & \cos v_{\rm frg} & \! 0 \\
0 & 0 & \! 1
\end{array}
\! \right] \!\cdot\! \left[\! \begin{array}{c}
U_{\rm R} \\
U_{\rm T} \\
U_{\rm N}
\end{array}
\!\right] \!\!.\! \!
\end{equation}
The ecliptical components of the companion's orbital velocity vector at time
$t_{\rm frg}$ are
\begin{equation}
\left[\! \begin{array}{c}
\dot{x}_{\rm frg}^{\prime} \\
\dot{y}_{\rm frg}^{\prime} \\
\dot{z}_{\rm frg}^{\prime}
\end{array}
\! \right] \!=\! \left[\! \begin{array}{c}
\dot{x}_{\rm frg} \\
\dot{y}_{\rm frg} \\
\dot{z}_{\rm frg}
\end{array}
\! \right] \!+\! \left[\! \begin{array}{c}
U_{\rm x} \\
U_{\rm y} \\
U_{\rm z}
\end{array}
\! \right] \!,
\end{equation}
and its orbital velocity $V_{\rm frg}^{\prime}$:
\begin{equation}
V_{\rm frg}^{\prime} = \sqrt{ \left(\dot{x}_{\rm frg}^{\prime}\right)^{\!2}
 \!\!+\! \left(\dot{y}_{\rm frg}^{\prime} \right)^{\!2}
 \!\!+\! \left(\dot{z}_{\rm frg}^{\prime} \right)^{\!2}}
\end{equation}
The companion's orbital eccentricity, $e^{\prime}$, is equal to
\begin{equation}
e^{\prime} = \sqrt{1 + p^{\prime} \!\left[\!\left(\:\!\!\frac{V_{\rm
 frg}^{\prime}}{k}\! \right)^{\!\!2} \!\!-\!\frac{2}{r_{\rm frg}} \right]},
\end{equation}
where $k$ is the Gaussian gravitational constant,
\begin{equation}
p^{\prime} = \frac{\Re_{\rm xy}^2\!+\!\Re_{\rm yz}^2\!+\!\Re_{\rm zx}^2}{k^2},
\end{equation}
and
\begin{equation}
\Re_{\rm xy} \!=\! \left|\! \begin{array}{cc}
x_{\rm frg} \! & \! y_{\rm frg} \\
\dot{x}_{\rm frg}^{\prime} \! & \! \dot{y}_{\rm frg}^{\prime}
\end{array}
\!\!\right| \! , \, \Re_{\rm yz} \!=\! \left|\! \begin{array}{cc}
y_{\rm frg} \! & \! z_{\rm frg} \\
\dot{y}_{\rm frg}^{\prime} \! & \! \dot{z}_{\rm frg}^{\prime}
\end{array}
\!\!\right| \! , \, \Re_{\rm zx} \!=\! \left|\! \begin{array}{cc}
z_{\rm frg} \! & \! x_{\rm frg} \\
\dot{z}_{\rm frg}^{\prime} \! & \! \dot{x}_{\rm frg}^{\prime}
\end{array}
\!\!\right| \! .
\end{equation}
The companion's perihelion distance is \mbox{$q^{\prime} = p^{\prime}/(1 \!+\!
e^{\prime})$} and its true anomaly $v_{\rm frg}^{\prime}$ at $t_{\rm frg}$ is
computed from 
\begin{eqnarray}
\sin v_{\rm frg}^{\prime} & = & \frac{\sqrt{p^{\prime}}}{kr_{\rm frg}e^{\prime}}
 \left( x_{\rm frg}\dot{x}_{\rm frg}^{\prime} \!+\! y_{\rm frg}\dot{y}_{\rm
 frg}^{\prime} \!+\! z_{\rm frg}\dot{z}_{\rm frg}^{\prime} \right) \! ,
 \nonumber \\[0.1cm]
\cos v_{\rm frg}^{\prime} & = & \frac{1}{e^{\prime}} \! \left( \!
 \frac{p^{\prime}}{r_{\rm frg}} \! - \! 1 \! \right) \! .
\end{eqnarray}
The eccentric anomaly at $t_{\rm frg}$, $E_{\rm frg}^{\prime}$, then comes
out to be
\begin{equation}
 E_{\rm frg}^{\prime} = 2 \arctan \! \left( \! \sqrt{\frac{1 \!-\!
 e^{\prime}}{1 \!+\! e^{\prime}}} \tan {\textstyle \frac{1}{2}} v_{\rm
 frg}^{\prime} \! \right) \!
\end{equation}
and the companion's perihelion time is finally derived from the relation
\begin{equation}
t_{\pi}^{\prime} = t_{\rm frg} - \frac{E_{\rm frg}^{\prime} \!- e^{\prime}
 \sin E_{\rm frg}^{\prime}}{k} \! \left( \!\frac{q^{\prime}}{1\!-\!
 e^{\prime}} \! \right)^{\! \frac{3}{2}} \!\! .
\end{equation}
Inserting from Equation (13) into (4), we obtain the temporal separation
between the companion and the primary at perihelion, $\Delta t_\pi$, the
quantity we were looking for.

\begin{table}[t]
\vspace{-0.17cm}
\noindent
\begin{center}
{\footnotesize {\bf Table 2}\\[0.1cm]
{\sc Separation Velocity Needed for 76.274 Days Gap Between\\Perihelion
 Arrival Times of C/1988 F1 and C/1988 J1\\(Heliocentric Osculating
Approximation).}\\[0.2cm]
\begin{tabular}{c@{\hspace{0.5cm}}c@{\hspace{0.4cm}}c@{\hspace{0.5cm}}c@{\hspace{0.3cm}}c}
\hline\hline\\[-0.25cm]
\multicolumn{3}{@{\hspace{-0.15cm}}c}{Parent comet's fragmentation} &
\multicolumn{2}{@{\hspace{-0.05cm}}c}{Separation velocity} \\[-0.06cm]
\multicolumn{3}{@{\hspace{-0.15cm}}c}{\rule[0.6ex]{4.65cm}{0.4pt}}
 & \multicolumn{2}{@{\hspace{-0.05cm}}c}{\rule[0.6ex]{3.35cm}{0.4pt}}\\[-0.06cm]
distance & \multicolumn{2}{@{\hspace{-0.3cm}}c}{time before perihelion}
 & radial, & transverse, \\
$r_{\rm frg}$\,(AU) & (yr) & (units\,of\,$P)$ & $U_{\rm R}$\,(m\,s$^{-1}$)
 & $U_{\rm T}$\,(m\,s$^{-1}$) \\[0.08cm]
\hline\\[-0.2cm]
200 & $-$227.5 & $-$0.0183 & +4.30 & {\hspace{-0.148cm}}+14.95 \\
300 & $-$431.2 & $-$0.0346 & +1.80 & +7.76 \\
400 & $-$688.3 & $-$0.0552 & +0.94 & +4.80 \\
500 & $\;$\llap{$-$}1002.5 & $-$0.0805 & +0.56 & +3.25 \\
600 & $\;$\llap{$-$}1381.6 & $-$0.1109 & +0.35 & +2.33 \\
700 & $\;$\llap{$-$}1840.0 & $-$0.1477 & +0.23 & +1.72 \\
800 & $\;$\llap{$-$}2404.1 & $-$0.1929 & +0.16 & +1.30 \\
900 & $\;$\llap{$-$}3129.0 & $-$0.2511 & +0.11 & +0.97 \\
\llap{1}000 & $\;$\llap{$-$}4176.3  & $-$0.3352 & +0.06\rlap{7} & +0.70 \\
\llap{1}073\rlap{.74$^{\rm a}$} & $\;$\llap{$-$}6230.0 & $-$0.5000
 & +0.03\rlap{4} & +0.43 \\
\llap{1}000 & $\;$\llap{$-$}8283.7 & $-$0.6648 & +0.01\rlap{9} & +0.28 \\
900 & $\;$\llap{$-$}9331.0 & $-$0.7489 & +0.01\rlap{4} & +0.22 \\
800 & $\!\!\!\:$\llap{$-$}10055.9 & $-$0.8071 & +0.01\rlap{1} & +0.18 \\
700 & $\!\!\!\:$\llap{$-$}10620.0 & $-$0.8523 & +0.00\rlap{94} & +0.15 \\
600 & $\!\!\!\:$\llap{$-$}11078.4 & $-$0.8891 & +0.00\rlap{79} & +0.13 \\
500 & $\!\!\!\:$\llap{$-$}11457.5 & $-$0.9195 & +0.00\rlap{66} & +0.10 \\
400 & $\!\!\!\:$\llap{$-$}11771.7 & $-$0.9448 & +0.00\rlap{55} & +0.08\rlap{2}\\
300 & $\!\!\!\:$\llap{$-$}12028.8 & $-$0.9654 & +0.00\rlap{45} & +0.06\rlap{1}\\
200 & $\!\!\!\:$\llap{$-$}12232.5 & $-$0.9817 & +0.00\rlap{34} & +0.04\rlap{0}\\
100 & $\!\!\!\:$\llap{$-$}12381.4 & $-$0.9937 & +0.00\rlap{23}
 & +0.02\rlap{0}\\[0.05cm]
\hline\\[-0.2cm]
\multicolumn{5}{l}{\parbox{8.1cm}{$^{\rm a}$\,{\scriptsize Aphelion.}}}
%
%
\end{tabular}}
\end{center}
\end{table}

We note that, as byproducts of the computations, we also determined the
effects of the companion's separation velocity on the perihelion distance,
\mbox{$\Delta q = q^{\prime} \!-\! q$}, and eccentricity \mbox{$\Delta e =
e^{\prime} \!-\!  e$}.  It is similarly possible to derive the separation
velocity's effects on the remaining elements --- the inclination, the
longitude of the ascending node, and the argument of perihelion.

For C/1988 F1 and C/1988 J1 these computations show in Table 2 that a gap
of 76 days in the time of arrival at perihelion is achievable with a
separation velocity of $\sim$1~m~s$^{-1}$ in the direction away from the
Sun at a heliocentric distance of barely 400~AU, less than 700 yr before
the 1987 perihelion.  The same effect is also achievable with a separation
velocity of 1~m~s$^{-1}$ in the direction perpendicular to the radial
direction in the orbital plane at a heliocentric distance of less than
900~AU some 3000~yr before the 1987 perihelion.  Only the component of the
separation velocity that is normal to the orbital plane cannot bring about
a gap of this magnitude.

\begin{table}[t]
\vspace{-0.17cm}
\noindent
\begin{center}
{\footnotesize {\bf Table 3}\\[0.1cm]
{\sc Comparison of Effects of Separation Velocity\\at 402.8 AU from the Sun and
Planetary Perturbations\\on Orbital Elements of Comet C/1988 F1\\(Epoch 1987
November 21.0 TT).}\\[0.08cm]
\begin{tabular}{l@{\hspace{0.15cm}}c@{\hspace{0.3cm}}c@{\hspace{0.3cm}}c@{\hspace{0.3cm}}c}
\hline\hline\\[-0.2cm]
Increment & Rigorous
 & \multicolumn{3}{@{\hspace{0cm}}c}{Separation velocity
 (m~s$^{-1}$) alone}\\[-0.03cm]
in orbital & integration
 & \multicolumn{3}{@{\hspace{0cm}}c}{\rule[0.6ex]{5.05cm}{0.4pt}}\\[-0.03cm]
element & of orbit$^{\rm a}$ & $U_{\rm R}\!=\!+1.00$ & $U_{\rm T}\!=\!+5.02$
        & $U_{\rm N}\!=\!+0.10$ \\[0.05cm]
\hline\\[-0.15cm]
$\Delta t_\pi$\,(day) &  +80.3136\, & +82.1665\, & +82.1665\, & +0.0007\, \\
$\Delta \omega$   &  +0$^\circ\!$.0004 & +0$^\circ\!$.0029 & $-0^\circ\!$.2186
                  &  $-0^\circ\!$.0125 \\
$\Delta \Omega$   & $-0^\circ\!$.0038 & $\;\;\:0^\circ\!$.0000
                  & $\;\;\:0^\circ\!$.0000 & +0$^\circ\!$.0272 \\
$\Delta i$        & $-0^\circ\!$.0025 & $\;\;\:0^\circ\!$.0000
                  & $\;\;\:0^\circ\!$.0000 & $-0^\circ\!$.0444 \\
$\Delta q$\,(AU)  & +0.000073 & +0.000003 & +0.106562 & \llap{+}0.000001 \\
$\Delta e$      & +0.000014\rlap{2} & $-$0.000004\rlap{4} & $-$0.000196\rlap{6}
                & $\;\:$0.0000000 \\[0.05cm]
\hline\\[-0.2cm]
\multicolumn{5}{l}{\parbox{8.15cm}{$^{\rm a}$\,{\scriptsize Increments are the
sums of an effect due to a separation velocity of \mbox{$U_{\rm R} \!=\! +1.00$
m s$^{-1}$} and the planetary perturbations.{\vspace{0.2cm}}}}}
\end{tabular}}
\end{center}
\end{table}

It is noted in Table 2 that, based on the radial component of the separation
velocity, the fragmentation of the parent of C/1988~F1 and C/1988~J1 most
probably occurred between 300~AU and 900~AU and not longer than 0.25 the
orbital period before the 1987 perihelion, possibly as recently as the 14th
century AD.  If it had occurred much earlier, the radial component would
have to have been more than an order of magnitude smaller than the transverse
component, an unlikely scenario.

Since we neglected the planetary perturbations, we checked the accuracy of
our results against a rigorous orbit-determination run.  We integrated the set
of elements of C/1988~F1 from Table~1 back in time to 1300 January 1, when
the comet was 402.8~AU from the Sun, added exactly 1~m~s$^{-1}$ to the radial
component of its orbital velocity, and integrated forward in time to the epoch
of 1987 November 21.  The effects on the elements are shown in column~2 of
Table 3.  The perihelion passage came out to take place on 1988 February
18.2664 TT; the computations based on Equations~(4)--(13) show a perihelion
time of 1988 February 20.1137 TT, differing by $\sim$2\% of the 80.3~day gap
and indicating that the effect of planetary perturbations is almost two orders
of magnitude {\it smaller\/} than the separation-velocity effect.

Table 3 summarizes the test results for the effects of separation-velocity not
only on the perihelion time, but on the other elements as well.  We note from
columns~2 and 3 that the perihelion time $t_\pi$ is the only orbital element
for which the effect of the radial component of the separation velocity
dominates the planetary perturbations.  The tabulated values of $\Delta \Omega$
and $\Delta i$ are the net effects due to the perturbations, because no change
can be triggered by a radial separation velocity in a direction off the
orbital plane that these elements reflect.

In the penultimate column of Table 3 we list the orbital increments that would
be triggered by a transverse component of the separation velocity generating
the same effect in the perihelion time as a 1~m~s$^{-1}$ radial component.
Striking are the large changes in $\omega$ and $q$, which are not observed.
The last column shows the increments due to a small normal component of the
separation velocity.  The exceptional similarity of the angular elements and
perihelion distance of the orbits of C/1988~F1 and C/1988~J1 (Table~1)
suggests that the transverse and normal components of the separation velocity
must have been very small compared to the radial component $U_{\rm R}$ and
almost certainly much smaller than 0.1~m~s$^{-1}$.

\begin{figure}[t] 
\vspace{0.2cm}
\hspace{-0.2cm}
\centerline{
\scalebox{0.73}{
\includegraphics{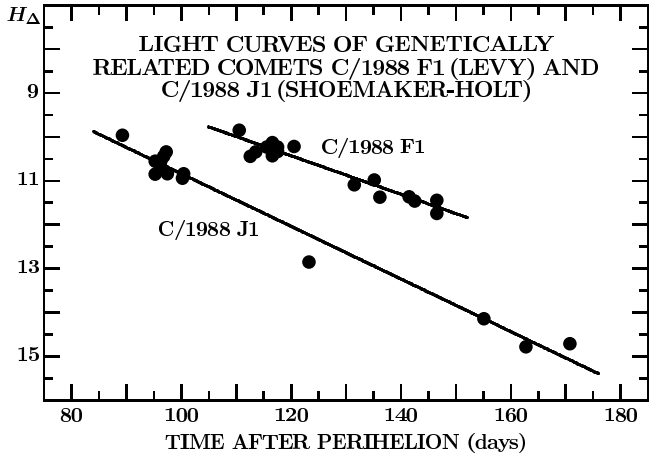}}} 
\vspace{-0.05cm}
\caption{The light curves of the genetically related comets C/1988 F1 and
C/1988 J1. Plotted against time (reckoned from perihelion of each comet) are
the total magnitudes $H_\Delta$, normalized to a unit geocentric distance,
approximately corrected for personal and instrumental effects, and reduced to
the photometric system of D.\,Levy, who made most of the plotted observations.
The data were collected from a number of issues of the {\it International
Comet Quarterly\/} and from the {\it Minor Planet Circulars\/} [only a
selection of magnitudes referred to as total (T)].{\vspace{0.35cm}}}
\end{figure}

\subsubsection{Effects of a nongravitational acceleration}

Our reconstruction of the light curves of C/1988~F1 and C/1988~J1 is shown
in Figure 1, its caption describing the sources of brightness data used.  The
plot leaves no doubt that C/1988~F1, the leading fragment, was intrinsically
brighter than C/1988~J1 by more than 1~mag; the latter object was also fading
at a steeper rate, making the brightness ratio of C/1988~F1 to C/1988~J1 to
increase with time.  Thus, the relative positions of the two fragments in the
orbit are consistent with the fainter (and presumably less massive) one having
been decelerated relative to the brighter (and presumably more massive) one.
Accordingly, the temporal separation of 76~days between both comets could have
been affected to a degree by a differential nongravitational acceleration.
We now investigate this problem in more detail.

We begin by assuming that the common parent~of~com\-ets C/1988 F1 and
C/1988~J1 had moved in a gravitational orbit identical with that of C/1988~F1,
fragmenting on its way from aphelion to perihelion at a time $t_{\rm frg}$,
when it was at a heliocentric distance $r_{\rm frg}$ (Section 2.1.1).  If the
primary fragment, C/1988~F1, arrived at perihelion at time $t_\pi$, the length
of the orbital arc that it traveled between $t_{\rm frg}$ and $t_\pi$ is equal
to
\begin{equation}
\ell = \!\! \int_{t_{\rm frg}}^{t_\pi} \!\! V_{\rm gr} (t) \, dt,
\end{equation}
where $V_{\rm gr}(t)$ is the orbital velocity of the primary's gravitational
motion at time $t$.  We now assume that the orbital motion of the companion
C/1988~J1 has been --- ever since the secession from the parent (at a zero
separation velocity) --- affected by a nongravitational acceleration pointing
in a general direction against the orbital motion.  During the period of time
from $t_{\rm frg}$ to $t_\pi$ the orbital arc traveled by the companion is
accordingly shorter by $\Delta \ell$,
\begin{equation}
\ell - \Delta \ell = \!\! \int_{t_{\rm frg}}^{t_\pi} \!\! V_{\rm ng}(t) \, dt,
\end{equation}
where $V_{\rm ng}(t)$ is the orbital velocity of the companion's
nongravitational motion.  Let $V_{\rm ng}(t)$ be a sum of a gravitational
part, which is equal to $V_{\rm gr}(t)$, the orbital velocity of the
primary fragment from Equation~(14), plus the nongravitational part,
\mbox{$\Delta\!\!\:V_{\rm ng}(t) = V_{\rm ng}(t) \!-\! V_{\rm gr}(t) < 0$},
which is equal to an effect of a nongravitational acceleration in the
direction of the orbital-velocity vector integrated from $t_{\rm frg}$ to
$t$.  Subtracting Equation~(15) from Equation~(14), introducing a ratio
\mbox{$z = q/r$} (where $r$ is the heliocentric distance at time $t$) as
a new integration variable, and approximating the gravitational orbit by
a parabola, we readily find
\begin{equation}
\Delta \ell = - \frac{q^{\frac{3}{2}}}{k\sqrt{2}} \int_{z_{\rm frg}}^{z_\star}
 \!\!\Delta\:\!\! V_{\rm ng}(z) \, z^{-\frac{5}{2}} (1\!-\!  z)^{-\frac{1}{2}}
 \, dz,
\end{equation}
where $k = 0.0172021$ AU$^{\frac{3}{2}}$\,day$^{-1}$ is the Gaussian
gravitational constant, $q$ is again the perihelion distance of the primary's
orbit, \mbox{$z_{\rm frg} = q/r_{\rm frg}$}, and \mbox{$z_\star = q/r_\star$},
where $r_\star$ is the heliocentric distance of the companion at the time the
primary is at perihelion.  The companion still has \mbox{$\Delta t_\pi =
76.274$}~days to go to its perihelion and the condition that $z_\star$ should
satisfy is
\begin{equation}
\Delta t_\pi = \frac{q^{\frac{3}{2}}\sqrt{2}}{3k} \, z_\star^{-\frac{3}{2}}
 (1 \!+\!2z_\star) \sqrt{1 \!-\! z_\star}.
\end{equation}
Next we consider separately the radial, $j_{\rm R}$, and transverse,
$j_{\rm T}$, components of the nongravitational acceleration in the orbital
plane of the {\bf RTN} right-handed coordinate system (Section 2.1.1).
In the Marsden et al.'s (1973) Style II formalism, they are written in the
following form:
\begin{equation}
\left[ \!
\begin{array}{c}
j_{\rm R}(r) \\
j_{\rm T}(r)
\end{array}
\!\! \right] = \left[ \!\!
\begin{array}{c}
A_1 \\
A_2
\end{array}
\!\! \right] g(r),
\end{equation}
where $g(r)$ is the standard empirical nongravitational law that the formalism
employs,
\begin{equation}
g(r) = 0.1113\,(r/r_0)^{-2.15}\!\left[1\!+\!(r/r_0)^{5.093}\right]^{-4.6142}\!,
\end{equation}
with a scaling distance $r_0 = 2.808$ AU appropriate for an acceleration
driven by the sublimation of water ice.  The law is normalized to
\mbox{$g(1\,{\rm AU}) = 1$}, so that the parameters $A_1$ and $A_2$ are
the radial and transverse accelerations at a unit heliocentric distance.
\begin{table*}[t]
\vspace{0.1cm}
\begin{center}
{\footnotesize {\bf Table 4}\\[0.08cm]
{\sc  Orbital Elements of Comets C/1988 A1 (Liller), C/1996 Q1 (Tabur),
and C/2015 F3 (SWAN) (Equinox J2000.0).}\\[0.05cm]
\begin{tabular}{l@{\hspace{0.55cm}}c@{\hspace{0.45cm}}c@{\hspace{0.85cm}}c}
\hline\hline\\[-0.22cm]
Quantity/Orbital element & Comet C/1988 A1$^{\rm a}$ & Comet C/1996 Q1
                         & Comet C/2015 F3 \\[0.08cm]
\hline\\[-0.25cm]
Osculation epoch (TT) & 1988 Mar.\,20.0 & 1996 Nov.\,13.0$\:\!^{\rm b}$
 & 2015 Feb.\,27.0 \\
Time of perihelion passage $t_\pi$\,(TT) & 1988 Mar.\,31.11442
 & 1996 Nov.\,3.53123$\:\pm\:$0.00274 & 2015 Mar.\,9.35613$\:\pm\:$0.00077 \\
Argument of perihelion $\omega$ & 57$^\circ\!$.38762
 & 57$^\circ\!$.41221$\:\pm\:$0$^\circ\!$.00314 
 & 57$^\circ\!$.56668$\:\pm\:$0$^\circ\!$.00121 \\
Longitude of ascending node $\Omega$ & 31$^\circ\!$.51540
 & 31$^\circ\!$.40011$\:\pm\:$0$^\circ\!$.00102
 & 31$^\circ\!$.63895$\:\pm\:$0$^\circ\!$.00042 \\
Orbit inclination $i$ & 73$^\circ\!$.32239
 & 73$^\circ\!$.35626$\:\pm\:$0$^\circ\!$.00119
 & 73$^\circ\!$.38679$\:\pm\:$0$^\circ\!$.00023 \\
Perihelion distance $q$ & 0.8413332 & \,\,\,0.8398052$\:\pm\:$0.0000139
                                    & \,\,\,0.8344515$\:\pm\:$0.0000121 \\
Orbital eccentricity $e$ & 0.9965647 & \,\,\,0.9986821$\:\pm\:$0.0001399
                                     & \,\,\,0.9964470$\:\pm\:$0.0000348 \\
\hspace*{2.75cm}osculation & 3832.7$\:\pm\:$9.4
                                & 16\,100$\:\pm\:$2600 
                                &  3599$\:\pm\:$53$\;\;\,$ \\[-0.259cm]
Orbital period (yr)
 $\!\!\left\{ \raisebox{1.5ex}{}\raisebox{-1.5ex}{} \right.$\\[-0.259cm]
\hspace*{2.75cm}original$\:\!^{\rm c}$ & 2933.0 & 10\,500 & 3300 \\[0.05cm]
\hline\\[-0.26cm]
Orbital arc covered by observations & 1988 Jan.\,12--1988 July 14
          & 1996 Aug.\,21--1996 Oct.\,18 & 2015 Mar.\,24--2015 May 28 \\
Number of observations employed & 100 & 199 & 283 \\
Root-mean-square residual & $\pm$0$^{\prime\prime}\!.8$
 & $\pm$1$^{\prime\prime}\!$.0 & $\pm$0$^{\prime\prime}\!$.8 \\
Orbit-quality code$^{\rm d}$ & 1B & 2B & 2A \\
Reference & Green (1988) & JPL database (Orbit 15)$^{\rm e}$ & Williams
 (2015) \\[0.05cm]
\hline\\[-0.26cm]
\multicolumn{4}{l}{\parbox{14cm}{\scriptsize $^{\rm a}$\,The errors of the
elements other than the orbital period are unavailable.}} \\[-0.08cm]
\multicolumn{4}{l}{\parbox{14cm}{\scriptsize $^{\rm b}$\,After integration
of the elements from an initial nonstandard epoch of 1996 Sept.\,18.0
TT.}} \\[-0.05cm]
\multicolumn{4}{l}{\parbox{17.2cm}{\scriptsize $^{\rm c}$\,With respect to
the barycenter of the Solar System; rigorous orbit integration gives for
$t_\pi$ at the previous return the date of $-$944 June 20.}}\\[-0.08cm]
\multicolumn{4}{l}{\parbox{14cm}{\scriptsize $^{\rm d}$\,Following the
classification system introduced by Marsden et al.\ (1978).}}\\[-0.05cm]
\multicolumn{4}{l}{\parbox{14cm}{\scriptsize $^{\rm e}$\,See {\tt
http://ssd.jpl.nasa.gov/sbdb.cgi}.}}\\[0.05cm]
\end{tabular}}
\end{center}
\end{table*}

The integrated contribution from the nongravitational acceleration to the
magnitude of the orbital-velocity vector, \mbox{$\Delta\:\!\! V_{\rm ng}$}
(now reckoned positive in the direction opposite the orbital motion), is
expressed, in a parabolic approximation, for any time $t$ (\mbox{$t_{\rm
frg} < t < t_\pi$}) along a preperihelion orbital arc by
\begin{equation}
\Delta\:\!\!V_{\rm ng}(t) = A_1\!\!\int_{t_{\rm frg}}^{t}\!\!g[r(t)]
 \sqrt{1\!-\!z}\,dt
\end{equation}
from the radial component and
\begin{equation}
\Delta\:\!\!V_{\rm ng}(t) = -A_2\!\!\int_{t_{\rm frg}}^{t} \!\!g[r(t)]
 \sqrt{z}\,dt
\end{equation}
from the transverse component.  Inserting Equations (20) and (21) into Equation
(16) and again replacing time with the ratio $z$ as an integration variable, we
find in the case that the effect is due entirely to the radial component
of the nongravitational acceleration:
\begin{equation}
\Delta \ell = \frac{A_1 q^3}{2k^2} \!\! \int_{z_{\rm frg}}^{z_\star} \!\!\!
 z^{-\frac{5}{2}} (1 \!-\!z)^{-\frac{1}{2}} \, dz\!\!\int_{z_{\rm frg}}^{z} \!
 \!\!g\!\left(\!\frac{q}{\zeta}\!\right) \zeta^{-\frac{5}{2}} \, d\zeta
\end{equation}
and similarly in the case that it is due entirely to the transverse component:
\begin{equation}
\Delta \ell =\!-\frac{A_2 q^3}{2k^2} \!\! \int_{z_{\rm frg}}^{z_\star} \!\!\!
 z^{-\frac{5}{2}} (1 \!-\!z)^{-\frac{1}{2}} \, dz\!\!\int_{z_{\rm frg}}^{z} \!
 \!\!g\!\left(\!\frac{q}{\zeta}\!\right) \zeta^{-2}(1\!-\!\zeta)^{-\frac{1}{2}}
 \, d\zeta.
\end{equation}
These expressions determine the parameters $A_1$ and~$A_2$, since the traveled
orbital-arc length $\Delta \ell$ is, in an adequate parabolic approximation,
a function of only $q$ and $z_\star$:
\begin{eqnarray}
\Delta \ell & = & k\!\!\int_{t_\pi}^{t_\pi^\prime}\!\!\!\!\!\:\sqrt{\frac{2}{r}}
\, dt = q \!\! \int_{z_\star}^{1} \! \frac{dz}{z^2 \sqrt{1\!-\!z}} =
2q \!\! \int_{0}^{\sqrt{z_\star^{-1}-1}} \!\!\!\! \sqrt{1 \!+\!x^2} \:dx
 \nonumber \\[0.15cm]
  & = & q \! \left( \! \frac{\sqrt{1 \!-\! z_\star}}{z_\star} + \log_{\rm e}\!
 \frac{1 \!+\! \sqrt{1 \!-\! z_\star}}{\sqrt{z_\star}} \right) \!.
\end{eqnarray}
With \mbox{$\Delta t_\pi = 76.274$}~days and \mbox{$q = 1.17418$}~AU,~we~find
\mbox{$z_\star = 0.70860$} (or \mbox{$r_\star = 1.65705$}~AU) from Equation~(17)
and \mbox{$\Delta \ell = 1.60358$}~AU from Equation (24).

The nongravitational parameters $A_1$ and $A_2$ are now functions of only the
heliocentric distance at fragmentation, $r_{\rm frg}$.  There is no measurable
contribution from the heliocentric distances much greater than the scaling
distance $r_0$ [Equation (19)], as the nongravitational law $g(r)$ falls off
precipitously at those distances.  This expectation is fully corroborated by
numerical integration of Equations~(22) and (23), resulting{\vspace{-0.03cm}}
in the nongravitational parameters \mbox{$A_1 = +0.00642$}~AU~day$^{-2}$
and \mbox{$A_2 = -0.00575$ AU day$^{-2}$} for any \mbox{$r_{\rm frg}
\mbox{\gapeq} 3.6$ AU}.  For smaller $r_{\rm frg}$ the magnitudes of
$A_1$ and $A_2$ are greater still.  The inferred minimum values $|A_1|$
and $|A_2|$ are found to exceed the gravitational acceleration of the Sun
at 1~AU by a factor of $\sim$20 or more and are meaningless.  We conclude
that the nongravitational effects in the motions of C/1988~F1 and C/1988~J1
cannot account for the gap of $\sim$76~days between their perihelion times.

These results, too, were confronted with rigorous orbital computations, which
showed in the first place that a positive radial nongravitational acceleration
($A_1 > 0$) would in fact make the companion C/1988~J1 pass through perihelion
{\it before\/} the primary C/1988~F1.  This is so because by reducing the
orbital velocity, this acceleration increases the perihelion distance and
significantly shortens the orbital period.

With reference to a negative radial nongravitational acceleration, the
rigorous orbital computations suggested that if a fragmentation event occurred,
for example, at 402.8~AU before perihelion, an acceleration described by
\mbox{$A_1 = -10^{-6}$ AU day$^{-2}$}, about 0.3\% of the Sun's gravitational
acceleration that would delay the perihelion time by merely 0.044~day, should
already be judged as unacceptably high in magnitude, because it would ---
contrary to the trivial differences between the orbital elements of C/1988~F1
and C/1988~J1 --- increase the argument of perihelion by 0$^\circ\!$.137 and
decrease the perihelion distance by 0.0014~AU.  If nearly comparable with the
Sun's gravitational acceleration, these nongravitational effects would make
the companion's orbit strongly hyperbolic.

Similarly, the transverse nongravitational acceleration described by
\mbox{$A_2 = -10^{-6}$ AU day$^{-2}$}, which would delay the passage through
perihelion by 0.24~day, is already unacceptably high in magnitude, because it
would increase the argument of perihelion by 0$^\circ\!$.227 and reduce the
perihelion distance by 0.0018~AU.  If nearly comparable with the Sun's
gravitational acceleration, these nongravitational effects would transform
the companion into a short-period comet of a small perihelion distance.

\subsection{A Trio of Genetically Related Comets}
Another comet that was discovered in early 1988 was C/1988~A1 (Liller), a
fairly, but not extraordinarily, active one.  We had to wait for more than
8~years to appreciate its special status.  In August of 1996, a newly
discovered comet C/1996~Q1 (Tabur) turned out to have an orbit nearly as
similar to that of C/1988~A1 as was the orbit of C/1988~J1 to that of
C/1988~F1.  And just months before this writing, yet another object was
discovered, C/2015~F3 (SWAN), also with an orbit nearly identical to those of
C/1988~A1 and C/1996~Q1.  The sets of orbital elements for the three comets
are compared in Table~4.  It is unfortunate that their quality is very uneven.
The best determined orbit is that of C/1988~A1, which was astrometrically
measured both before and after perihelion and the observed orbital arc was
just about six months.  Although the formal errors of the elements were not
--- except for the orbital period (or the semimajor axis) --- published,
their uncertainty is not greater than a few units of the last or penultimate
decimal place.  The second best determined orbit is that of C/2015~F3, whose
story of discovery and follow-up observation is peculiar (Green 2015).
Although its SWAN Lyman-alpha images were detected as early as 2015 March 3,
6~days before perihelion, the astrometric and brightness observations did
not commence until March 24, more than two weeks after perihelion.  The
delay notwithstanding, it was astrometrically measured over a period of
slightly more than two months.  By contrast, C/1996~Q1, whose orbit is least
well determined, was discovered more than 10~weeks before perihelion, but the
astrometric observations had to be terminated some eight weeks later, still
before perihelion, because of the loss of the nuclear condensation.  The
orbital arc's length covered by the astrometry was a few days short of two
months.

\begin{figure}[t] 
\vspace{0.18cm}
\hspace{-0.17cm}
\centerline{
\scalebox{0.441}{
\includegraphics{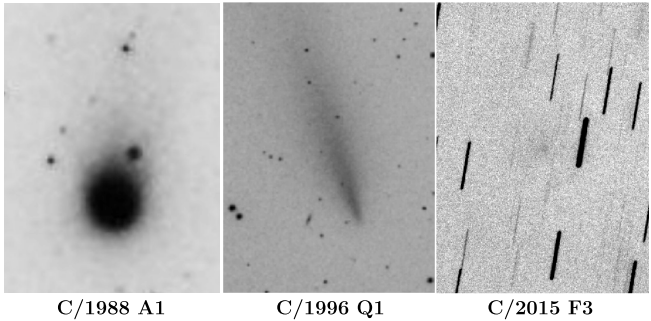}}} 
\vspace{-0.1cm}
\caption{Dramatically different appearance of three genetically related
comets.  All three images are 14$^\prime$ along a diagonal.  The
image of comet C/1988~A1  is a 12-minute exposure obtained by M.\,Oates
from the suburbs of Manchester, England, with a 6.7-cm f/4.5 camera on
1988 April 23.974~UT (see {\tt http://www.manastro.
co.uk/members/contrib/moates/metcomet/liller88.htm}),~about 24~days after
perihelion.  The comet was 1.31~AU~from~the~Earth and 0.95~AU from the
Sun. --- The image of comet C/1996~Q1 is a 5-minute exposure{\vspace{-0.07cm}}
by H.\,Miku\v{z} with a 36-cm f/6.8 Schmidt-Cassegrain telescope of the
\v{C}rni Vrh Observatory near Idrija,~Slo\-ve\-nia.  At the time of
{\vspace{-0.025cm}}observation, 1996 November 9.709~UT (see {\tt
http://www.observatorij.org/}), the comet was 6~days past perihelion, 0.99~AU
from the Earth and 0.85~AU from the Sun.  --- The image of comet C/2015~F3
is a sum of ten 3-minute exposures obtained by G.\,Masi remotely with a 43-cm
f/6.8 robotic unit of the Virtual Telescope Project, Ceccano, Italy, on
2015 May 18.944 UT (see {\tt http://www.virtualtelescope.eu/2015/05/19/
comet-c2015-f3-swan-image-18-2015}), some 70~days after perihelion.  At
this time the comet was 1.16~AU from the Earth and 1.48~AU from the Sun.
(Credits:\ M.\,Oates; \v{C}rni Vrh Ob\-serva\-tory; Virtual Telescope
Project, Bellatrix Astronomical Observatory.{\vspace{0.35cm}})}
\end{figure}

In spite of being fragments of a common parent, the three comets had a very
different appearance, morphology, and behavior.  This is illustrated by
both the sample images in Figure~2 and by the light curves in Figure~3.

\begin{figure*} 
\centerline{
\scalebox{0.73}{
\includegraphics{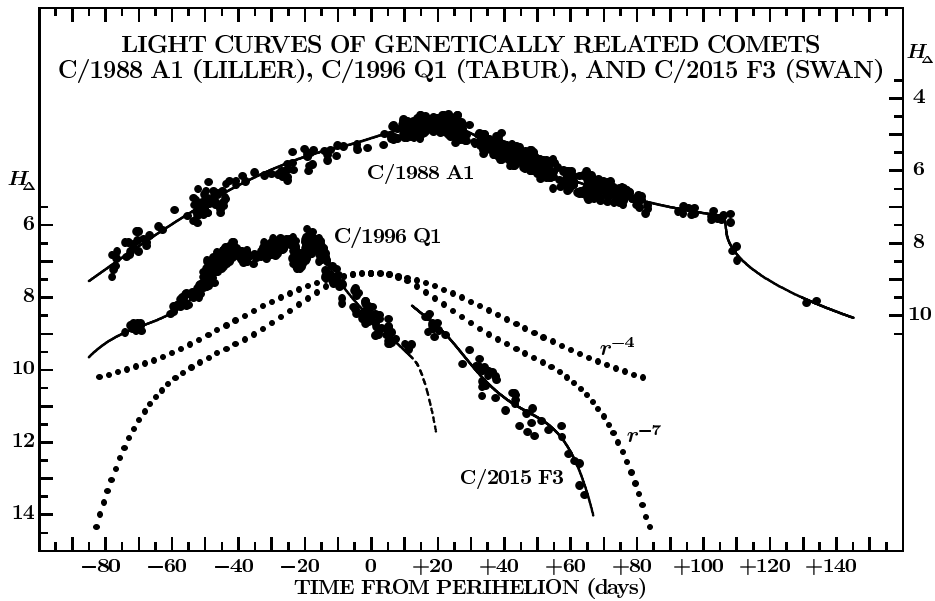}}} 
\vspace{-0.05cm}
\caption{Light curves of three genetically related comets C/1988~A1, C/1996
Q1, and C/2015 F3.  Plotted against time (reckoned from perihelion of each
comet) are the total magnitudes $H_\Delta$, normalized to a unit geocentric
distance, approximately corrected for personal and instrumental effects, and
reduced to the photometric system of experienced visual observers.  The
magnitude scale on the right applies to C/1988~A1, the scale on the left,
shifted by 2~magnitudes upward, to C/1996~Q1 and C/2015~F3.  The data were
collected from issues of the {\it International Comet Quarterly\/}; for
C/2015~F3, from the ICQ website and the {\it Minor Planet Circulars\/} [only
selected magnitudes referred to as total (T)].  The light curve of C/1996~Q1
does not show an unconfirmed observation by a single observer, made 20~days
after perihelion (Marsden 1996), which was nominally inconsistent with a nearly
{\vspace{-0.05cm}}simultaneous nondetection by another observer and with the
comet's systematic fading trend.  The dotted curves show the variations
following the laws $r^{-4}$ and $r^{-7}$, respectively.{\vspace{0.45cm}}}
\end{figure*}

Comet C/1988 A1, presented in Figure 2 as it looked near its peak intrinsic
brightness, displays a coma of at least 3$^\prime$, equivalent to 170\,000~km,
in diameter and appears to be healthy and active.  The light curve in
Figure~3 shows that its post-perihelion fading was, at least at heliocentric
distances smaller than 2~AU, substantially less steep than its preperihelion
brightening and that the intrinsic brightness peaked some three weeks after
perihelion.  Overall, the light curve was fairly smooth, again especially
within 2~AU of the Sun.

The behavior of C/1996 Q1 was, in several respects, contrary to that of
C/1988~A1.  The image of C/1996~Q1 in Figure~2, from November 9, six days
after perihelion, consists of a headless long narrow tail, without
any nuclear condensation, which explains the termination of astrometric
observations long before this image was taken.  The light curve in Figure~3
is quite erratic, displaying at least three outbursts between 40 and
20~days before perihelion.  The first of them is seen to have
begun more than 50 days before perihelion and for more than 10~days in this
span of time C/1996~Q1 was intrinsically {\it brighter\/} than C/1988~A1
at the same heliocentric distance (note the shift in the brightness scale
in Figure~3).  The last of the three outbursts was accompanied by major
morphological changes in the comet's head, with the condensation disappearing
within a few days following the event's peak.  It was at this time, about
16~days before perihelion, that all astrometric observations terminated.  The
light curve shows a continuous steep decline of brightness ever since the
peak of the third outburst.  The last confirmed magnitude estimates were
obtained about two weeks after perihelion, but extremely faint ``ghost''
images of the comet's residual tail were marginally detected\footnote{See,
e.g., {\tt http://www.observatorij.org}.} as late as mid-January of 1997.

Comparison of C/2015 F3 with C/1996 Q1 suggests only one common trait:\ both
comets were fading much more rapidly than C/1988~A1, obviously the primary,
most massive fragment of their common parent.  Otherwise C/2015~F3 differed
from C/1996~Q1 significantly.  First of all, C/2015~F3 survived perihelion
seemingly intact.  In fact, relative to perihelion, this comet was~not even
discovered in the SWAN images (Green 2015) by~the time C/1996~Q1 had already
disintegrated.  Second, the light curve of C/2015~F3 is in Figure~3 located
above~the light curve of C/1996~Q1.  Third, no prominent tail was observed to
have survived the head of C/2015~F3, as illustrated in Figure~2 by one of the
images in its advanced phase of evolution.  It was not until some 50 or so
days after perihelion that this comet began to lose its nuclear condensation,
which resulted in an accelerated fading (Figure~3).  And, fourth, there is no
evidence that this rapid fading was triggered by an outburst, as it was in
the case of C/1996~Q1.

\subsubsection{Fragmentation of the parent}
With the previous return to perihelion computed to have occurred in 945 BCE
(note c in Table 4), we find that both C/1996~Q1 and C/2015~F3 are likely to
have separated from C/1988~A1 in their common parent's fragmentation event or
events in a general proximity of that perihelion.  Equation (2) implies that
the gaps of 3139.42~days between the perihelion times of C/1988~A1 and
C/1996~Q1 {\vspace{-0.04cm}}and 9839.24~days between C/1988~A1 and C/2015~F3
require separation velocities (in m~s$^{-1}$) along the orbital-velocity
vector of, respectively,
\begin{eqnarray}
\Delta \!\!\:V & = & 0.092 \sqrt{r_{\rm frg}/q} \;\;\;
 {\rm for\:C/1996\:Q1\,vs\:C/1988\:A1},\;\;\;
 \nonumber \\
\Delta \!\!\:V & = & 0.289 \sqrt{r_{\rm frg}/q} \;\;\;
 {\rm for\:C/2015\:F3\,vs\:C/1988\:A1}.\;\;\;
\end{eqnarray}
Since the direction of the separation velocity vector {\boldmath $V$}$_{\!\rm
sep}$ generally differs from the direction of the orbital-velocity
vector, {\boldmath $V$}$_{\!\rm frg}$, at the time of fragmentation, it is
always \mbox{$V_{\rm sep}$ = {\boldmath $\!|V$}$_{\!\rm sep}| \geq \Delta
\!\!\:V$}.  If $\theta$ is an angle between{\nopagebreak} the two vectors,
then \mbox{$\cos \theta = \Delta \!\!\:V/V_{\rm sep}$} and a statistically
averaged angle $\langle \theta \rangle$ is given by integrating over a
hemisphere centered on the direction of the orbital-velocity vector,
\begin{equation}
\frac{\pi}{2}\,\langle \cos \theta \rangle = \!\!\int_{0}^{\frac{\pi}{2}} \!\!
 \sin \theta \cos \theta \,d\theta = \frac{1}{2} ,
\end{equation}
so that $\langle \cos \theta \rangle = 1/\pi$ and a statistically averaged
separation velocity \mbox{$\langle V_{\rm sep} \rangle = \pi \Delta \!\!\:V$}.

Fragmentation events are most likely to occur at, but are not limited to,
times not too distant from the perihelion time.  Considering, for example,
that the parent comet of C/1988~A1, C/1996~Q1, and C/2015~F3 split at a time
of up to a year around perihelion, equivalent to heliocentric distances of
less than $\sim$5~AU, an overall range of implied velocities $\Delta \!\!\:V$
for the two companions is from $\sim$0.1 to $\sim$0.7~m~s$^{-1}$ and a range
for the separation velocities $V_{\rm sep}$ is from $\sim$0.3 to
$\sim$2~m~s$^{-1}$, in perfect harmony with expectation (Section 2.1.1).

\subsubsection{Comparison of C/1988 A1 and C/1996 Q1 in terms of their physical
 properties}

The manifest sudden termination of activity of comet C/1996 Q1 about 16~days
before perihelion is independently documented by the water production data
listed in Table~5.  The numbers also show that the production of water from
C/1996~Q1 was apparently stalling for at least 10 days before its termination,
that 10~days before perihelion the water production of C/1988~A1 was more than
10~times as high as that of C/1996~Q1, and that 43 days after perihelion
C/1988~A1 produced about as much water as C/1996~Q1 19~days before perihelion.

The image of C/1996 Q1 in Figure 2 is one of three that Fulle et al.\ (1998)
used to study the comet's dust tail.  The employed method (Fulle 1989) allowed
them to{\nopagebreak} determine, among others, the production rate of dust as
a function of time.  For a bulk density of dust particles of 1~g~cm$^{-3}$
and a product of their geometric albedo and phase function (with the phase
angles of 65$^\circ$ to 75$^\circ$ at the imaging times) of 0.02, Fulle et
al.'s effort resulted in four similar dust production curves, one of which is
plotted in Figure~4.  Because the employed method has a tendency to smooth
short-term variations, minor outbursts are suppressed.  This is the reason why
the curve in the figure does not show the steep increase in the amount of dust
in the comet's atmosphere around September~14, 50~days before perihelion (an
event that was called attention to by Lara et al.\ 2001), and the terminal
outburst one month later shows up as a relatively small bulge.

\begin{table*}
\vspace{0.1cm}
\begin{center}
{\footnotesize {\bf Table 5}\\[0.06cm]
{\sc  Water Production Rates from Comets C/1988 A1 (Liller) and C/1996 Q1
 (Tabur).}\\[0.06cm]
\begin{tabular}{c@{\hspace{0.7cm}}c@{\hspace{0.5cm}}c@{\hspace{0.2cm}}c@{\hspace{0.6cm}}c@{\hspace{0.9cm}}c@{\hspace{0.9cm}}l}
\hline\hline\\[-0.24cm]
      & Time from  & Distance & Production &        & Active         & \\
      & perihelion & from$\:$Sun & rate$\:$of$\:$H$_2$O  & Method
      & area$^{\rm b}$ & \\
Comet & (days) & (AU) & (10$^5$\,g/s) & used$^{\rm a}$ & (km$^2$)
      & Reference \\[0.06cm]
\hline\\[-0.22cm]
C/1988 A1 & $-$10\rlap{$\:\!^{\rm c}$} & 0.86 & 35 & OH\,(radio) & 5.5 
          & Crovisier et al.\,(2002) \\
          & +40          & 1.12 & 13\rlap{.7} & OH\,(UVB) & 3.8
          & A'Hearn et al.\,(1995) \\[0.1cm]
C/1996 Q1 & $-$30\rlap{$\:\!^{\rm d}$} & 1.01 & 16 & OH\,(radio) & 3.5
          & Crovisier et al.\,(2002) \\
      & $-$28 & 0.99 & 10 & H$_2$O$^+$(red) & 2.1 & Jockers et al.\,(1999) \\
          & $-$26\rlap{$\:\!^{\rm e}$} & 0.97 & 14 & OH\,(radio) & 2.8
          & Crovisier et al.\,(2002) \\
          & $-$23\rlap{$\:\!^{\rm f}$} & 0.94 & 11 & OH\,(radio) & 2.1
          & Crovisier et al.\,(2002) \\
          & $-$19                  & 0.92 & 12\rlap{.6} & H\,(EUV) & 2.3
          & M\"{a}kinen et al.\,(2001) \\
 & $-$10\rlap{$\:\!^{\rm g}$} & 0.86 & $\:\!\!\!<$3 & OH\,(radio)
          & \llap{$<$}0.5 & Crovisier et al.\,(2002) \\[0.05cm]
\hline\\[-0.22cm]
\multicolumn{7}{l}{\parbox{14.35cm}{\scriptsize $^{\rm a}$\,Water production
{\vspace{-0.03cm}}rate assumed 1.1 times hydroxyl productioni rate; OH\,(radio)
= from observation of hydroxyl emission at 18~cm with a radiotelescope at
Nan\c{c}ay;{\vspace{-0.07cm}} OH\,(UVB) = from narrow-band photometry of
hydroxyl emission at 3085~{\AA}; H$_2$O$^+$\,(red) = from{\vspace{-0.04cm}}
column-density maps of ionized water emission at 6153~{\AA}; H\,(EUV) = from
maps of Lyman-alpha emission of atomic hydrogen at 1216~{\AA} taken with the
SWAN imager on board SOHO.}}\\[-0.03cm]
\multicolumn{7}{l}{\parbox{14.35cm}{\scriptsize $^{\rm b}$\,Effective
sublimation area, when the Sun is at zenith.}}\\[-0.05cm]
\multicolumn{7}{l}{\parbox{14.35cm}{\scriptsize $^{\rm c}$\,Average from data
taken between 1988 March 17 and 24.}}\\[-0.09cm]
\multicolumn{7}{l}{\parbox{14.35cm}{\scriptsize $^{\rm d}$\,Average from data
taken between 1996 October 2 and 5.}}\\[-0.05cm]
\multicolumn{7}{l}{\parbox{14.35cm}{\scriptsize $^{\rm e}$\,Average from data
taken between 1996 October 7 and 9.}}\\[-0.09cm]
\multicolumn{7}{l}{\parbox{14.35cm}{\scriptsize $^{\rm f}$\,Average from data
taken between 1996 October 10 and 12.}}\\[0.02cm]
\multicolumn{7}{l}{\parbox{14.35cm}{\scriptsize $^{\rm g}$\,Average from data
taken between 1996 October 20 and 27.{\vspace{0.14cm}}}}
\end{tabular}}
\end{center}
\end{table*}

Comparison with the water production data in Figure~4 suggests a water-to-dust
mass production rate ratio of $\sim$15, so that C/1996~Q1 was an exceptionally
dust-poor comet.  Integrating the dust production curve from 100 days before
perihelion to 25 days after perihelion, Fulle et al.\ (1998) determined that
over this span of time the comet lost \mbox{$1.2 \times \!  10^{12}$}\,g of
dust.  An exponential extrapolation over the rest of the orbit (outlined by
the dashed curve in Figure 4) does not increase the mass to more than
\mbox{$1.4 \times \! 10^{12}$}\,g.  Assumptions of an overall constant
water-to-dust mass production rate ratio (15:1) and of water dominance among
volatile species [cf.\ Womack \& Suswal's (1996) unsuccessful search for
carbon monoxide] led to a conservative lower limit of \mbox{$2 \times \!
10^{13}$}\,g for this comet's total mass in the case of complete
disintegration.  Possible survival of a limited number of boulder-sized
fragments cannot significantly affect this estimate.

Fulle et al.\ (1998) alleged that the nucleus of comet C/1996~Q1 did not
fragment and that its activity ceased due to mantle formation, seasonal
variations, or depletion of volatile species, including water.  These authors
argued that the fairly high water production rate in the first half of October,
a month before perihelion (cf.\ Table~5), suggested the nuclear diameter was
greater than 0.7~km.  We remark that the implied bulk density would then have
been close to 0.1~g~cm$^{-3}$.  Contrary to Fulle et al.\ (1998), Wyckoff
et al.\ (2000) concluded from their observations with large telescopes that
the nucleus probably began to fragment around October~7, some four weeks
before perihelion, and that during the next several weeks it completely
dissipated.

Because the same method of dust-tail analysis was also applied to C/1988~A1
(Fulle et al.\ 1992), it is possible to compare the two genetically related
comets in terms of dust production.  The images of C/1988~A1 selected for
this study were taken with a 106-cm Cassegrain telescope at the Hoher List
Observatory in mid-May 1988 (Rauer \& Jockers 1990) and the adopted product
of the geometric albedo and the phase function was 0.06, that is, three times
higher than for C/1996~Q1, while the phase angle was 50$^\circ$.  Normalizing
to the same geometric albedo, the results show for C/1988~A1 a dust-mass loss
of \mbox{$1.4 \times \!10^{14}$}\,g integrated between 200~days before
perihelion and 40~days after perihelion; the total loss of dust over one
revolution about the Sun can be estimated at more than \mbox{$2 \times
\!10^{14}$}\,g.  Keeping in mind the factor of three, comparison of the dust
production curve with the water production 10 days before perihelion (Table~5)
suggests a water-to-dust mass production rate ratio of about~0.2.  Similarly,
at the time of the post-perihelion water production data point in the table,
the ratio is about 0.3.  Thus, with grains as dark as in C/1996~Q1, C/1988~A1
was a truly dust-rich comet, about 60 times dustier than C/1996~Q1, and its
total loss of mass per revolution was near \mbox{$3 \times\! 10^{14}$}\,g,
some 15~times higher than C/1996~Q1's.

\begin{figure}[b] 
\vspace{0.3cm}
\hspace{-0.25cm}
\centerline{
\scalebox{0.57}{
\includegraphics{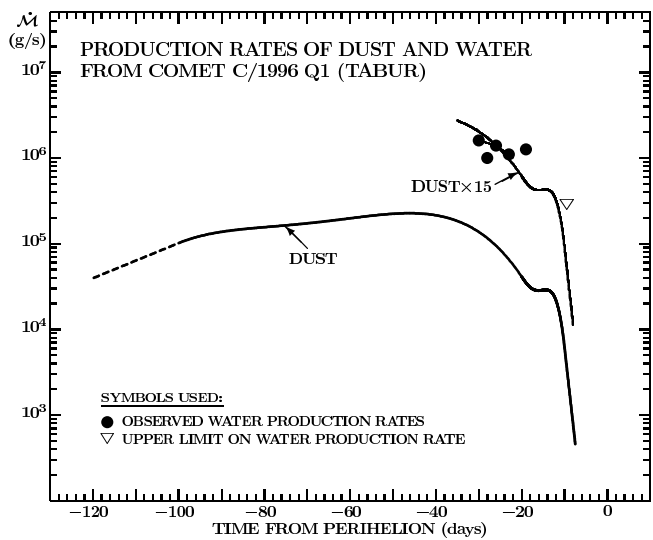}}} 
\vspace{-0.05cm}
\caption{Dust and water production from C/1996 Q1 as a function of time.
The thick curve is one of four similar models of dust emission, derived by
Fulle et al.\ (1998) from their analysis of three CCD images taken by
H.~Miku\v{z} on 1996 October~31 and on November~2 and 9 (see footnote earlier
in Section 2.2).  The employed method has a tendency to smooth short-term
variations in the production rate of dust, so that minor outbursts are
suppressed.  The dashed part of the curve is an exponential extrapo\-lation
of dust production for a period of time ending 100~days before perihelion.
The water production rates from Table~5 (solid circles; the three of them
that are connected together come from the radio observations) and an upper
limit on the production rate (an open triangle) are consistent with the thin
curve, which represents a mass production rate 15 times higher than the dust
rate.{\vspace{-0.05cm}}}
\end{figure}

It should be remarked that a strikingly different gas-to-dust production ratio
for the two genetically related comets was noticed by Kawakita et al.\ (1997)
after they compared the results of their spectroscopic observations of
C/1996~Q1, made between 1996 September 14 and October 16, with the results for
C/1988~A1 published earlier by Baratta et al.\ (1989) and by A'Hearn et al.\
(1995).  Because the data on the water and dust production of C/1996~Q1
(Crovisier et al.\ 2002; M\"{a}kinen et al.\ 2001; Fulle et al.\ 1998) were not
yet available, Kawakita et al.\ could determine this comet's gas-to-dust ratio
{\vspace{-0.03cm}}only~very approximately.  From a continuum flux at 5200--5300
{\AA} in their spectra they calculated the dust production's {\it Af}$\rho$
proxy (A'Hearn et al.\ 1984) and from C$_2(\Delta v\!=\!0$) band the production
rate $Q({\rm C}_2)$.  They found that in the same range of heliocentric distance
the ratio of \mbox{$Q({\rm C}_2)/{\it Af}\!\rho$} for C/1996~Q1 was four times
greater than for C/1988~A1.  Employing Fink \& DiSanti's (1990) technique for
deriving the water production rate from the intensity of{\vspace{-0.03cm}} the
forbidden [OI]$^1\!D$ line of atomic oxygen at 6300~{\AA}, Kawakita et al.\
computed the ratio $Q({\rm H}_2{\rm O})/{\it Af}\!\rho$ for C/1996~Q1 at the
same heliocentric distance at which it was known for C/1988~A1 from A'Hearn et
al.\ (1995), found that for C/1996~Q1 the ratio came out to be $\sim$15 times
higher; they concluded that the nucleus of the parent comet must have been
strongly inhomogeneous.\footnote{Kawakita et al.\ (1997) did not publish the
value of the water production rate that they obtained at 1.15~AU before
perihelion, which is the reason why we do not plot their result in Figure~4.
From their ratio of the water production rate to {\it Af}$\rho$ they inferred
a gas-to-dust mass ratio of 8.5, consistent, within a factor of two, with the
fit in Figure~4.}

The issue of whether the uneven dust content in the two comets correlates with
the chemical composition of volatile species was addressed by Turner \& Smith
(1999).  Comparing their own results for C/1996~Q1 from 1996 September 19--20
with those by A'Hearn et al.\ (1995) for C/1988~A1, they showed that the
production ratios C$_2$/CN, C$_3$/CN, and NH/CN were practically identical.
The only exception was the abundance of OH relative to these molecules,
which in C/1996~Q1 was an order of magnitude {\it lower\/}.  Turner \& Smith
admitted that their OH results, derived from the observations{\vspace{-0.07cm}}
of the \mbox{(0-0)} band at 3085~{\AA}, may have been burdened by large
errors.\footnote{Errors greatly exceeding one order of magnitude are certainly
implied by comparison not only with the water production rates based on a
variety of October observations, but also with the already discussed results
by Kawakita et al.\ (1997), who began to make their spectroscopic observations
five days before Turner \& Smith's run.  The latter's hydroxyl production
rates, equivalent to water production rates of \mbox{$2.3$--$3.4 \times \!
10^4$}\,g~s$^{-1}$ on September 19--20 are about an order of magnitude lower
than the dust production rate at the time, while the water production derived
by Kawakita et al.\ was very much above the dust curve.  For this reason,
Turner \& Smith's inferred water data points are not plotted in Figure~4.}
Ignoring this discrepancy, the authors arrived at a conclusion that their
results are consistent with a homogeneous composition for the ices in the
nucleus of the parent comet.

Mineralogical comparison of the dust content in the comets C/1988~A1 and
C/1996~Q1 is unfortunately not possible because the necessary data are not
available for the first object.  Examining the spectrophotometric data for
C/1996~Q1 in the thermal infrared between 7.6~$\mu$m and 13.2~$\mu$m,
Harker et al.\ (1999) found that a mineralogical model that best fits the
spectra obtained on October~8--10, more than three weeks before perihelion,
is a mix of grains of crystalline olivine 2~$\mu$m in diameter and amorphous
pyroxene 6~$\mu$m in diameter.  On the other hand, Turner \& Smith (1999)
allowed for the possible existence of gradually destroyed grains containing
water ice in order to explain the blue color of the continuum in the optical
spectrum at small distances from the nucleus on its sunward side.

It is too early to compare C/1988~A1 and C/1996~Q1 with C/2015~F3.  However,
the very facts that this new member of the group was first spotted due to
its Lyman-alpha images of atomic hydrogen but appeared fairly faint and with
no prominent dust tail to the ground observers suggests that, like C/1996~Q1,
this comet was most probably water-rich and dust-poor.

\section{The Lick Object of 1921 August 7}
Equipped with much information on the genetically related comet pairs and
groups, we are now ready to attack the problem of the celebrated 1921 Lick
object, whose detection caused much world-wide stir for a long time after
the initial report was issued but whose identity remains a mystery to this
day.

The story began with a party that gathered at the residence of
W.\,W.\,Campbell, then Director of the Lick Observatory, on Mount Hamilton
to watch the colorful setting Sun on 1921 August 7.  As the Sun's disk was
passing through the horizon haze, a member of the party, which also included
H.\,N.\,Russell, asked a question on the identity of a bright star-like
body 6~solar diameters (1 solar diameter at the time = 31$^\prime\!$.55)
to the east of the Sun.  As subsequent scrutiny by Campbell and Russell
eliminated the suspected bright planets, the two astronomers agreed that
a short message should be dispatched at once to the Harvard College
Observatory for immediate dissemination (Campbell 1921a, 1921b) and
expanded reports of the phenomenon prepared later for scientific journals
(Campbell 1921c, 1921d; Russell 1921).
 
Stellar even when briefly viewed through binoculars, brighter than Venus would
have been if seen in the same position and circumstances, and sharing the Sun's
motion in the sky, the object was located about 3$^\circ$ east and 1$^\circ$
south of the Sun (Campbell 1921c), at \mbox{R.A. = 9$^{\rm h}$\,22$^{\rm
m}\!$}, \mbox{Dec. = +15$^\circ\!$.5} (Campbell 1921d).  While no equinox
was explicitly stated, the equinox in use at the time was generally 1900.0.
The quoted coordinates put the Lick object ahead of the Sun by 3$^\circ\!$.2
in right ascension and below the Sun by 0$^\circ\!$.9 in declination, in
slightly better agreement with the offset estimate in declination (Campbell
1921c) than would the same coordinates at the equinox of the date.  As for
the observation time, we find that~at the top of Mount Hamilton the Sun set
at 3:14~UT (August 8), not 2:50~UT as claimed by Pearce (1921).  The time
derived by us includes the effects of the observing site's elevation above
sea level and atmospheric refraction and refers to the last contact with
the true horizon.
 
The object was unsuccessfully searched for at Lick both at sunrise and sunset
on August~8 and at sunrise on August~9 (Campbell 1921d).  Because of its high
galactic latitude ($\sim$40$^\circ$), the object was thought{\vspace{-0.03cm}}
to be more probably the head of a comet than a nova.\footnote{An additional
argument is that it is extremely unlikely for a nova to flare up at a position
nearly coinciding with that of the Sun.  On the other hand, the brightness of
comets is known to increase dramatically near the Sun.}  However, we ruled
out a member of the Kreutz sungrazer system, because at this time of the
year it should approach the Sun from the southwest.\footnote{See {\tt
http://www.rkracht.de/soho/c3kreutz.htm}.{\vspace{-0.16cm}}} Subsequently we
learned that a Kreutz sungrazer was eliminated as a potential candidate long
ago (Ashbrook 1971; Baum 2007).
 
Although Campbell and his Lick party were not --- as described below --- the
first to detect the object, they were the first to report it.  This is why
it is often referred to as the 1921 Lick object.

A number of reports appeared in worldwide response, but only a few of them
included helpful temporal and positional information.  We next list the more
interesting observations in the order of decreasing relevance.

An elaborate account of the circumstances of one such observation, originating
with Nelson Day (1921a, 1921b), lieutenant of Royal Naval Reserve, and made at 
Fern\-down, Dorset, England, was published by Markwick (1921), based on their
correspondence within weeks of the sighting.  This report conveys that, with
others, Nelson Day saw, in the evening of August 7, a bright object, allegedly
of magnitude $-$2, at an angular distance of 4$^\circ$ from the Sun, whose
elevation above the horizon was about 8$^\circ$ and bearing approximately
S\,45$^\circ$\,W relative to the object, referring presumably, as in ground
navigation, to the horizontal coordinate system (Section 4.3).  Although the
time of observation was not stated, Markwick estimated it at 19:00~UT.  We
calculate that at Ferndown the Sun was 8$^\circ$ above the horizon at 18:44~UT
and the sunset (the time of last contact) occurred at 19:44 UT.  Markwirk
judged this account credible and we tend to concur, with the exception of the
brightness that must be grossly underestimated. In Section~4.2 we argue that
an object of apparent magnitude $-2$ cannot be detected in daylight with the
naked eye 4$^\circ$ from the Sun.

Another report, by Emmert (1921a, 1921b), described a daylight observation
from Detroit, MI, on August 6 at 22:50 UT:\ the object was found to be
5$^\circ$ east of the Sun in azimuth and within 0$^\circ\!$.5 at the same
elevation above the horizon as the Sun, for which Emmert gave an azimuth
of 90$^\circ$ and an elevation of 15$^\circ$.  We compute that at the given
time of observation, the Sun's azimuth was 94$^\circ\!$.8 and its elevation
19$^\circ\!$.9.  The accuracy of Emmert's estimates thus cannot compete
with Campbell's, and his account does not offer credible information on the
object's absolute position, a conclusion that is supported by Pearce (1921).
At best, Emmert's observation offers a very approximate estimate of the
object's angular distance from the Sun more than one day before Campbell et
al.'s observation.  It should be remarked that the publication of Emmert's
(1912a) original report was dated as late as 1921 October 14 and that it
obviously was not available in time for publication in the preceding issue,
dated September~13, still more than 5 weeks after the event.  It appears
that Emmert tried to reconstruct his observation from memory long after he
made it, an effort that is notoriously unreliable.  By contrast, even though
communicated via a third person (thus involving extra delays in mailing),
the original letter of Nelson Day's (1921a) observation made a September 8
issue of the journal.  Their very unequal weights notwithstanding, Campbell's,
Nelson Day's, and Emmert's accounts offer a fairly consistent scenario for
the object's sunward apparent motion over a period of more than 28~hours.
 
Generally supportive, but of a still lesser weight, was another report from
England, by Fellows (1921a, 1921b, 1921c).  Observing on August 7 from
Wolverhampton, in the West Midlands, he detected with his binoculars a bright
object, elongated in the direction of the Sun and of a distinct reddish hue,
at about 20:30 UT, more than 30 minutes after sunset (for which we compute
19:53 UT as the time of last contact).  Fellows remarked that the object,
seen by him for only a few minutes, was very low above the horizon and
he judged it to be about 6$^\circ$ from the Sun and a little south of it.
Because the Sun had already set, the reported angular distance cannot be
more than an extremely crude guess and hardly of any diagnostic value.

Further reports appear to be even less relevant.  Observations of transient
luminous bands at the K\"{o}nigstuhl and Sonneberg Observatories and elsewhere
(Wolf 1921a; Hoffmeister 1921) after midnight UT from August~8 to~9, although
interpreted by some in the literature as the Lick object's tail, must have
been unrelated, because such phenomena were also observed during the night of
August 5/6 (Wolf 1921a).  Hoffmeister (1921) noted that the bands resembled
auroral streamers.  Wolf (1921b) was more inclined to a possible correlation
only because he suspected that the Lick object may also have been observed in
daylight at Plauen, Germany, on August 7.816 UT (at a distance of 27$^\circ$
away from the Lick position), in which case it would have to have been very
close to the Earth.  Since that object was different (and in all probablity
identical with Jupiter; Pearce 1921), an association between the Lick object
and the luminous bands can safely be ruled out.

An observation of similar phenomena was reported by Kanda (1922).
Between 9:35 and 9:55 UT on August~9 he noticed a band-like object
2$^\circ$ broad extending from about \mbox{R.A. = 9$^{\rm h}$\,48$^{\rm m}$},
\mbox{Dec. = +14$^\circ$} to \mbox{R.A. = 10$^{\rm h}$\,42$^{\rm m}$},
\mbox{Dec. = +30$^\circ$}.  He was observing from Simo-Sibuya, near Tokyo,
Japan, and speculated that the band could be a tail of the Lick object.
If this feature was located outside the Earth's atmosphere, it would have
hardly escaped attention of Campbell (1921d), who under superior observing
conditions at Lick saw nothing at sunrise on August 9 (the first contact
at 13:11 UT), about 3$\frac{1}{2}$~hr after Kanda's reported sighting.
Several additional reports could under no circumstances refer to the Lick
object and there is no point in listing them here.

\section{Searching for Potential Candidates of\\a Genetically Related Comet}
The fundamental idea in our quest for determining the identity and origin of
the 1921 Lick object is to search for a long-period comet, with which the
object could share the common orbit and thereby be genetically associated.
The task is to develop an approach strategy based on what we learned about
the comet pairs and groups in the preceding sections of this paper in order
to streamline and optimize the search.

A necessary condition for a genetically related comet is that its motion,
adjusted for a time of perihelion passage, should emulate the Lick object's
motion and therefore closely match its observed or estimated
%
%
position(s).  The issue is the degree of uncertainty that is introduced by
this orbital emulation, because the errors involved define the degree of
tolerance that is to be maintained in judging each candidate's chances.

An estimate of this uncertainty is illustrated on the known pair and trio of
the genetically related comets investigated, respectively, in Sections~2.1 and
2.2.  It makes no difference whether the incurred errors are established by
emulating a companion's orbit with a primary's orbit, or vice versa.  Choosing
the first option, we selected an arbitrary time near perihelion of the
companion and calculated its equatorial coordinates from its own set of orbital
elements.  Next, we tried to match this artificial position with the primary's
orbit, using the companion's optimized perihelion time.  This procedure
required the introduction of a correction $\Delta t_\pi$ to minimize the total
residual, $\Delta \Pi$, consisting of the residuals in right ascension,
$\Delta$R.A., and declination, $\Delta$Dec.  This exercise provided us with
estimates of the errors that ought to be tolerated when an effort is made to
emulate the companion's motion with the motion of the primary.

The results, in Table~6, indicate that for the pair of C/1988~F1 and C/1988~J1,
whose orbits in Table~1 are seen to differ hardly at all (as a result of a
relatively recent separation of the fragments; Section~2.1.1), the astrometric
error is well below 1$^\prime$ and the optimized perihelion time agrees with
the companion's to better than 0.01~day.  For the trio of the primary C/1988~A1
and the companions C/1996~Q1 and C/2015~F3, whose orbital elements in Table~4
differ often in the third significant digit (due to the fragments' separation
during their parent comet's previous return to the Sun; Section~2.2.1), the
astrometric errors are near 8$^\prime$ and the perihelion times differ by
up to several units of 0.01~day.  This appears to be the order of errors that
are introduced in a comet pair or group by emulating a member's motion by
the motion of another member.  We adopt accordingly that uncertainties in the
astrometric positions of the Lick object that ought to be tolerated in our
search for its genetically related comet are up to several arcminutes.

\begin{table}[t]
\vspace{0.1cm}
\begin{center}
{\footnotesize {\bf Table 6} \\[0.08cm]
{\sc Errors in Companion's Perihelion Time and Equatorial Coordinates and
 Position Introduced by Emulating Companion's Motion with Primary's
 Motion.}\\[0.1cm]
\begin{tabular}{c@{\hspace{0.4cm}}c@{\hspace{0.3cm}}c@{\hspace{0.2cm}}c@{\hspace{0.3cm}}c@{\hspace{0.35cm}}c}
\hline\hline\\[-0.25cm]
\multicolumn{2}{@{\hspace{0.05cm}}c}{Fragment}
 & \multicolumn{4}{@{\hspace{-0.15cm}}c}{Errors introduced by primary's
 orbit} \\[-0.05cm]
\multicolumn{2}{@{\hspace{0.05cm}}c}{\rule[0.7ex]{3.25cm}{0.4pt}}
 & \multicolumn{4}{@{\hspace{-0.15cm}}c}{\rule[0.7ex]{4.8cm}{0.4pt}} \\[-0.05cm]
primary & companion & $\Delta t_\pi$(days) & $\Delta$R.A. & $\Delta$Dec.
 & $\Delta \Pi$ \\[0.1cm]
\hline \\[-0.26cm]
C/1988 F1 & C/1988 J1 & +0.004 & +0$^\prime\!\!$.2 & $\;\;\,$0$^\prime\!\!$.0
 & 0$^\prime\!\!$.2 \\
C/1988 A1 & C/1996 Q1 & $-$0.059 & $-$5.8 & +5.1 & 7.7 \\
C/1988 A1 & C/2015 F3 & +0.013 & +6.5 & +5.1 & 8.3 \\[0.04cm]
\hline\\[-0.08cm]
\end{tabular}}
\end{center}
\end{table}

\subsection{Constraints Based on Campbell et al.'s Account}

An important point is raised by Campbell's (1921d) statement that the Lick
object was brighter than Venus in the same position and circumstances.
Since any small elongation, $E$, from the Sun is reached by Venus either
near the superior or inferior conjunction, we consulted the phase function
of the planet to derive its apparent brightness for those two scenarios.
With \mbox{$E = 3^\circ\!.3$} and the Earth's distance from the Sun of
\mbox{$r_\oplus = 1.01387$ AU} at the time, Venus' phase angle at its mean
heliocentric distance would have been 4$^\circ\!.7$ near the superior
conjunction and 175$^\circ\!$.3 near the inferior conjunction.

A history of the planet's phase function investigations from a standpoint
of {\it The Astronomical Almanac\/}'s needs was summarized by Hilton (2005).  A
recently updated phase curve, the only one that accounts for forward-scattering
effects of sulfuric-acid droplets in the planet's upper atmosphere and covers
all angles from 2$^\circ$ to 179$^\circ$, was published by Mallama et al.\
(2006).  Based on their results we find that at \mbox{$E = 3^\circ\!.3$} the
$V$ magnitude of Venus is $-$3.9 and $-$3.7 near the superior and inferior
conjunction, respectively.  Incorporating a minor color correction to convert
the $V$ magnitude to the visual magnitude (e.g., Howarth \& Bailey 1980) and
interpreting Campbell et al.'s statement to imply that the Lick object was at
least $\sim$0.5~magnitude brighter than Venus would have been, we adopted that
the object's apparent visual magnitude was about \mbox{$H_{\rm app} = -4.3$}
or possibly brighter.

Campbell et al.'s remarks on the Lick object allow us to examine the
constraints on its magnitude $H_0$, normalized to unit heliocentric and
geocentric distances (sometimes called an {\it absolute magnitude\/}),
as a function of the unknown heliocentric distance $r$ (in AU), on the
assumption that the object's brightness varied with heliocentric distance
as $r^{-n}$,
\begin{equation}
H_0 = -4.3 - 5 \log \Delta - 2.5 \, n \log r - 2.5 \log \Phi(\varphi),
\end{equation}
where $n$ is a constant and $\Delta$, $\varphi$, and $\Phi(\varphi)$ are,
respectively, the object's geocentric distance (in AU), phase angle, and
phase function.  The distances $r$ and $\Delta$ are constrained by the
observed elongation angle $E$,

\begin{equation}
r^2 = r_{\!\oplus}^2 + \Delta^2 - 2 r_\oplus \Delta \cos E,
\end{equation}
where $R_\oplus$ is again the earth's heliocentric distance at the time of
Campbell et al.'s observation.  Because the phase function depends on the
object's unknown dust content and the small elongation implies strong
effects of forward scattering by microscopic dust at the object's geocentric
distances \mbox{$\Delta \mbox{\lapeq} 0.95$ AU}, we constrained the phase
function by employing two extreme scenarios:\ dust-free and dust-rich.  In
the first case we put \mbox{$\Phi(\varphi) \equiv 1$} at all phase angles,
while in the second case we approximated $\Phi(\varphi)$ with the
Henyey-Greenstein law as modified by Marcus (2007) but normalized to a zero
phase angle \mbox{$[\Phi(0^\circ) = 1]$} rather than to 90$^\circ$.

We varied $\Delta$ from 0.001~AU to 2~AU to obtain a plot of the absolute
magnitude $H_0$ against $r$ in Figure~5 for three values of the exponent $n$;
a dust-free case is on the left, a dust-rich case on the right.  Although
the models differ from one another enormously at \mbox{$\Delta < 1$ AU}, some
properties of the plotted curves are independent, or nearly independent, of
the phase law.

\begin{figure*}[t]  
\vspace{0.15cm}
\begin{center}
\leftline{
\hspace{0.8cm}
\scalebox{0.6}{
\includegraphics{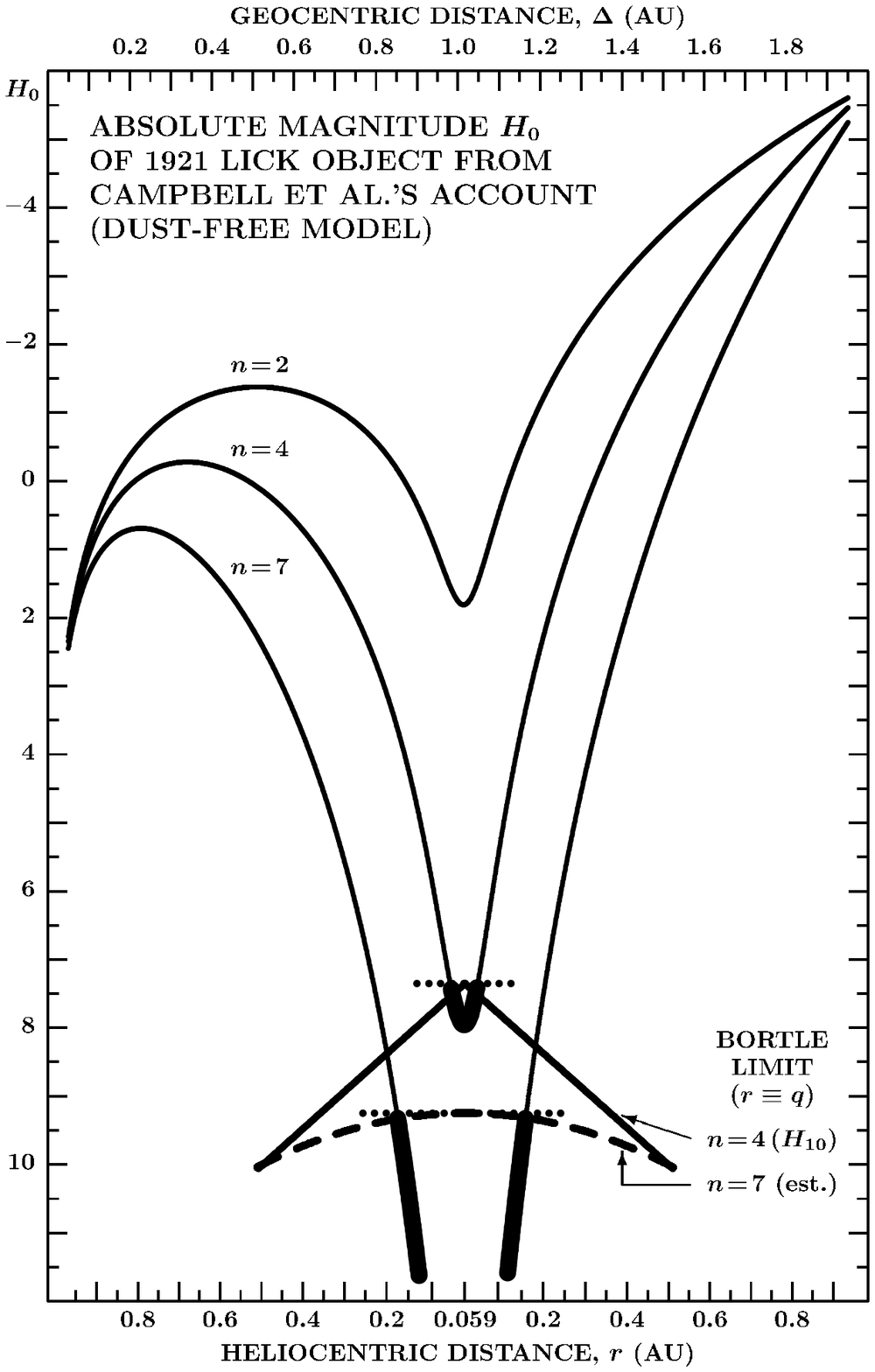}}} 

\vspace{-12.01cm}

\hspace{-2.1cm}
\rightline{
\scalebox{0.6}{
\includegraphics{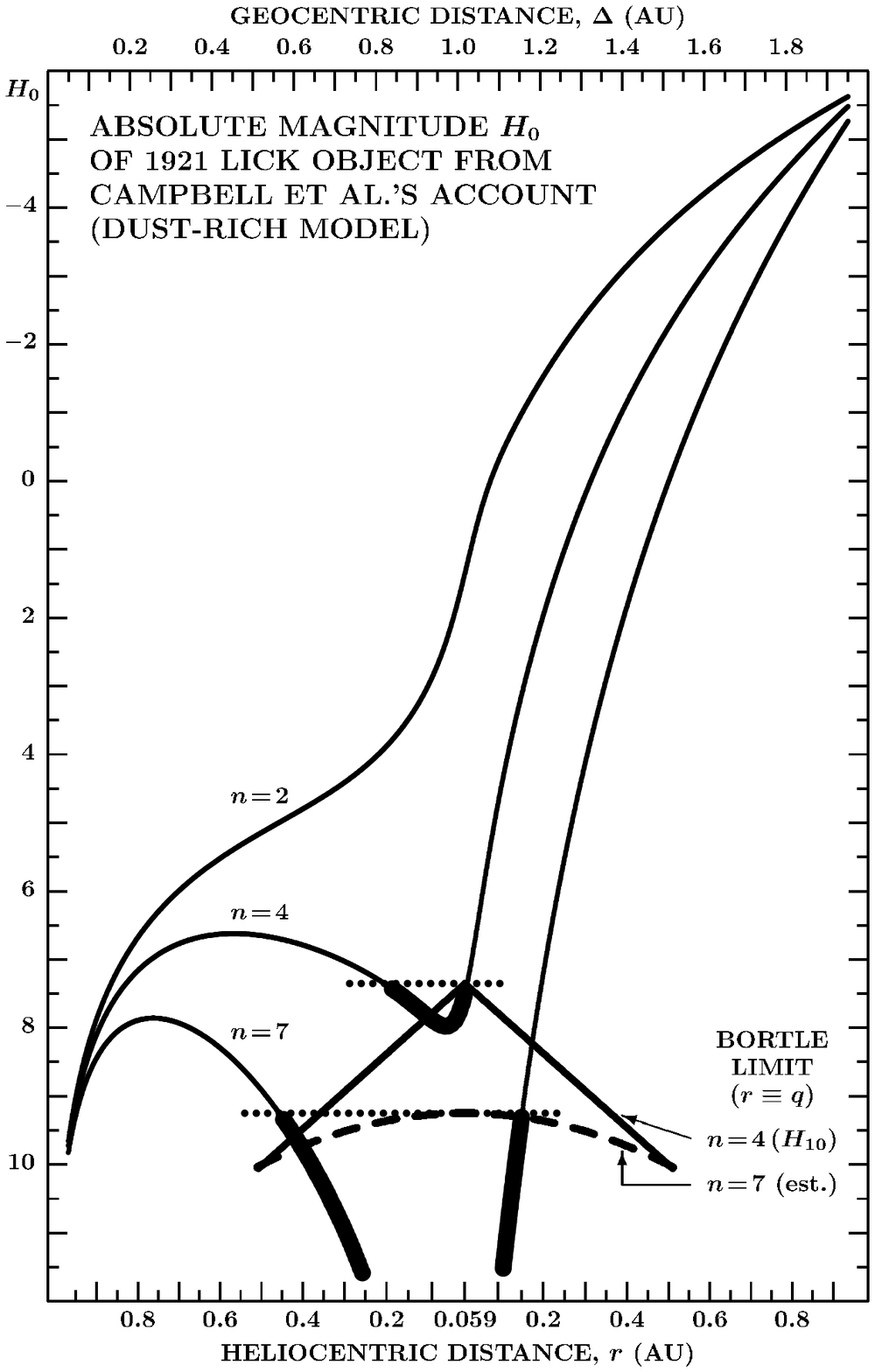}}} 
\vspace{-0.05cm}
\caption{Absolute magnitude $H_0$ of the 1921 Lick object as a function of its
assumed heliocentric distance at the time of Campbell et al.'s observation
of sunset on Mount Hamilton on August 8.1347 UT.  The $H_0$ variations are
plotted for three different photometric laws $r^{-n}$, with \mbox{$n = 2$, 4,
and 7} and for two dust-content scenarios:\ dust-free (the panel on the left)
and dust-rich.  The enormous divergence that the two display for geocentric
distances smaller than 1~AU is due to strong effects of forward scattering
by microscopic dust at large phase angles implied by the object's small
elongation. Also plotted for \mbox{$n = 4$} is Bortle's (1991) survival limit
\mbox{$H_{10} \equiv H_0(n \!=\! 4)$} for comets with perihelion distances
\mbox{$q \leq 0.51$ AU}, when the plotted $r$ is equal to $q$.  For \mbox{$r >
q$} the $H_{10}$ value of surviving comets lies along a horizontal (examples
for \mbox{$q = 0.059$ AU} shown by dots) that passes through the Bortle curve
at \mbox{$r = q$}.  For \mbox{$n = 7$} we estimated the survival limit (dashed
curve) on certain assumptions (see the text).  The heavy portions of the
magnitude curves for \mbox{$n = 4$} and \mbox{$n = 7$} show the ranges of
heliocentric distance within which the object is unlikely to have survived.
Statistically, about 70\% of comets with the intrinsic magnitude brighter
than the Bortle limit survive, and vice versa.  The magnitude limit for
\mbox{$q = 0$} is 7 for \mbox{$n = 4$} and it is estimated at 9.3 for
\mbox{$n = 7$}.{\vspace{0.2cm}}}
\end{center}
\end{figure*}

Rather odd-looking in Figure 5 are the magnitude curves for \mbox{$n = 2$},
typical of dynamically new comets arriving from the Oort Cloud (Whipple
1978), which generally imply an unacceptably high intrinsic brightness of
the Lick object.  In the dust-free scenario the solutions are way off, while
in the dust-rich model the acceptable options imply a very small geocentric
distance \mbox{$\Delta < 0.2$ AU}.  But such scenarios are ruled out by
(i)~being inconsistent with Campbell et al.'s observation of the object's
stellar appearance; (ii)~failing to explain why the object was not observed
systematically for a longer period of time around August~7 (at an expected
rate of motion of up to several degrees per day); and (iii)~explicitly
contradicting Campbell's failure to find it on the following days at the
times of sunrise and/or sunset.  It is therefore highly unlikely that the
Lick object was a dynamically new comet and it was not near the Earth.

Campbell's (1921d) unsuccessful search for the object on August~8--9 (Mount
Hamilton time) suggests that by then it may have disintegrated.  If
it was a member of a comet pair and {\it not\/} a dynamically new comet, it
should, like the members of the C/1988~A1 group, belong to Whipple's (1978)
class III (with an orbital period on the order of thousands of years), for
which Whipple found an average \mbox{$\langle n \rangle = 4.7 \pm 0.8$} before
perihelion.  Not unexpectedly, the preperihelion light curves of C/1988~A1 and
C/1996~Q1 (up to the onset of terminal fading) in Figure~3 show that they, too,
are fairly consistent with \mbox{$n = 4$} or slightly higher.  The magnitude
curves for \mbox{$n = 4$} in Figure~5 are thus expected to approximate class
III comets more closely than the other curves.  The possibility that the Lick
object disintegrated near perihelion, just as did C/1996~Q1, suggests that it
could~be a companion to the pair's more massive member, which was likely to
have arrived at perihelion long {\it before\/} the Lick object.  Regardless
of the dust content, Figure~5 shows that the minimum on the magnitude curves
for \mbox{$n = 4$} is \mbox{$(H_0)_{\rm min} \simeq 8$}.  Accordingly, the
Lick object should have had a small perihelion distance, \mbox{$q \mbox{\lapeq}
0.1$ AU}, and the small elongation reflected its true proximity to the Sun.
Since the minimum absolute magnitude varies with $n$ approximately as
\mbox{$(H_0)_{\rm min} = 3.09 \, n \!- \! 4.3$}, it equals $\sim$10 for
\mbox{$n = 4.7$}.  We remark that C/1996~Q1 had \mbox{$H_0 \simeq 6.6$}
before fading (Figure~3), being intrinsically about 0.7~magnitude fainter
than C/1998~A1 at $\sim$1~AU from the Sun.

The curve for \mbox{$n = 7$} is presented in Figure~5 as an example of an
unusually steep light curve --- displayed, e.g., by the sungrazing comet
C/2011~W3 (Sekanina \& Chodas 2012) --- and to illustrate a degree of
nonlinearity in the plot of the Lick object's $H_0$ magnitude.  Figure~3 shows
that the $r^{-7}$ law is  not consistent with the preperihelion light curve
of either C/1988~A1 or C/1996~Q1 and that it could be a plausible match only
to the post-perihelion light curve of C/2015~F3.

To learn more about the Lick object's possible disintegration, we recall that
a failure of long-period comets to survive perihelion was investigated, among
others, by Bortle (1991).  For such comets with perihelion distances smaller
than 0.51~AU he found that the chance of disintegration depends on the
intrinsic brightness.  By fitting the same $r^{-4}$ law to the light curves
of the numerous comets that he examined, Bortle estimated that there was
a 70\% probability of disintegration for the comets that were intrinsically
fainter than the absolute magnitude $H_{10}^{\rm surv}$ (equivalent in our
notation to $H_0$ for \mbox{$n = 4$} and \mbox{$\Phi = 1$}).  This survival
limit is according to Bortle related to the perihelion distance $q$ (in AU)
by \mbox{$H_{10}^{\rm surv} = 7 \!+\! 6 \,q$}.  It is plotted in Figure~5
against $r$, thus assuming the equality between the plotted $r$ and the
perihelion distance.  For \mbox{$r > q$}, $H_{10}$ lies along a horizontal
line that passes through the Bortle curve at \mbox{$r = q$}.  The Bortle limit
suggests that comets intrinsically brighter than \mbox{$H_{10} = 7$} have an
increasingly higher probablility to survive, even if their perihelia are
within, say, 0.1~AU of the Sun.

On certain assumptions, the Bortle limit can readily be extended to photometric
laws $r^{-n}$ with \mbox{$n \neq 4$}.  For example, if an orbital arc covered
by the magnitude observations, employed to determine the absolute magnitude,
is centered on a heliocentric distance $r_{\rm aver}$ for which \mbox{$\log
r_{\rm aver} = \log(q + 0.5)$}, the survival limit $H_{2.5n}^{\rm surv}$ is
related to $H_{10}^{\rm surv}$ by \mbox{$H_{2.5n}^{\rm surv} = H_{10}^{\rm
surv} + (10 \!-\!  2.5 n) \log (q + 0.5)$}.  In Figure~5 we display, as a
dashed curve, a survival limit $H_{17.5}^{\rm surv}$ for the magnitude
curve of \mbox{$n = 7$}.  For perihelion distances smaller than 0.5~AU,
$H_{17.5}^{\rm surv}$ is fainter than the $H_{10}^{\rm surv}$ and its
variation with the perihelion distance is much flatter.

While the minimum on the magnitude curve provides a lower bound to the Lick
object's absolute brightness, the Bortle limit offers a sort of an upper
bound, if the object's disintegration is considered as very likely.  Figure~5
shows that for \mbox{$n = 4$} the two bounds almost coincide, leaving a
narrow allowed range from 8.1 to 7.3.  In the case of a dust-poor comet,
such as was C/1996~Q1 (Section~2.2.2), the magnitude curve for \mbox{$n = 4$}
in Figure~5 suggests that the Lick object's perihelion distance did not exceed
0.07~AU, while the same curve in the case of a dust-rich comet allows
a perihelion distance of up to 0.20~AU.  For the unlikely magnitude curves
with \mbox{$n = 7$} the limits on the perihelion distance would be,
respectively, 0.17~AU and 0.46~AU.

Given that the above bounds are not absolute, a search for the Lick object's
pair member should allow for some leeway.  However, it ought to be pointed out
that if the apparent brightness assigned to the object, based on Campbell's
comparison to Venus, should be increased to, say, magnitude $-$4.5 or even
$-$5, the curves in Figure~5 would move up and further reduce a range of
conforming perihelion distances, compensating for any intrinsic brightness
increase caused by softening the Bortle survival limit's constraint.

There is circumstantial evidence, to be discussed below, that the Lick
object's brightness, as observed by Campbell et al.\ at sunset on August~7,
was a result of more complex temporal variations than a simple power law of
heliocentric distance can explain.  If this indeed was so, our conclusion
that, based on Campbell et al.'s account, the object was truly in close
proximity of the Sun (and not only in projection onto the plane of the sky)
should further be strengthened.

\subsection{Daytime Observations}
The Lick object's observations by Nelson Day (Markwirk 1921), on August~7,
8.5~hr before Campbell et al.'s sighting, by Emmert (1921a, 1921b), on
August~6, another 20~hr earlier, and possibly by others, were made with the
naked eye in broad daylight, with the Sun a number of degrees above the
horizon.  This kind of visual observation has both advantages and disadvantages
over twilight observations, such as that made by Fellows (1921a, 1921b).  An
obvious advantage is that the angular distance of the object from the Sun is
readily observed.  A disadvantage that has implications for the Lick object's
physical behavior is a potentially enormous uncertainty in the estimated
brightness.

In Section 3 we already hinted that Nelson Day's~report of the Lick object's
apparent magnitude of~$-$2 at 4$^\circ$ from the Sun on August~7 must be
grossly underrated. The issue of daylight visual-magnitude estimates of
bright celestial objects was addressed by Bortle (1985, 1997), who
presented a formula for a limiting magnitude of~an object just detectable
with 8-cm binoculars as a function of its elongation from the Sun.
At 4$^\circ$ from the Sun one could observe with this instrument an object
as faint~as $-$1.3.  Furthermore, on a website Bortle posted a list of
threshold visual magnitudes for daytime observations with the unaided
eye.\footnote{J.\ E.\ Bortle's one page document, titled ``Judging
Potential Visibility of Daylight Comets,'' is available from a website
{\tt http://www.eagleseye.me.uk/DaylightComets.pdf}.} From this document
it follows, for example, that an object of magnitude $-$4.0 is just
visible with the unaided eye if more than 5$^\circ$ from~the Sun, while an
object of $-$5.5 or somewhat brighter is rather easily visible when the
Sun is blocked.  Since, in reference to the Ferndown observation, Markwirk
(1921) conveyed that the Lick object looked ``striking,'' it must have been
at this time quite a bit brighter than magnitude $-$4, possibly even
brighter than $-$5.

Essentially the same constraint applies to Emmert's sighting on August~6.
Emmert estimated the Lick object's brightness by asserting that it ``was
fully as bright as Venus in twilight{\vspace{-0.05cm}} at her greatest
brilliancy'' (Emmert 1921a),\footnote{From Mallama et al.'s (2006) phase
curve we find that at Earth's and Venus' mean heliocentric distances, the
peak apparent magnitude of Venus is \mbox{$V = -4.81$} at a phase angle of
123$^\circ\!$.55 and elongation of 37$^\circ\!$.07.} while elsewhere he
claimed that it ``shone altogether too bright for Venus'' (Emmert 1921b).
The~two statements are not incompatible, as the latter refers implicitly
to the planet's expected brightness (fainter than $-$4; Section~4.1) at
the time of observation; the former statement implies a magnitude just
shy of $-5$.

We cannot rule out that the Lick object was getting fainter as it was
approaching the Sun, which brings to mind the instances of post-outburst
terminal fading of disintegrating comets, including C/1996~Q1 (Figure~3).
It is possible that both Emmert and Nelson Day saw the Lick object during
its final outburst, while Campbell's (1921d) observation was made along
the outburst's subsiding branch and his unsuccessful search on the
following days after the event was over.

\subsection{Astrometric Data and Other Tests in Search for\\a Genetically
Related Comet}
We were now ready to undertake a search for a comet (or comets), whose orbit
could serve as a basis for fitting the limited data on the Lick object, as
described in Section~3.  Performed in Equinox~J2000, the search rested on the
following set of diagnostic tests of the evidence:

(1) The most credible data were provided by Campbell (1921d):\
{\vspace{-0.05cm}}the object's position for 1921 August 8.1347 UT,
\mbox{R.A.(J2000) = 9$^{\rm h}$27$^{\rm m}\!$.5}, \mbox{Dec.(J2000) =
+15$^\circ$04$^\prime$}, whose uncertainty is estimated at $\pm$10$^\prime$.

(2) The observation by Nelson Day on 1921 August 7.7806 UT (Markwick 1921),
placing the object 4$^\circ$ from the Sun, was judged less reliable than
Campbell's and was employed as a second, less diagnostic test, with an
estimated uncertainty of $\pm$30$^\prime$.  The somewhat cryptic depiction
of the object-Sun orientation (Section~3) suggests, if correctly interpreted,
the object's equatorial coordi\-nates of \mbox{R.A.(J2000)\,=\,9$^{\rm
h}$29$^{\rm m}\!$.6}, \mbox{Dec.(J2000)\,=\,+16$^\circ$29$^\prime$}, with
an estimated uncertainty of more than $\pm$30$^\prime$.

(3) The observation by Emmert (1921a, 1921b) was judged to furnish
information that was much less credible than Campbell's and less credible
than Nelson Day's.  We employed the object's reported angular distance from
the Sun, 5$^\circ$ on 1921 August 6.9514 UT, as the only piece of data worth
testing; its uncertainty was estimated at $\pm$1$^\circ$ at best.

To explain the Lick object's sudden appearance, disappearance, and other
circumstances, it was judged helpful if a genetically related comet and
its orbit satisfied four additional constraints:

(4) The most likely reason for the object's sudden appearance was thought
to be its unfavorable trajectory in the sky during its long approach to
perihelion:\ from behind the Sun, along a narrow orbit that would entail
a small elongation from the Sun over an extended period of many weeks
before near-perihelion discovery.

(5) A very small perihelion distance, \mbox{$q < 0.1$ AU}, was preferred
because it would most straightforwardly account for the object's brief
brightening, thus aiding the argument in the previous point.  It would also
expose the object to a highly perilous environment near perihelion, thus
increasing the chances of its disintegration and thereby its abrupt
disappearance (Section 4.1).

(6) Since in comet pairs and groups the chance of near-perihelion demise
was limited to companions (Section 2), which typically arrived at perihelion
much {\it later\/} than the primary members, one would expect that the
Lick object was a companion to the pair's primary and, accordingly, that
the primary would more probably have arrived at perihelion well {\it
before\/} 1921.

(7) Expressed by Equation (2), the correlation between (i)~a velocity at which
the fragments begin to separate after their parent's breakup near perihelion
and (ii)~a temporal gap between their observed arrivals to perihelion one
revolution later, favors, for a typical gap of dozens of years, fragments of
a comet with an orbital period on the order of thousands of years.  Thus, such
a comet is more likely to be genetically related to the Lick object than
a comet with a shorter or longer orbital period; we address the preferred
range for our test purposes in Section~5.

Having described the tests and constraints that the comets potentially
associated with the Lick object were expected to satisfy, we next compiled an
initial list of selected candidates.  We allowed a wide range of orbital
periods, $P$, by letting the statistically averaged separation velocity,
$\langle V_{\rm sep} \rangle$ [for the definition, see the text below
Equation~(26)] to vary from 0.1~to 3~m~s$^{-1}$; the difference in the orbital
period, $\Delta P$, between the primary and the companion, to range from 3
to 300~yr; and the heliocentric distance at fragmentation, $r_{\rm frg}$,
to span 0.01 to 10~AU.  Expressing $\Delta V$ from Equation~(2) in terms of
$\langle V_{\rm sep} \rangle$, the orbital periods range from 100 to
100\,000~yr.  Requiring conservatively that the perihelion distance not
exceed $\sim$0.25~AU, we assembled all comets from Marsden \& Williams'
(2008) catalog that satisfied these conditions (the orbital period as
derived from the original semimajor axis; for C/1769~P1 computed from an
osculating value by R.\ Kracht), and with orbits of adequate precision
(a minimum of five decimals in $q$ and three in the angular elements).
For the period of 2008--2015 we consulted the JPL Small-Body Database
Search Engine.\footnote{See
{\tt http://ssd.jpl.nasa.gov/sbdb\_query.cgi}.{\vspace{0.23cm}}}

\begin{table}[t]
\vspace{0.17cm}
\begin{center}
{\footnotesize {\bf Table 7}\\[0.07cm]
{\sc  Potential Candidates for a Comet That Could Be\\Genetically Related to
the Lick Object.}\\[0.1cm]
\begin{tabular}{l@{\hspace{0.8cm}}c@{\hspace{0.8cm}}c@{\hspace{1cm}}c}
\hline\hline\\[-0.24cm]
          & Perihelion & Orbital              & Incli- \\[-0.03cm]
Candidate & distance,  & period,$^{\rm a}$    & nation, \\[-0.03cm]
comet      & $q$\,(AU) & $P_{\rm orig}$(yr)   & $i$   \\[0.07cm]
\hline\\[-0.2cm]
C/1533 M1  & 0.2548 & \ldots\ldots & 149$^\circ\!\!$.59 \\
C/1577 V1  & 0.1775 & \ldots\ldots & 104.85 \\
C/1582 J1  & 0.1687 & \ldots\ldots & 118.54 \\
C/1593 O1  & 0.0891 & \ldots\ldots & $\;\:$87.91 \\
C/1665 F1  & 0.1065 & \ldots\ldots & 103.89 \\
C/1668 E1  & 0.0666 & \ldots\ldots & 144.38 \\
C/1680 V1  & 0.0062 & 9530         & $\;\:$60.68 \\
C/1689 X1  & 0.0644 & \ldots\ldots & $\;\:$63.20 \\
C/1695 U1  & 0.0423 & \ldots\ldots & $\;\:$93.59 \\
C/1733 K1  & 0.1030 & \ldots\ldots & $\;\:$23.79 \\[0.1cm]
C/1737 C1  & 0.2228 & \ldots\ldots & $\;\:$18.33 \\
C/1743 X1  & 0.2222 & \ldots\ldots & $\;\:$47.14 \\
C/1758 K1  & 0.2154 & \ldots\ldots & $\;\:$68.30 \\
C/1769 P1  & 0.1228 & 2100\rlap{$^\ast$} & $\;\:$40.73 \\
C/1780 U2  & 0.0993 & \ldots\ldots & 126.18 \\
C/1816 B1  & 0.0485 & \ldots\ldots & $\;\:$43.11 \\
C/1821 B1  & 0.0918 & \ldots\ldots & 106.46 \\
C/1823 Y1  & 0.2267 & \ldots\ldots & 103.82 \\
C/1826 U1  & 0.0269 & \ldots\ldots & $\;\:$90.62 \\
C/1827 P1  & 0.1378 & \ldots\ldots & 125.88 \\[0.1cm]
C/1831 A1  & 0.1259 & \ldots\ldots & 135.26 \\
C/1844 Y1  & 0.2505 & 7590         & $\;\:$45.57 \\
C/1847 C1  & 0.0426 & 8310         & $\;\:$48.66 \\
C/1851 U1  & 0.1421 & \ldots\ldots & $\;\:$73.99 \\
C/1859 G1  & 0.2010 & \ldots\ldots & $\;\:$95.49 \\
C/1865 B1  & 0.0258 & \ldots\ldots & $\;\:$92.49 \\
C/1874 D1  & 0.0446 & \ldots\ldots & $\;\:$58.89 \\
C/1901 G1  & 0.2448 & \ldots\ldots & 131.08 \\
C/1905 X1  & 0.1902 & \ldots\ldots & $\;\:$43.65 \\
C/1917 F1  & 0.1902 & 143  & $\;\:$32.68 \\[0.1cm]
C/1927 X1  & 0.1762 & 14\,600       & $\;\:$84.11 \\
C/1945 W1  & 0.1944 & \ldots\ldots & $\;\:$49.48 \\
C/1947 X1  & 0.1100 & 3720         & 138.54 \\
C/1948 L1  & 0.2076 & 83\,100       & $\;\:$23.15 \\
C/1948 V1  & 0.1354 & 21\,500       & $\;\:$23.12 \\
C/1953 X1  & 0.0723 & \ldots\ldots & $\;\:$13.57 \\
C/1959 Q2  & 0.1655 & \ldots\ldots & $\;\:$48.26 \\
C/1961 O1  & 0.0402 & 44\,900       & $\;\:$24.21 \\
C/1967 M1  & 0.1783 & \ldots\ldots & $\;\:$56.71 \\
C/1970 B1  & 0.0657 & \ldots\ldots & 100.18 \\[0.1cm]
C/1975 V1  & 0.1966 & 16\,300       & $\;\:$43.07 \\
C/1985 K1  & 0.1063 & \ldots\ldots & $\;\:$16.28 \\
C/1987 W1  & 0.1995 & \ldots\ldots & $\;\:$41.62 \\
C/1988 P1  & 0.1646 & \ldots\ldots & $\;\:$40.20 \\
C/1996 B2  & 0.2302 & 16\,400       & 124.92 \\
C/1998 J1  & 0.1532 & \ldots\ldots & $\;\:$62.93 \\
C/2002 V1  & 0.0993 & 9080         & $\;\:$81.71 \\
C/2004 F4  & 0.1693 & 2690         & $\;\:$63.16 \\
C/2004 R2  & 0.1128 & \ldots\ldots & $\;\:$63.17 \\
C/2004 V13 & 0.1809 & \ldots\ldots & $\;\:$34.81 \\[0.08cm]
\hline\\[-0.23cm]
\multicolumn{4}{l}{\parbox{7.5cm}{\scriptsize $^{\rm a}$\,The asterisk
indicates the original orbit was derived by the second author;
parabolic orbits are shown by dots.}} \\[0.16cm]
%
%
\end{tabular}}
\end{center}
\end{table}
\begin{figure*}[ht] 
\vspace{0.15cm}
\hspace{-0.2cm}
\centerline{
\scalebox{0.735}{
\includegraphics{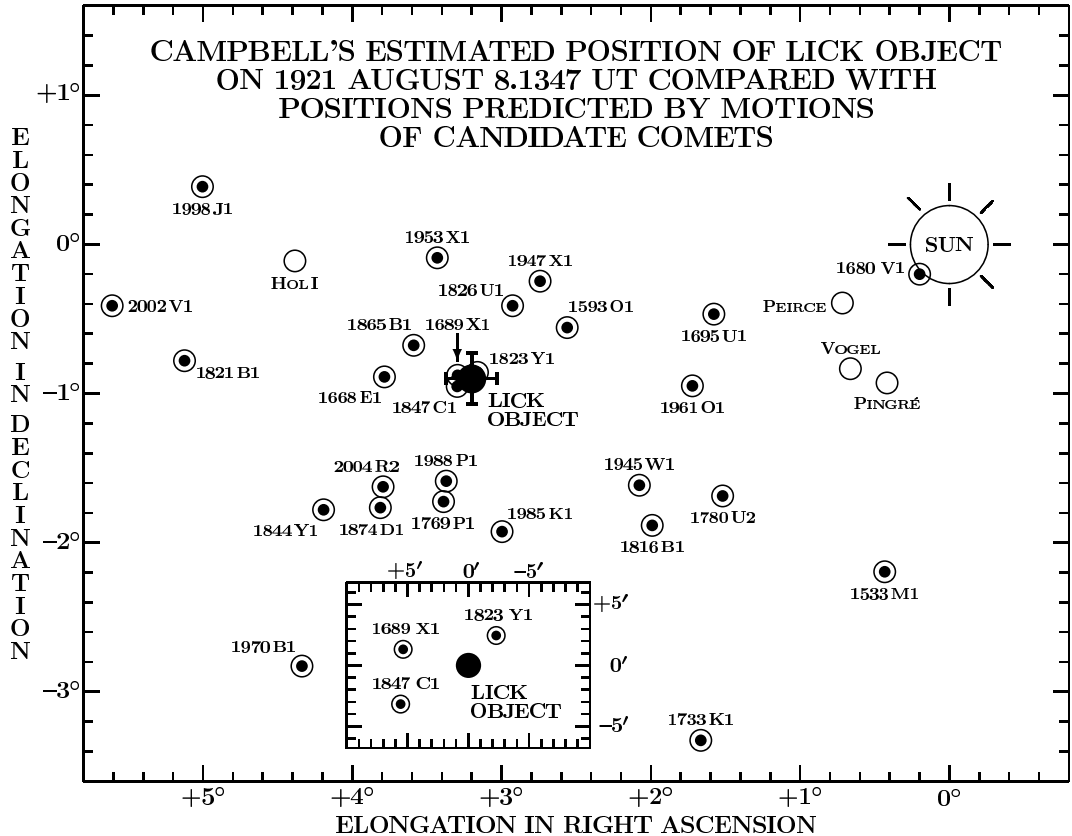}}} 
\vspace{-0.05cm}
\caption{Position of the Lick object relative to the Sun on 1921 August 8.1347
UT (solid circle), reported by Campbell (1921d), and its predicted positions
(circled dots), derived by emulating the object's motion with the motions of
potentially related comets, listed in Table~7.  The plot contains all 26 comets
(of the 50) that fit its limits.  Three comets --- C/1823~Y1, C/1689~X1, and
C/1847~C1 (Hind) --- match the Lick object's position to within 10$^\prime$.
For C/1689~X1 we show not only the result from the cataloged orbit (circled
dot), but also from the other available orbits (open circles), marked by the
name of the computer (see also Tables 10 and 11).  The size of the Sun is
drawn to scale.  On a scale expanded by a factor of nearly five, the inset
displays the three comets, whose orbits match the position of the Lick object
most closely.  The width of the inset window equals the estimated uncertainty
of the Lick object's position.{\vspace{0.6cm}}} 
\end{figure*}

The candidate comets are presented in Table~7.  In the Marsden-Williams
catalog we found merely 14 objects that satisfied our selection constraints,
including C/1917~F1 (Mellish), a parent body of an extensive meteor-stream
complex (e.g., Neslu\v{s}an \& Hajdukov\'a 2014).  We found no candidate
comets from 2005 to 2015.  Since comets with an orbital period between 100
and 100\,000~yr may easily appear among those for which only a parabolic
approximation is available, we extended the set of candidates by adding
small-perihelion comets with parabolic orbits.  Even though we argued that
the comets before 1921 should be the more attractive candidates [see the
point (6) above], we allowed the comets that arrived after 1921 to be
subjected to the same tests.  We ended up with a total of 50 candidate
comets.

\section{Results of a Search for a Comet Potentially Related to the Lick
Object}
Since we assigned the greatest weight to the position of the Lick object
on August 8.1347~UT, as reported by Campbell (1921d), test~1 was applied
first.  The object's position relative to the Sun is in Figure~6 compared
with the positions predicted by the orbits of 26 of the 50 candidate
comets listed in Table~7.  The remaining 24 lie outside the plot's limits.
There are three comets whose cataloged orbits match the Lick object's
position estimated by Campbell highly satisfactorily, well within 10$^\prime$,
its expected uncertainty (Section~4.3).  In chronological order, the
three objects are C/1689~X1, C/1823~Y1, and C/1847~C1 (Hind), all of them
great, intrinsically bright comets.  Based on this evidence alone, the Lick
object could be genetically related to any one of the three.  We note that
they all preceded the arrival of the Lick object, as preferred by test 6.
\begin{table}[b]
\vspace{0.35cm}
\begin{center}
{\footnotesize {\bf Table 8} \\[0.08cm]
{\sc List of 10 Comets with Orbits That Most Closely\\Match Campbell's
Position for the Lick Object.}\\[0.1cm]
\begin{tabular}{l@{\hspace{0.45cm}}c@{\hspace{0.1cm}}c@{\hspace{0cm}}c@{\hspace{0.25cm}}c@{\hspace{0.35cm}}c@{\hspace{0.35cm}}c}
\hline\hline\\[-0.24cm]
 & \multicolumn{2}{@{\hspace{-0.02cm}}c}{Offset}
 & Time\,diff. & \multicolumn{2}{@{\hspace{-0.15cm}}c}{Distance\,(AU)\,to}
 & \\[-0.05cm]
Candidate & \multicolumn{2}{@{\hspace{-0.02cm}}c}{\rule[0.7ex]{1.6cm}{0.4pt}}
          & $t_{\rm obs} \!-\! t_\pi$
          & \multicolumn{2}{@{\hspace{-0.15cm}}c}{\rule[0.7ex]{2.18cm}{0.4pt}}
          & Phase \\[-0.05cm]
comet     & Dist. & $\;$P.A. & (days) & Sun & Earth & angle \\[0.1cm]
\hline \\[-0.22cm]
C/1823 Y1 & 3$^\prime\!\!$.4 & 317\rlap{$^\circ$} & $-$2.266 & 0.2410 & 0.7784
 & 166$^\circ\!\!$.0 \\
C/1689 X1 & 5.5 & $\;\:$76 & $-$1.285 & 0.1050 & 0.9263 & 144.8 \\
C/1847 C1 & 6.4 & 119 & $-$0.667 & 0.0682 & 1.0426 & $\;\:$63.4 \\
C/1865 B1 & \llap{2}6.9 & $\;\:$60 & $-$0.713 & 0.0713 & 0.9821 & 114.6 \\
C/1826 U1 & \llap{3}3.6 & 331 & +0.479 & 0.0530 & 1.0049 & $\;\:$98.2 \\
C/1668 E1 & \llap{3}4.1 & $\;\:$89 & +1.021 & 0.0933 & 1.0743 & $\;\:$47.7 \\
C/1988 P1 & \llap{4}2.5 & 166 & +0.261 & 0.1649 & 1.1626 & $\;\:$23.8 \\
C/1593 O1 & \llap{4}3.5 & 298 & $-$0.177 & 0.0897 & 1.0894 & $\;\:$31.4 \\
C/1947 X1 & \llap{4}7.9 & 325 & $-$0.652 & 0.1152 & 1.1170 & $\;\:$25.1 \\
C/1953 X1 & \llap{5}0.5 & $\;\:$16 & $-$0.202 & 0.0736 & 1.0535
          & $\;\:$55.7 \\[0.07cm]
\hline\\[-0.7cm]
\end{tabular}}
\end{center}
\end{table}
\begin{table*}[t]
\vspace{0.15cm}
\begin{center}
{\footnotesize {\bf Table 9} \\[0.06cm]
{\sc Tests for Candidate Comets That Could Be Genetically Related to the
 Lick Object.}\\[0.1cm]
\begin{tabular}{c@{\hspace{0.7cm}}l@{\hspace{0.8cm}}c@{\hspace{0.6cm}}c@{\hspace{0.6cm}}c@{\hspace{0.6cm}}c}
\hline\hline\\[-0.24cm]
& & Condition
  & \multicolumn{3}{@{\hspace{0cm}}c}{Potential genetically related
    comet} \\[-0.05cm]
Test & & (observation/
  & \multicolumn{3}{@{\hspace{0cm}}c}{\rule[0.7ex]{5.8cm}{0.4pt}}\\[-0.05cm]
No. & Lick object's test description & desirability) & C/1689 X1 & C/1823 Y1
  & C/1847 C1 \\[0.05cm]
\hline \\[-0.2cm]
1 & Observation 1921 August 8.1347 UT: & & & & \\
  & {\hspace{0.3cm}}Elongation$^{\rm a}$       & 3$^\circ\!$.3 & 3$^\circ\!$.4
  & 3$^\circ\!$.3 & 3$^\circ\!$.4 \\
  & {\hspace{0.3cm}}Elongation in R.A.$^{\rm a}$ & \llap{+}3$^\circ\!$.2
  & \llap{+}3$^\circ\!$.3 & \llap{+}3$^\circ\!$.2 & \llap{+}3$^\circ\!$.3 \\
  & {\hspace{0.3cm}}Elongation in Dec.$^{\rm a}$  & \llap{$-$}0$^\circ\!$.9
  & \llap{$-$}0$^\circ\!$.9 & \llap{$-$}0$^\circ\!$.9
  & \llap{$-$}1$^\circ\!$.0 \\[0.1cm]
2 & Observation 1921 August 7.7806 UT: & & & & \\
  & {\hspace{0.3cm}}Elongation$^{\rm b}$          & $\sim$4$^\circ\;\;$ & 4$^\circ\!$.5
  & 3$^\circ\!$.2 & 4$^\circ\!$.0 \\
  & {\hspace{0.3cm}}Elongation in R.A.$^{\rm c}$ & (+4$^\circ\!$.0)
  & \llap{+}3$^\circ\!$.9 & \llap{+}2$^\circ\!$.5 & \llap{+}4$^\circ\!$.0 \\
  & {\hspace{0.3cm}}Elongation in Dec.$^{\rm c}$ & (+0$^\circ\!$.4)
  & \llap{$-$}2$^\circ\!$.3 & \llap{$-$}2$^\circ\!$.0
  & \llap{$-$}0$^\circ\!$.1 \\[0.1cm]
3 & Observation 1921 August 6.9514 UT: & & & & \\
  & {\hspace{0.3cm}}Elongation$^{\rm d}$ & $\sim$5$^\circ\;\;$ & 7$^\circ\!$.4 & 4$^\circ\!$.5
  & 5$^\circ\!$.3 \\[0.1cm]
4 & Approach trajectory in the sky: & & & & \\ [0.1cm]
  & & small enough & & & \\[-0.045cm]
  & {\hspace{0.3cm}}Maximum elongation at \mbox{$r < 2$ AU}
  & to preclude & 85$^\circ$ & 49$^\circ$ & 46$^\circ$ \\[-0.045cm]
  & & \llap{ea}rly\,detecti\rlap{on} & & & \\[0.03cm]
  & & should be & steadily & steadily & steadily \\[-0.205cm]
  & {\hspace{0.3cm}}Overall trend in elongation with time & & & & \\[-0.205cm]
  & & decreasing & decreasing & decreasing\rlap{$^{\rm e}$}
    & decreasing \\[0.03cm]
  & & from behind & from behind & from behind & from behind \\[-0.205cm]
  & {\hspace{0.3cm}}Approach direction in space & & & & \\[-0.205cm]
  & & the Sun & the Sun & the Sun & the Sun \\[0.1cm]
5 & Perihelion distance & $<$0.1 AU & 0.064 AU & 0.227 AU & 0.043 AU \\[0.07cm]
  & & years earlier to& 231.7 yr & 97.7 yr & 74.4 yr \\[-0.2cm]
6 & Perihelion arrival relative to Lick object &  & & & \\[-0.2cm]
  & & fit\,condition\,(2) & earlier & earlier & earlier \\[0cm]
  & & between $\sim$2000 & & & \\[-0.2cm]
7 & Orbital period (since previous return) & & unknown
  & $\sim$2100 yr$^{\rm f}$ & $\sim$8300 yr \\[-0.2cm] 
  & & and $\sim$50\,000 yr & & & \\[0.03cm]
\hline\\[-0.26cm]
\multicolumn{6}{l}{\parbox{15.7cm}{\scriptsize $^{\rm a}$\,Estimated
uncertainty of $\pm$10$^\prime$.}}\\[-0.07cm]
\multicolumn{6}{l}{\parbox{15.7cm}{\scriptsize $^{\rm b}$\,Estimated
uncertainty of $\pm$30$^\prime$.}}\\[0.03cm]
\multicolumn{6}{l}{\parbox{15.7cm}{\scriptsize $^{\rm c}$\,Condition subject
{\vspace{-0.03cm}}to validity of employed interpretation of reported position
(Sections~3 and 4.3); if valid, estimated uncertainty of more than
$\pm$30$^\prime$.}}\\[-0.05cm]
\multicolumn{6}{l}{\parbox{15.7cm}{\scriptsize $^{\rm d}$\,Estimated
uncertainty of $\pm$1$^\circ$ at best.}}\\[-0.05cm]
\multicolumn{6}{l}{\parbox{15.7cm}{\scriptsize $^{\rm e}$\,Except in close
proximity of perihelion, starting about 2.5~days before the
passage.}}\\[-0.04cm]
\multicolumn{6}{l}{\parbox{15.7cm}{\scriptsize $^{\rm f}$\,See Section
5.2.{\vspace{0.14cm}}}}
\end{tabular}}
\end{center}
\end{table*}

Table 8 presents the ten top candidates for a comet genetically related to
the Lick object in the order of increasing offset.  Columns~2 and 3 list,
respectively, the offset and the position angle from the Lick object's position
estimated by Campbell (1921d); column~4 shows whether the observation, at time
$t_{\rm obs}$, was made before~or after the candidate's perihelion, $t_\pi$;
columns 5 and 6 provide, respectively, the heliocentric and geocentric distances
at $t_{\rm obs}$, describing the role of geometry; and the last column lists the
phase angle, which allows one to assess the effect of forward scattering on the
object's observed brightness.  Starting with the fourth candidate, C/1865~B1,
the chances of a genetic relationship with the Lick object are highly unlikely,
as from there on the offsets exceed the estimated uncertainty by more than a
factor of two.  We list these comets to accentuate the existence of a
remarkable gap between the first three entries and the rest.

In the following we focus on the three promising candidates, whose comparison
in terms of all seven tests formulated in Section~4.3 is offered in Table~9.
Although the range of orbital periods of the candidate comets in Table~7
spanned three orders of magnitude, the orbital period of a likely candidate
is now expected to be more tightly constrained, because the lower limit to
the heliocentric distance at fragmentation and the temporal gap between the
arrival times of the candidate comets and the Lick object are known.  Test~7
in Table~9 reflects this tightening of the range of orbital periods, whose
crude limits span a range from $\sim$2000 to $\sim$50\,000~yr.  We are now
ready to inspect each of the three contenders separately in chronological
order.

\subsection{Comet C/1689 X1}
Close examination of the history of this comet's orbit determination shows
that the study by Holetschek (1891, 1892), whose ``best'' orbital set was
cataloged by Marsden \& Williams (2008) and used by us, was preceded by
orbital investigations of Vogel (1852a, 1852b), B.\,Peirce (Kendall 1843),
and Pingr\'e (1784).\footnote{Converted to the equinox of J2000, the orbital
sets from all four investigations are conveniently summarized by G.\,W.\,Kronk
at a website {\tt http://cometography.com/orbits\_17th.html}.} There are very
dramatic differences among the computed orbits, which are traced to different
interpretations of the crude observations.  Although the comet was discovered
about a week before perihelion (Kronk 1999) and was later observed also on the
island of Ternate, Indonesia (Pingr\'e 1784) and in China (Struyck 1740;
Pingr\'e 1784; Hasegawa \& Nakano 2001), its approximate positions could only
be constructed from the accounts reported on four mornings after perihelion by
the observers at Pondicherry, then a French territory in India (Richaud 1729),
and at Malacca, Malaysia (de B\`eze \& Comilh 1729).\footnote{The name of de
B\`eze's co-worker was misspelled in the original French publication as {\it
Comille\/}; no one by that name ever worked with de B\`eze, yet the error
propagated through the literature as it was copied over and over again for
nearly three centuries [Kronk (1999) misspells both de B\`eze and Comilh].
All three mentioned observers of C/1689~X1, Richaud, de B\`eze, and Comilh
were among 14~French Jesuits sent by Louis~XIV as ``royal mathematicians''
to King of Siam to assist him in implementing his intention to found a
scientific academy and an observatory (e.g.,\,Ud\'{\i}as 2003; Hsia
2009).{\vspace{-0.06cm}}} The difficulties encountered in an effort to
determine this comet's orbit were befittingly summarized by Plummer (1892),
including the intricacies that Holetschek (1891, 1892) got involved with in
order to accommodate a constraint based on a remark by de B\`eze and Comilh
that the comet reached a peak rate of more than 3$^\circ$ per day in
apparent motion between 1689 December 14 and 15.  Unable to find an orbital
solution that would satisfy this constraint, Holetschek himself eventually
questioned the report's credibility, considering his preferred orbital
solution (based on the adopted positions from December 10, 14, and 23 and
on some experimentation with the observation time on the 10th) as a
compromise.

\begin{table}[t]
\vspace{0.12cm}
\begin{center}
{\footnotesize {\bf Table 10}\\[0.1cm]
{\sc Sets of Parabolic Orbital Elements of C/1689 X1\\Derived by
 J. Holetschek (Equinox J2000)}.\\[0.1cm]
\begin{tabular}{l@{\hspace{0.18cm}}c@{\hspace{0.18cm}}c@{\hspace{0.18cm}}c}
\hline\hline\\[-0.22cm]
Quantity/Orbital element & Catalog & {\sc Hol\,I} & {\sc Hol\,II} \\[0.08cm]
\hline\\[-0.25cm]
Perihelion time $t_\pi$\,(1689 Nov.\,UT) & 30.659 & 30.545 & 28.461 \\
Argument of perihelion $\omega$ & $\;\:$78$^\circ\!$.134 & 163$^\circ\!$.857
 & 143$^\circ\!$.535 \\
Longitude of ascending node $\Omega$ & 283$^\circ\!$.754
 & $\;\:$67$^\circ\!$.841 & $\;\:$51$^\circ\!$.682 \\
Orbit inclination $i$ & $\;\:$63$^\circ\!$.204 & 143$^\circ\!$.211
 & 135$^\circ\!$.308\\
Perihelion distance $q$ & 0.06443 & 0.09281 & 0.16673 \\[0.07cm]
\hline\\[-0.25cm]
\end{tabular}}
\end{center}
\end{table}
\begin{table}[b]
\vspace{0.35cm}
\begin{center}
{\footnotesize {\bf Table 11} \\[0.1cm]
{\sc Orbits of C/1689 X1 Emulating Campbell's Position\\for the Lick
 Object.}\\[0.1cm]
\begin{tabular}{l@{\hspace{0.35cm}}c@{\hspace{0.2cm}}c@{\hspace{0cm}}c@{\hspace{0.25cm}}c@{\hspace{0.35cm}}c@{\hspace{0.35cm}}c}
\hline\hline\\[-0.24cm]
 & \multicolumn{2}{@{\hspace{-0.02cm}}c}{Offset}
 & Time$\:$diff. & \multicolumn{2}{@{\hspace{-0.15cm}}c}{Distance\,(AU)\,to}
 & \\[-0.05cm]
Computer/ & \multicolumn{2}{@{\hspace{-0.02cm}}c}{\rule[0.7ex]{1.6cm}{0.4pt}}
          & $t_{\rm obs} \!-\! t_\pi$
          & \multicolumn{2}{@{\hspace{-0.15cm}}c}{\rule[0.7ex]{2.18cm}{0.4pt}}
          & Phase \\[-0.05cm]
designation & Dist. & P.A. & (days) & Sun & Earth & angle \\[0.1cm]
\hline \\[-0.24cm]
Catalog       & $\;\:$5$^\prime\!$.5 & 76\rlap{$^\circ$} & $-$1.285 & 0.1050
 & 0.9263 & 144$^\circ\!\!$.8 \\
{\sc Hol} I  & 85.4 & 56 & $-$0.098 & 0.0930 & 0.9596 & 123.5 \\
{\sc Hol} II & \llap{3}71.6 & 50 & $-$0.517 & 0.1682 & 0.9257 & 117.3 \\
Vogel    & \llap{1}52.3 & \llap{2}72 & +0.081 & 0.0211 & 1.0060 & 111.2 \\
Peirce   & \llap{1}52.1 & \llap{2}82 & +0.279 & 0.0386 & 1.0493 & $\;\:$22.9 \\
Pingr\'e & \llap{1}67.0 & \llap{2}70 & +0.989 & 0.0948 & 1.1065
         & $\;\:$11.5 \\[0.1cm]
\hline\\[-0.7cm]
\end{tabular}}
\end{center}
\end{table}

Once the condition of a peak rate of apparent motion is dropped, there is no
longer any reason to prefer this compromise solution over other available
orbital sets.  Accordingly, we examined whether or not the orbits by Pingr\'e
(1784), by Peirce (Kendall 1843; Cooper 1852), by Vogel (1852a, 1852b), and the
alternative orbits by Holetschek (1891) himself matched Campbell's position
for the Lick object as closely as did the cataloged orbit for C/1689~X1.
The two alternative orbits by Holetschek (1891) that we considered are
compared to the cataloged orbit in Table~10.  The first set of elements,
referred to as ``{\sc Hol I}'' in Table~10, was a key solution that Holetschek
derived before he began a complicated data manipulation.  The second set,
referred to as ``{\sc Hol~II}'' in Table~10, is one of three considered by
him --- in the course of another effort to accommodate the high rate of
apparent motion --- to avoid the tendency of the computed path to have
strayed to the northern hemisphere before perihelion, and the only of
the three to have the perihelion distance smaller than 0.25~AU (thus
complying with our selection rules).  It is somewhat peculiar that
Holetschek's orbit in the Marsden-Williams catalog is the only set of
elements (among the six we consider for C/1689~X1) that shows the comet's
motion to be direct; the rest are all retrograde, with the inclinations
between 110$^\circ$ and 150$^\circ$.

In Table 11, organized in the same fashion as Table~8, the various orbits for
C/1689~X1 are compared in terms of the quality of matching the Lick object's
position published by Campbell.  The table shows that due to enormous
uncertainties in the orbit determination, the very small offset in Figure~6 and
Table~8 from the cataloged set of elements must be judged as purely fortuitous.
Accordingly, a genetic relationship between the Lick object and C/1689~X1
is on account of the orbital uncertainties entirely indeterminable.  Arguments
that tests~3 and 4 in Table~9 are unfavorable to the relationship are --- while
correct --- under these circumstances unnecessary.

\subsection{Comet C/1823 Y1}
This was an intrinsically bright comet.  On the assumption that its brightness,
normalized to a unit geocentric distance and corrected for a phase effect
according to the law by Marcus (2007), varied with heliocentric distance as
$r^{-n}$, a least-squares solution to several magnitude estimates collected
by Holetschek (1913) yields an absolute magnitude of \mbox{$H_0 = 3.70 \pm
0.22$} and \mbox{$n = 3.62 \pm 0.43$}.  At first sight it is surprising that
the comet was not discovered before perihelion, but a straightforward
explanation is provided by Figure~7, which shows that the approach to
perihelion was from the high southern declinations:\ the culprit was the
lack of comet discoverers in the southern hemisphere in the early 19th
century.  For two months, from the beginning of 1823 September to the
beginning of November, the comet was nearly stationary relative to the Sun,
at elongations between 55$^\circ$ and 62$^\circ$ and at heliocentric
distances from 2.2~AU to 1.1~AU.

The comet was poorly placed for observation in an early post-perihelion period
as well, as it took three weeks before its elongation increased to 40$^\circ$,
at which time it was discovered as a naked-eye object\footnote{According to
Olbers (1824a), in the morning of 1824 January 5 the brightness of C/1823~Y1
equaled that of a star of magnitude 3; this was the comet's first brightness
estimate (Holetschek 1913).} independently by several observers.

The most comprehensive orbit for this comet to date was computed by Hnatek
(1912) from hundreds of astrometric observations made between 1823 December~30
and 1824 March~31.  The perturbations by five planets, Venus through Saturn,
were accounted for.  An unusual feature of Hnatek's study, marring its
scientific value, was that while he computed both the most probable parabola
and the most probable ellipse (with an osculation orbital period of 9764~yr)
in the process of orbit refinement, he presented only a parabolic solution as
the comet's final orbit, dropping the last normal place (because of large
residuals) and failing to demonstrate that the error of the eccentricity
exceeded its deviation from unity.
\begin{figure*}[t]  
\vspace{0.15cm}
\begin{center}
\leftline{
\hspace{-0.1cm}
\scalebox{0.99}{
\includegraphics{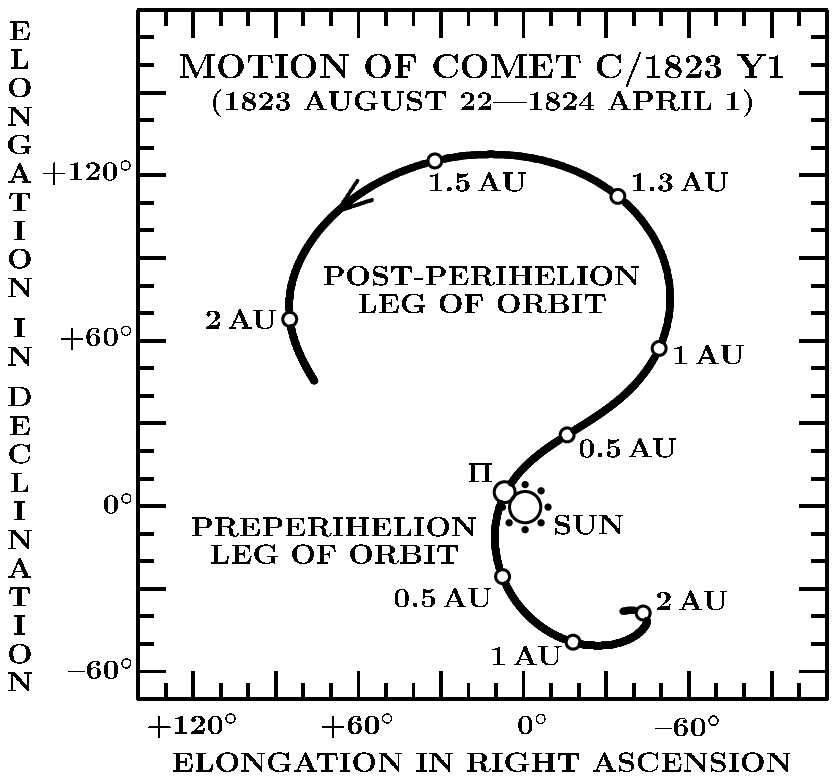}}} 

\vspace{-7.72cm}

\hspace{-0.5cm}
\rightline{
\scalebox{0.99}{
\includegraphics{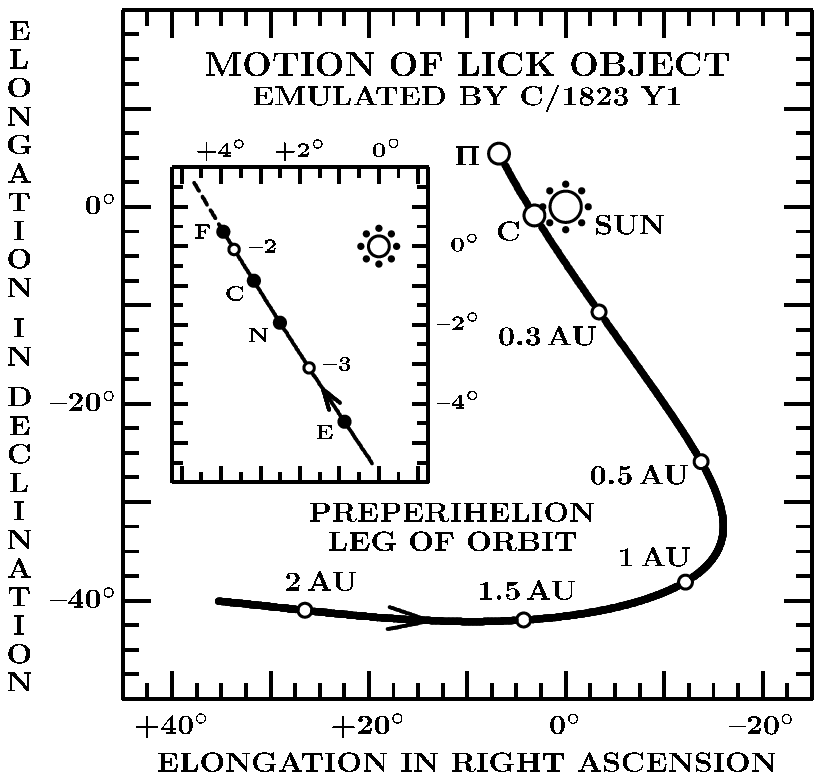}}} 
\vspace{-0.02cm}
\caption{Apparent motions, relative to the Sun, of comet C/1823 Y1 (left) and
of the Lick object, emulated by the comet's motion (right).  The scale of the
plot on the right is $\sim$3.6 times the scale on the left; $\Pi$ is the
perihelion point on, respectively, 1823 December 9.9 (left) and 1921 August
10.4 (right).  Heliocentric distances at the comet's positions and the Lick
object's predicted positions are depicted by the open circles. The elongation
is reckoned negative to the south and the west of the Sun. --- {\sf Left}:\
The comet's motion from 1823 August 22 (110~days before perihelion) to 1824
April 1 (113~days after perihelion).  The comet was almost stationary relative
to the Sun for two months, from early September to early November.  --- {\sf
Right}:\ The Lick object's emulated motion from 1921 May 1 (101 days before
perihelion) to perihelion; C is the object's predicted location at the time
of Campbell et al.'s observation on 1921 August 8.135 UT.  Closeup of the
object's emulated motion near perihelion is shown in the inset, whose scale
is 4 times the scale of the panel and in which the size of the Sun is drawn
to the scale.  The small open circles show the object's predicted positions
3 and 2 days before perihelion, respectively; C, N, and E refer to the object's
locations at the times of Campbell et al.'s, Nelson Day's, and Emmert's
observations, respectively; and F is the first of the three occasions, at
sunrise on August 8 at Lick, on which Campbell failed to recover the
object.{\vspace{0.05cm}}}
\end{center}
\end{figure*}

Before deciding whether or not to redetermine the orbit with increased
attention to its possible ellipticity, we noticed numerous accounts in the
literature of the comet's unusual apperance over a period from 1824 January 22
through 31, around the time of the Earth's transit across the comet's orbit
plane on January 24.0~UT.  At least five observers (Hansen 1824; Harding 1824;
Olbers 1824b; von Biela 1824; Gambart 1825) reported that the comet displayed,
next to its ordinary tail in the antisolar direction, a second extension that
pointed sunward and was referred to by both Harding and Olbers as an {\it
anomalous\/} tail.  Exhibited since by many other comets, most memorably by
C/1956~R1 (Arend-Roland) and C/1973~E1 (Kohoutek), this type of comet-dust
feature is nowadays customarily known as an {\it antitail\/} and its dynamics,
governed by the conservation-of-orbital-momentum law, and appearance, controled
in large part by conditions of the projection onto the plane of the sky, are
well understood (for a review, see Sekanina et al.\ 2001 and the references
therein).  The appearance of an antitail is clearly a signature of a comet
rich in large-sized dust, but it is not diagnostic of its orbital category.
The statistics show that both dynamically new comets, arriving from the Oort
Cloud, and dynamically old comets (with the orbital periods on the order of
thousands to tens of thousands of years) did show an antitail in the past.
Indeed, as Marsden et al.\ (1973, 1978) showed, both C/1956~R1 and C/1973~E1
were dynamically new comets, as were several additional objects, e.g.,
C/1895~W1 (Perrine), C/1937~C1 (Whipple), C/1954~O1 (Voz\'arov\'a), or, quite
recently, C/2011~L4 (PanSTARRS); while other antitail-displaying objects,
such as C/1844~Y1, C/1961~O1 (Wilson-Hubbard), C/1969~T1 (Tago-Sato-Kosaka),
or, recently, C/2009~P1 (Garradd) and C/2014~Q2 (Lovejoy) were {\it not\/}
dynamically new comets.  As part of a broader antitail investigation, the
feature in C/1823~Y1 was investigated by Sekanina (1976) and found to
consist of submillimeter-sized and larger dust grains released from the
nucleus before perihelion as far from the Sun as Jupiter, if not farther
away still.
%

Given the orbit ambivalence in the antitail statistics, we decided to go ahead
with the orbit redetermination, in order to ascertain whether C/1823~Y1 is in
compliance with test~7 of Table~9.  Employing a code {\it EXORB7\/} developed
by A.\ Vitagliano, the second author performed the computations, using selected
consistent astrometric observations made between 1824 January 2 and March~31,
which were re-reduced with the help of comparison-star positions from the {\it
Hipparcos\/} and {\it Tycho~2\/} catalogs.  The results are described in the
Appendix; here we only remark that the original semimajor axis came out to be
\mbox{$(1/a)_{\rm orig} = +0.006077\,\pm\,0.000471$}~(AU)$^{-1}$, implying a
true orbital period of $\sim$2100~yr and demonstrating that the comet had
not arrived from the Oort Cloud.

Having determined that C/1823 Y1 moved in an orbit that did not rule out its
potential association with the Lick object (test~7), we next examined the
comet's compliance with the other constraints summarized in \mbox{Table~9}.
A parameter that is a little outside the proper range is the perihelion
distance (test~5).  The Lick object's motion along the orbit of C/1823~Y1
should have two implications for the observing geometry at the time of Campbell
et al.'s observation, as seen from Table~8:\ (i)~the object would have been
only 0.78~AU from the Earth and (ii)~its apparent brightness should have been
strongly affected by forward scattering, if the object was as dust rich as
C/1823~Y1.  We reckon that for an $r^{-4}$ law of variation, the Lick object
would have appeared to Campbell et al.\ $\sim$11.5~mag brighter than was its
absolute magnitude $H_0$.  With an estimated apparent visual magnitude of
$-$4.3, \mbox{$H_0 \approx 7.2$}.  With an $r^{-5}$ law, the absolute
brightness would drop to \mbox{$H_0 \approx 8.7$}.  If the Lick object was
caught by Campbell et al.\ in the midst of the terminal outburst's subsiding
branch, its absolute brightness in quiescent phase could have been fainter
still.

\begin{figure*}[t] 
\vspace{0.16cm}
\begin{center}
\leftline{
\hspace{-0.1cm}
\scalebox{0.99}{
\includegraphics{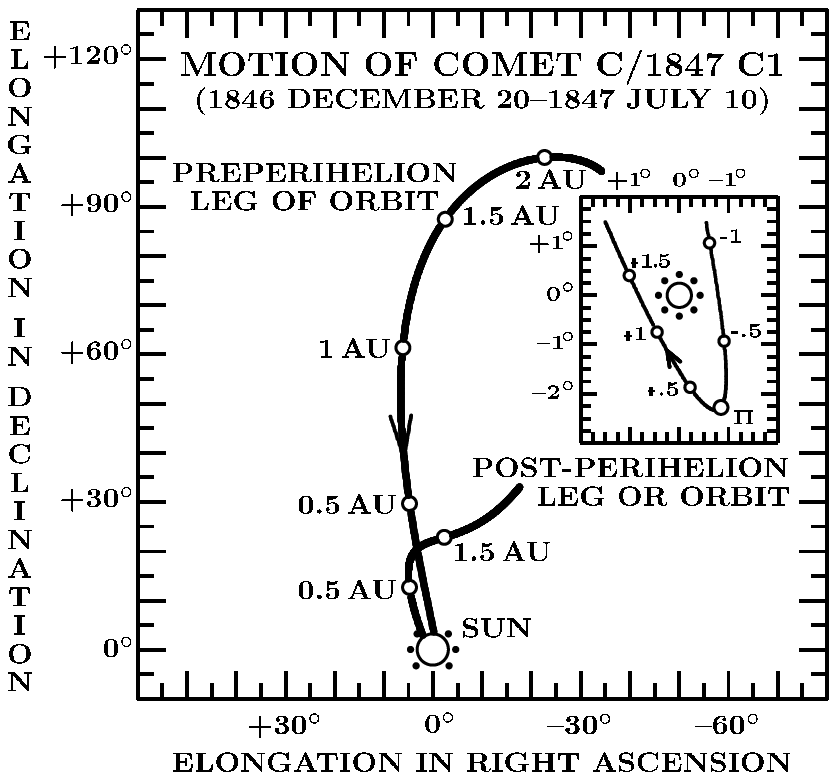}}} 

\vspace{-7.72cm}

\hspace{-0.5cm}
\rightline{
\scalebox{0.99}{
\includegraphics{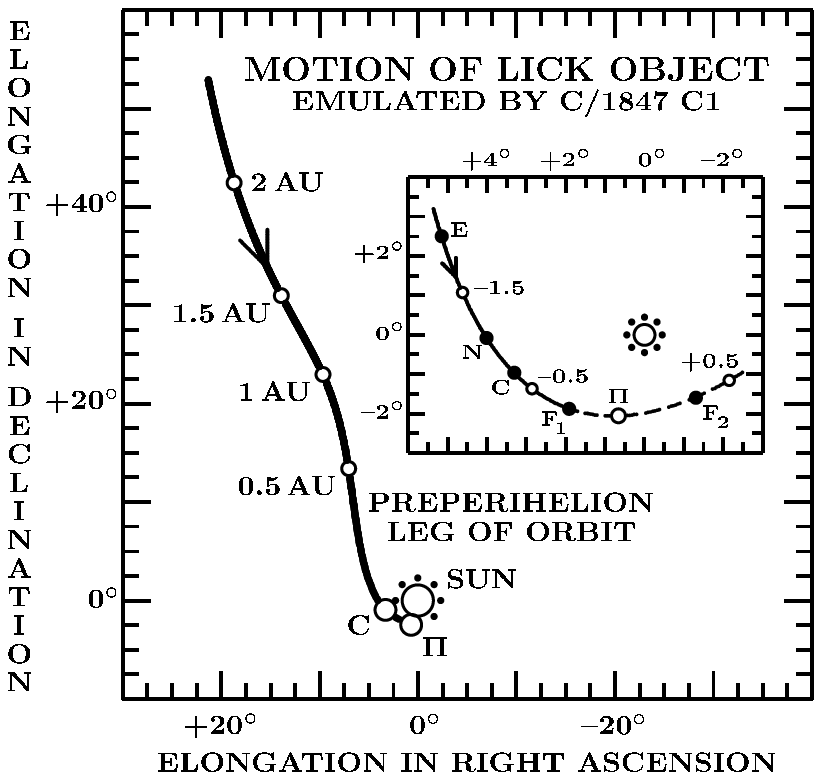}}} 
\vspace{0cm}
\caption{Apparent motions, relative to the Sun, of comet C/1847 C1 (left)
and of the Lick object, emulated by the comet's motion (right).  The scale
of the plot on the right is twice the scale on the left; $\Pi$ is the
perihelion point on, respectively, 1847 March 30.8 (left) and 1921 August
8.8 (right).  Heliocentric distances at the comet's positions and the Lick
object's predicted positions are depicted by the open circles.  The elongation
is reckoned negative to the south and the west of the Sun. --- {\sf Left}:\
The comet's motion from 1846 December 20 (101~days before perihelion) to
1847 July 10 (101~days after perihelion).  Unlike C/1823~Y1, C/1847~C1 was
approaching the Sun from the north.  Closeup of the comet's motion near
perihelion is displayed in the inset, whose scale is 10 times the scale of
the panel and in which the size of the Sun is drawn to the scale.  The small
open circles show the comet's positions 1 and 0.5~days before perihelion and
0.5, 1, and 1.5 days after perihelion.  --- {\sf Right}:\ The Lick object's
emulated motion from 1921 May 1 (100~days before perihelion) to perihelion;
C refers to the location of the object at the time of Campbell et al.'s
observation on 1921 August 8.135 UT.  Closeup of the Lick object's emulated
motion near perihelion is shown in the inset, whose scale is 4 times the
scale of the panel and in which the size of the Sun is drawn to the scale.
The small open circles show the object's predicted positions 1.5 and 0.5~days
before perihelion and 0.5 days after perihelion; C, N, and E refer to the
object's locations at the times of Campbell et al.'s, Nelson Day's, and
Emmert's observations, respectively; and F$_1$ and F$_2$ are the first two
occasions, at sunrise and sunset on August 8 at Lick, on which Campbell
failed to recover the object.{\vspace{0.25cm}}}
\end{center}
\end{figure*}

While we find that the range of absolute magnitude is plausible for a companion
of an intrinsically bright long-period comet, this model for the Lick object
has difficulties that involve other tests besides test~5.  The emulated
trajectory for the period of August 6--8 predicts that the object should have
moved from the southwest to the northeast (Figure~7), passing closest to the
Sun, at an elongation of 3$^\circ\!$.16, on August~7.867 UT, about 2~hours
after Nelson Day's observation.  As a result, the object's modeled elongation
was increasing with time when Campbell et al.\ detected it.  While this by
itself is not contradicted by the observation (because the direction of
the object's motion relative to the Sun during the several minutes of
monitoring at Lick could not be established), the predicted elongation of
3$^\circ\!$.2 at the time of the Ferndown observation is not consistent with
the observed 4$^\circ$ elongation and its estimated uncertainty of
$\pm$30$^\prime$ (test 2 in Table~9).

The Lick object's predicted motion after August~7 suggests that the
day-to-day observing conditions were improving at sunset, but remained poor
at sunrise.  Of the three occasions on which Campbell failed to recover the
object, the best opportunity presented itself unquestionably at sunset on
August~8, when the predicted location was about 5$^\circ\!$.3 from the Sun
in position angle of $\sim$70$^\circ$; the object would then still have
had some 30~hr to go to perihelion.  This hypothesis would thus indicate that
the object disintegrated more than 30~hr before perihelion.

Finally, we examine the implications of the nearly 100~yr gap between the
perihelion times of C/1823~Y1 and the Lick object for the separation
velocity at the time of their presumed common parent's fragmentation.
With the orbital period of 2110$\,\pm\,$260~yr (Table A-2), the relation
between the statistically averaged separation velocity [governed by
Equation~(26)], $\langle V_{\rm sep} \rangle$ (in m s$^{-1}$), and the
heliocentric distance at separation, $r_{\rm frg}$ (in AU), is
\begin{equation}
\langle V_{\rm sep} \rangle = 6.2_{-1.1}^{+1.5} \; r_{\rm frg}^{\frac{1}{2}}.
\end{equation}
We note that even if the parent comet split right at perihelion, the required
separation velocity would be close to 3~m~s$^{-1}$, already exceeding the
plausible upper limit (Section 2.1.1); and at \mbox{$r_{\rm frg} \simeq 2$--3
AU}{\vspace{-0.04cm}} it would reach a completely intolerable magnitude of
nearly 10~m~s$^{-1}$.  The only way to avoid this contradiction is to assume
that the Lick object survived not one but two (or more) revolutions about the
Sun, an option that is rather unattractive.  Accordingly, it appears that this
scenario fails to satisfy test~6.

We summarize by noting that while the hypothesis of a genetic relationship
between the Lick object and comet C/1823~Y1 passes tests 1, 3, and 7, it is,
strictly, incompatible with the constraints presented by tests 2, 5, 6, and,
to some degree, 4.
 
\subsection{Comet C/1847 C1 (Hind)}
This is the last of the top three candidates in Table~8 and the most promising
one, as argued below. Figure~8 shows that, unlike C/1823~Y1, this comet was
approaching the Sun from the high northern declinations; it was discovered by
Hind 52~days before perihelion (Bishop \& Hind 1847), when 1.49~AU from the
Sun, as a faint telescopic object.  The comet brightened steadily and was
detected with the naked eye more than three weeks before perihelion (Schmidt
1847a, 1847b; Bond 1847).  The discoverer observed the comet telescopically
in broad daylight on 1847 March 30, several hours before perihelion, within
2$^\circ$ of the Sun (Hind 1847).  As Figure~8 shows, the comet remained close
to the Sun during the first few weeks after perihelion.  As a result, there
were merely {\it three\/} astrometric positions available along the receding
branch of the orbit, made on April~22--24.  Interestingly, around April~12
the projection conditions were favorable for the appearance of an antitail
(Sekanina 1976); unfortunately, the comet was then only 15$^\circ$ from the
Sun.

Employing a total of 160 astrometric observations made between 1847 February~6
and April~24 and accounting for the perturbations by Mercury through Jupiter,
Hornstein (1870) derived an elliptical orbit with a period of \mbox{10\,220
$\pm$ 570}~yr for the osculation epoch at perihelion, implying a true orbital
period (i.e., a time interval between 1847 and the previous perihelion) of
\mbox{8310 $\pm$ 570}~yr (Marsden et al.\ 1978).  To find out whether we
could rely on Hornstein's computations, we derived two new sets of orbital
elements for C/1847~C1 based on the observations made only in London (critical
at the beginning of the orbital arc), Berlin (critical at its end), and Vienna.
Choice of two different cutoffs for the residuals of rejected observations
($\pm$10$^{\prime\prime}$ and $\pm$6$^{\prime\prime}$) yielded osculation
orbital periods of, respectively, \mbox{8190 $\pm$ 930}~yr and \mbox{14\,100
$\pm$ 2300}~yr.  Interestingly, Hornstein (1854a) estimated, in his earlier
study based on 145 astrometric observations from the same period of time, that
the orbital period was not shorter than 8000~yr and not longer than 14\,000~yr.
In that early study, Hornstein (1854a, 1854b) established the most probable
orbital period of 10\,818~yr for the osculation epoch at perihelion; the
details of that paper indicate the orbital period's rms error of $\pm$1510~yr.
Even though the quality of Hornstein's final set of orbital elements was
classified as only category~2B by Marsden et al.\ (1978), the numbers for the
orbital period appear to be consistent.  We conclude that Hornstein's results
are credible enough to indicate that the comet's true orbital period amounted
to about 8000~yr (with an uncertainty of several centuries) and was compatible
with the constraints of test~7 in Table~9.
%

As a companion to C/1847~C1, the Lick object was at the time of Campbell et
al.'s observation a little beyond the Sun, so that --- unlike in the C/1823~Y1
scenario --- its observed brightness was unaffected by forward scattering.
On the other hand, the heliocentric distance was now much smaller, which made
the object brighter.  Accounting for a phase effect (Marcus 2007), we find
that the object should have appeared to Campbell et al.\ about 10.7~mag
brighter than was its absolute magnitude with an $r^{-4}$ law but 13.6~mag
brighter with an $r^{-5}$ law.  The estimated apparent magnitude of $-$4.3
implies for the two laws an absolute magnitude of \mbox{$H_0 \approx 6.4$}
and 9.3, respectively.  We argue that, again, this is a plausible range of
absolute magnitude for a companion of an intrinsically bright long-period
comet, with the bright end of the range very close to the absolute magnitude
of C/1996~Q1, one of the companions to C/1988~A1 investigated in detail in
Section~2.2 (Figure~3).

Because the trajectory of the Lick object emulated by C/1847~C1 runs in
Figure~8 in a generally southwestern direction on August~6--8, the predicted
motion brought the object almost 2$^\circ$ south of the Sun in close proximity
of perihelion late on August~8.  Viewed from the Lick Observatory, the object
was still more than 2$^\circ$ below the horizon at sunrise and already about
2$^\circ$ below the horizon at sunset on August~8, so in this scenario
Campbell's failure to see it was due to unfavorable geometry and not
necessarily due to its disintegration.  However, by sunrise on August~9 the
object swayed already far enough to the west of the Sun that it rose before
the Sun, with a net difference of more than 2$^\circ$ in elevation.  Hence,
Campbell's failure to see the Lick object at sunrise on August~9, the last
occasion on which he was searching for it, cannot be explained by unfavorable
geometry.  However, the constraint on the object's disintegration time is
relaxed from $\sim$6~hr preperihelion to $\sim$18~hr {\it past\/} perihelion.

A review of the tests in Table 9 shows that they~all~are satisfied by
C/1847~C1.  Particularly gratifying is this scenario's correspondence with
the evidence, criteria, and requirements in tests~\mbox{1--4}.  The Lick
object's elongation on approach to the Sun is predicted to have dropped
below 50$^\circ$ at a heliocentric distance of more than 2~AU and to continue
decreasing with time until after perihelion.

The absolute magnitude of C/1847~C1 has{\vspace{-0.04cm}} not been well
determined.  For an $r^{-4}$ law, Vsekhsvyatsky (1958) gave \mbox{$H_{10} =
6.8$}, but the brightness estimates by Schmidt (1847a, 1847b) and by Bond
(1847) strongly suggest that between 1.1~AU and 0.5~AU from the Sun the comet
brightened less steeply with decreasing heliocentric distance.  Correcting a
few available magnitude estimates for the phase effect (Marcus 2007) and the
geocentric distance, we obtained by least squares \mbox{$H_0 = 5.41 \pm 0.13$}
and \mbox{$n = 2.02 \pm 0.31$}.  We note that Holetschek (1913), assuming a
brightness variation with an inverse square heliocentric distance derived from
a single estimate \mbox{$H_0 = 5.7$}, a value that, like Vseksvyatsky's, was
not corrected for a phase effect.  Because the phase angle varied between
55$^\circ$ and 90$^\circ$ during the relevant period of time, an average
phase correction should make the absolute magnitude brighter by $\sim$0.4 to
$\sim$0.7~mag, depending on the dust content.

Since Figures 1 and 3 show that companion comets have generally steeper light
curves than the primaries, the difference between the derived slope for the
light curve of C/1847~C1, as the pair's primary, and the assumed range of
slopes (\mbox{$n = 4$ to 5}) for the light curve of the Lick object, as the
companion to C/1847~C1, are not unrealistic.  In summary, we find that even
if the Lick object was not in terminal outburst when observed by Campbell et
al., its absolute magnitude was probably at least 1~mag --- and possibly
several magnitudes --- fainter than the absolute magnitude of C/1847~C1.  If
the object was in terminal outburst, its quiescent-phase absolute magnitude
was yet another magnitude or so fainter.

As the last point of our investigation of C/1847~C1 and the Lick object, as a
pair of long-period comets that are likely to have split apart from a common
parent near its perihelion some 8310$\,\pm\,$570~yr before 1847, is their
statistically averaged separation velocity $\langle V_{\rm sep} \rangle$
(Section~2.2.1), an issue that we focused on in discussing the fragmentation
of long-period comets in Sections~2.1 and 2.2.  With a temporal gap of 74.4~yr
between the arrivals of C/1847~C1 and the{\vspace{-0.04cm}} Lick object
(Table~9), the separation velocity (in m s$^{-1}$) is equal to
\begin{equation}
\langle V_{\rm sep} \rangle = 0.48_{-0.04}^{+0.05}\; r_{\rm frg}^{\frac{1}{2}}.
\end{equation}
If one assumes that the common parent of C/1847~C1 and the Lick object split
right at perihelion, the fragments separated at a rate{\vspace{-0.04cm}} of
\mbox{$\langle V_{\rm sep} \rangle = 0.1$ m s$^{-1}$}; if at a heliocentric
distance of \mbox{$r_{\rm frg} \simeq 1$ AU}, then \mbox{$\langle V_{\rm sep}
\rangle \simeq 0.5$ m s$^{-1}$};{\vspace{-0.04cm}} and if at \mbox{$r_{\rm frg}
\simeq 10$ AU}, \mbox{$\langle V_{\rm sep} \rangle \simeq 1.5$ m s$^{-1}$}.
This is a typical range of separation velocities as documented by the known
split comets (Section 2.1.1; Sekanina 1982).

In spite of the incompleteness of the data on the Lick object, we believe
that a fairly compelling case has been presented for its genetic relationship
with C/1847~C1, implying that their separation had occurred in the general
proximity of the previous perihelion in the 7th millennium BCE.  This example
demonstrates that the temporal gap between the members of a genetically related
pair or group of long-period comets could easily reach many dozens of years and
that both the primary and the companion can survive for millennia after the
parent comet's splitting.  Finally, it is nothing short of remarkable that a
premise of the membership in a comet pair can successfully be applied to
instances of unidentified interplanetary objects with poorly known motions.

\section{Conclusions}
The objectives of this investigation were (i)~to describe the dynamical
properties of the genetically related members of pairs or groups of
long-period comets other than the Kreutz system of sungrazers and (ii)~by
exploiting these findings to determine whether the intriguing and as yet
unidentified Lick object, discovered near the Sun at sunset on 1921 August~7,
happened to be a member of such a pair and to which long-period comet could
it be genetically related.  Our conclusions are as follows:

(1) The orbits of genetically related long-period comets are --- except for
the perihelion time --- nearly identical and among objects with dependably
determined motions their common origin cannot be disputed.

(2) Up to now, however, only two groups of such genetically related long-period
comets have been recognized:\ a pair of C/1988~F1 (Levy) and C/1988~J1
(Shoemaker-Holt); and a trio of C/1988~A1 (Liller), C/1996~Q1 (Tabur), and
C/2015~F3 (SWAN).

(3) On an example of the comet pair of C/1988~F1 and C/1988~J1 we affirm that,
of the two mechanisms that evoke a steadily increasing separation of fragments
of a split comet with time, an orbital-momentum increment acquired at the
parent's breakup is overwhelmingly more consequential for the motions of
genetically related long-period comets than is the role of an outgassing-driven
differential nongravitational acceleration.  Initial velocities (especially the
component along the radius vector) that set the fragments apart do not exceed
a few m~s$^{-1}$ and are equivalent to separation velocities of fragments of
the known split comets (Sekanina 1982).  This result, confirmed by numerical
experimentation, makes sense, as the nongravitational effects are utterly
trifling far from the Sun and become significant only along a short orbital
arc close to perihelion.

(4) The pair of C/1988 F1 and C/1988 J1 differs from the trio C/1988~A1,
C/1996~Q1, and C/2001~F3 in that the pair's parent fragmented only several
hundred AU from the Sun after having passed intact through aphelion on
its way to the 1987 perihelion, whereas the trio's parent broke up in
the general proximity of perihelion during its previous return to the Sun.

(5) We emphasize that the trio's members appear to have differed from one
another in their appearance, morphology, and physical behavior, as reflected
by their light curves.  Only C/1988~A1, presumably the most massive member,
showed no sign of imminent disintegration; C/1996~Q1, a dust-poor fragment,
lost its nuclear condensation already before perihelion and, appearing as
a headless tail, faded progressively afterwards; whereas C/2015~F3 was
quite prominent in the SWAN Lyman-alpha atomic-hydrogen images relative
to its optical brightness, implying tentatively a high water-to-dust
production ratio still to be confirmed by a future research; this most recent
fragment survived perihelion seemingly intact, but was fading steeply while
receding from the Sun.

(6) The experience with both the pair and the trio of genetically related
comets shows the orbits of either group's members so much alike that the
near-perihelion motion of one member can readily be emulated by a set
of orbital elements of another member, with the perihelion time being the
only parameter to be changed; the positions are found to fit within a few
tenths of an arc\-min for the pair of C/1988~F1 and C/1988~J1, and to better
than 10$^\prime$ for the trio of C/1988~A1, C/1996~Q1, and C/2015~F3.

(7) This remarkable match suggests that a meaningful search for a member of
an as yet unrecognized pair or group of genetically related comets can be
carried out by scouting around for a comet whose motion offers this orbital
correspondence, even when the astrometric data for the examined object are
not sufficient to compute a set of orbital elements in a standard fashion.

(8) This idea is tested on a celebrated bright object of 1921 August 7, whose
discovery near the Sun was made with the naked eye at sunset by a group of
people, including two prominent astronomers, from the premises of the Lick
Observatory.  However, the Lick object was searched for unsuccessfully at
sunrise on August~8 and 9 and at sunset on August~8.  It was definitely not
a Kreutz sungrazer and its identity has remained shrouded in mystery to this
day.

(9) The search for a comet that could be genetically related to the Lick
object was based primarily on the object's estimated position at sunset
on August~7, published by Campbell and believed to be accurate to about
$\pm$10$^\prime$; this was the primary test.

(10) Two independent, though less accurate observations made on 1921 August
6--7 as well as additional constraints --- such as the requirement of an
unfavorably oriented approach trajectory in the sky (at persistingly small
elongations) to explain the failure to detect the Lick object earlier; or
a preference for the related comet to have arrived at perihelion prior to
the Lick object --- were employed to develop additional tests.

(11) Brightness arguments strongly suggested that the Lick object's
close proximity to the Sun was not merely an effect of projection, but
that the actual distance between the object and the Sun was indeed very
small when observed at the Lick Observatory.

(12) Accordingly, among comets with perihelion distances not exceeding
0.25~AU (a conservative upper limit) we searched for a comet that could
be genetically related to the Lick object; we found three candidates that
satisfied the primary constraint based on the Campbell report: C/1689~X1,
C/1823~Y1, and C/1847~C1 (Hind).

(13) Close examination of the past work on the motion of C/1689~X1 showed
that it is essentially indeterminate; an apparent match of one of the
comet's many different orbits to the Lick object must be fortuitous.  The
orbit of C/1823~Y1 fails to satisfy up to four of the seven tests, including
a limit on the separation velocity at breakup.

(14) Only the orbit of C/1847~C1 is fully compatible with all applied
tests, suggesting that if the Lick object is a fragment of another
comet, C/1847~C1 is by far the most likely primary fragment of the same
parent body.

\begin{table*}[t]
\vspace{0.1cm}
\begin{center}
{\footnotesize {\bf Table A-1} \\[0.1cm]
{\sc Residuals from the Orbital Solution for C/1823 Y1 Based on 102 Astrometric
Observations\\Between 1824 January 2 and March 31 (Equinox J2000).}\\[0.15cm]
\begin{tabular}{l@{\hspace{0cm}}r@{\hspace{0.25cm}}c@{\hspace{0.2cm}}c@{\hspace{0.3cm}}l@{\hspace{0.15cm}\vline\hspace{0.15cm}}l@{\hspace{0.07cm}}r@{\hspace{0.25cm}}c@{\hspace{0.02cm}}c@{\hspace{0.3cm}}l@{\hspace{0.15cm}\vline\hspace{0.15cm}}l@{\hspace{-0.03cm}}r@{\hspace{0.25cm}}c@{\hspace{0.15cm}}c@{\hspace{0.3cm}}l}
\hline\hline\\[-0.35cm]
\multicolumn{2}{@{\hspace{0.15cm}}c}{\raisebox{0ex}[3ex]{Time of}}
& \multicolumn{2}{@{\hspace{-0.05cm}}c}{Residual in}
& & \multicolumn{2}{@{\hspace{-0.05cm}}c}{Time of}
& \multicolumn{2}{@{\hspace{-0.05cm}}c}{Residual in}
& & \multicolumn{2}{@{\hspace{-0.05cm}}c}{Time of}
& \multicolumn{2}{@{\hspace{0cm}}c}{Residual in}
& \\[-0.06cm]
\multicolumn{2}{@{\hspace{0.15cm}}c}{observation} &
\multicolumn{2}{@{\hspace{-0.05cm}}c}{\rule[0.7ex]{1.6cm}{0.4pt}} & Observ- &
\multicolumn{2}{@{\hspace{-0.05cm}}c}{observation} &
\multicolumn{2}{@{\hspace{-0.05cm}}c}{\rule[0.7ex]{1.6cm}{0.4pt}} & Observ- &
\multicolumn{2}{@{\hspace{-0.05cm}}c}{observation} &
\multicolumn{2}{@{\hspace{0cm}}c}{\rule[0.7ex]{1.6cm}{0.4pt}}
& Observ- \\[-0.04cm]
\multicolumn{2}{@{\hspace{0.15cm}}c}{1824 (UT)} & \,R.A. & \,Dec.
& atory &
\multicolumn{2}{@{\hspace{-0.05cm}}c}{1824 (UT)} & \,R.A. & \,Dec.
& atory &
\multicolumn{2}{@{\hspace{-0.05cm}}c}{1824 (UT)} & \,R.A. & \,Dec.
& atory \\[0.1cm]
\hline \\[-0.4cm]
\raisebox{0cm}[4ex]{Jan.} &  2.23975 & $-$8$^{\prime\prime}\!\!$.2 & $-$5$^{\prime\prime}\!\!$.3
& Paris &
Jan. & 15.22277 & $-$1$^{\prime\prime}\!\!$.7 &   +6$^{\prime\prime}\!\!$.1
& Marseilles &
Feb. &  7.24658 &   +4$^{\prime\prime}\!\!$.2 & $-$5$^{\prime\prime}\!\!$.1
& Marseilles \\
     &  2.28741 &   +5.1 & $-$2.4 & Greenwich &
     & 15.22806 & $-$0.7 & $-$2.9 & $\;\;\;\;$ " &
     &  7.88922 & $-$6.3 & $-$4.9 & Nikolayev \\
     &  2.28883 &   +9.1 &   +6.3 & $\;\;\;\;$ " &
     & 15.22888 & $-$0.2 & $-$4.3 & $\;\;\;\;$ " &
     & 10.74380 &   +0.1 & $-$8.1 & Vienna \\
     &  4.25867 & $-$3.4 &   +1.1 & Mannheim &
     & 15.23419 &   +2.3 &   +6.7 & $\;\;\;\;$ " &
     & 12.22860 & $-$3.1 & $-$0.8 & Marseilles \\
     &  5.19242 & $-$2.1 &   +2.3 & $\;\;\;\;$ " & 
     & 15.23506 &   +2.0 &   +8.4 & $\;\;\;\;$ " &
     & 17.76379 & $-$9.8 &   +0.7 & Vienna \\[0.1cm]
     &  6.19679 & $-$2.0 & $-$4.8 & $\;\;\;\;$ " &
     & 15.24029 &   +0.2 & $-$2.4 & $\;\;\;\;$ " &
     & 18.78077 & +10.5$\;\:$ & $-$1.6 & Mannheim \\
     &  6.21341 &   +6.2 & $-$6.2 & Vienna &
     & 16.03822 &   +0.4 & $-$4.6 & Nikolayev &
     & 22.90792 &   +6.2 &   +2.6 & $\;\;\;\;$ " \\
     &  7.16077 & $-$3.3 &   +8.3 & $\;\;\;\;$ " &
     & 16.04074 & $-$8.3 &   +4.4 & $\;\;\;\;$ " &
     & 24.80239 & $-$0.8 & $-$3.1 & Nikolayev \\
     &  7.16077 & $-$1.9 &   +4.2 & $\;\;\;\;$ " &
     & 18.08508 & $-$5.1 & $-$3.8 & $\;\;\;\;$ " &
     & 24.81977 &   +1.5 & $-$1.1 & $\;\;\;\;$ " \\
     &  7.19604 &   +1.0 & $-$0.2 & Marseilles &
     & 19.15802 & $-$1.6 & $-$8.1 & Marseilles &
     & 24.86462 &   +6.4 & $-$6.6 & $\;\;\;\;$ " \\[0.1cm]
     &  7.20727 & $-$4.8 &   +7.6 & $\;\;\;\;$ " &
     & 19.18314 &   +3.1 & $-$6.7 & $\;\;\;\;$ " &
     & 27.92409 & $-$5.9 & $-$3.4 & $\;\;\;\;$ " \\
     &  7.20815 & $-$4.3 &   +8.2 & $\;\;\;\;$ " &
     & 19.18422 &   +3.7 & $-$5.9 & $\;\;\;\;$ " &
     & 27.93371 &   +5.5 & $-$1.8 & Mannheim \\
     &  7.21940 &   +5.4 & $-$7.0 & $\;\;\;\;$ " &
     & 19.19462 &   +9.6 &   +4.1 & $\;\;\;\;$ " &
     & 27.94977 &   +1.8 &   +8.9 & Nikolayev \\
     &  7.21940 &   +2.9 & $-$8.9 & $\;\;\;\;$ " &
     & 24.99474 &   +6.9 & $-$0.5 & $\;\;\;\;$ " &
     & 27.95905 &   +5.7 &   +2.6 & $\;\;\;\;$ " \\
     &  9.20090 &   +2.3 & $-$6.4 & $\;\;\;\;$ " &
     & 27.04860 & +11.6$\;\:$ & +1.2 & Nikolayev &
     & 29.81069 & $-$0.7 & $-$2.8 & $\;\;\;\;$ " \\[0.1cm]
     &  9.22572 &   +7.4 & $-$1.5 & $\;\;\;\;$ " &
     & 30.85869 & +0.2 & +10.4$\;\:$ & Marseilles &
     & 29.82372 & $-$4.4 & $-$3.8 & $\;\;\;\;$ " \\
     &  9.25445 &   +0.4 & $-$4.4 & $\;\;\;\;$ " &
     & 30.85869 &   +0.6 & $-$1.7 & $\;\;\;\;$ " &
     & 29.83197 & $-$2.3 & $-$7.7 & $\;\;\;\;$ " \\
     &  9.25725 & $-$3.0 &   +1.2 & $\;\;\;\;$ " &
     & 30.85869 & $-$0.9 &   +6.8 & $\;\;\;\;$ " &
     & 29.84108 & $-$0.4 & $-$0.6 & $\;\;\;\;$ " \\
     &  9.25805 & $-$2.0 &   +2.8 & $\;\;\;\;$ " &
     & 30.86334 &   +2.7 &   +4.1 & $\;\;\;\;$ " &
Mar. &  2.93493 & $-$4.0 & $-$6.7 & Dorpat$^{\rm a}$ \\
     &  9.26293 &   +4.0 & $-$2.7 & $\;\;\;\;$ " &
     & 30.86334 &   +2.8 & $-$1.0 & $\;\;\;\;$ " &
     &  4.83756 & $-$2.9 & $-$3.4 & Nikolayev \\[0.1cm]
     & 10.24861 & $-$7.7 &   +2.8 & Mannheim & 
     & 30.86334 & $-$0.8 &   +2.6 & $\;\;\;\;$ " &
     &  4.84897 &   +3.1 &   +8.4 & $\;\;\;\;$ " \\
     & 12.22219 &   +3.7 & $-$6.4 & Marseilles &
     & 30.86712 &   +0.4 & $-$1.0 & $\;\;\;\;$ " &
     &  4.85972 & +10.4$\;\:$ & $-$0.2 & $\;\;\;\;$ " \\
     & 12.24189 & $-$4.5 & $-$6.8 & $\;\;\;\;$ " &
     & 30.86712 &   +2.0 & $-$0.5 & $\;\;\;\;$ " &
     &  4.86952 &   +6.0 &   +6.8 & $\;\;\;\;$ " \\
     & 12.24543 &   +1.7 & $-$6.8 & $\;\;\;\;$ " &
     & 30.86712 & $-$0.8 & $-$0.8 & $\;\;\;\;$ " &
     &  4.94224 &   +0.9 &   +2.4 & Mannheim \\
     & 12.24619 &   +3.9 & $-$6.7 & $\;\;\;\;$ " &
     & 30.87187 & $-$2.6 & $-$4.1 & $\;\;\;\;$ " &
     &  6.86113 & $-$7.7 & $-$3.3 & Nikolayev \\[0.1cm]
     & 13.20964 &   +5.3 & $-$1.3 & Mannheim &    
     & 30.87187 &   +0.9 &   +2.0 & $\;\;\;\;$ " &
     &  6.95187 & $-$5.9 &   +6.2 & $\;\;\;\;$ " \\
     & 13.22120 & $-$7.4 &   +6.3 & Marseilles &
     & 30.87187 & $-$0.8 &   +5.6 & $\;\;\;\;$ " &
     &  6.97363 & $-$7.7 & $-$3.0 & $\;\;\;\;$ " \\
     & 13.23014 & $-$8.0 &   +6.3 & $\;\;\;\;$ " &
Feb. &  1.85089 &   +1.4 & $-$7.0 & $\;\;\;\;$ " &
     & 18.77981 & $-$2.8 &   +1.9 & $\;\;\;\;$ " \\
     & 14.19640 &   +0.8 &   +8.3 & $\;\;\;\;$ " &
     &  1.87081 &   +4.1 & $-$10.3$\;\:$ & $\;\;\;\;$ " &
     & 18.78533 &   +2.2 &   +1.8 & $\;\;\;\;$ " \\
     & 14.20295 & $-$3.3 &   +6.7 & $\;\;\;\;$ " &
     &  2.75793 &   +3.2 &   +2.1 & Mannheim &
     & 20.79306 & $-$1.2 & $-$4.9 & $\;\;\;\;$ " \\[0.1cm]
     & 14.20384 & $-$2.6 &   +6.5 & $\;\;\;\;$ " &
     &  4.75398 & $-$8.3 & $-$5.3 & Vienna &
     & 20.79955 & $-$7.9 &   +6.8 & $\;\;\;\;$ " \\
     & 14.21009 &   +1.4 &   +4.5 & $\;\;\;\;$ " &
     &  4.75398 & $-$6.6 & $-$3.1 & $\;\;\;\;$ " &
     & 21.78829 &   +5.6 &   +8.4 & $\;\;\;\;$ " \\
     & 14.21106 &   +0.4 &   +4.2 & $\;\;\;\;$ " &
     &  4.80650 &   +8.1 &   +8.7 & Marseilles &
     & 22.78238 &   +2.0 &   +1.5 & $\;\;\;\;$ " \\
     & 14.21903 & $-$7.0 &   +2.8 & $\;\;\;\;$ " &
     &  4.82240 & $-$3.7 & $-$3.0 & $\;\;\;\;$ " &
     & 31.89001 &   +6.4 &   +4.6 & $\;\;\;\;$ " \\[0.1cm]
\hline\\[-0.15cm]
\multicolumn{15}{l}{\parbox{12cm}{\scriptsize $^{\rm a}$\,Nowadays:\ Tartu,
Estonia.{\vspace{0.95cm}}}}
\end{tabular}}
%
%
{\footnotesize {\bf Table A-2}\\[0.08cm]
{\sc New Orbital Elements for Comet C/1823 Y1
and Comparsion with Hnatek's Orbit\\(Equinox J2000.0).}\\[0.1cm]
\begin{tabular}{l@{\hspace{0.7cm}}c@{\hspace{0.6cm}}c}
\hline\hline\\[-0.22cm]
& & Differences \\[-0.04cm]
Quantity/Orbital element & New orbit & Hnatek's minus new orbit \\[0.08cm]
\hline\\[-0.2cm]
Osculation epoch (TT) & 1823 Dec.\,11.0 &  (1824 Feb.\,15.0) \\
Time of perihelion passage $t_\pi$ (TT) & 1823 Dec.\,9.94585$\:\pm\:$0.00063
 & $-$0.0116 \\
Argument of perihelion $\omega$ & $\;\:$28$^\circ\!$.2778$\:\pm\:$0$^\circ\!$.0160
 & +0$^\circ\!$.2073 \\
Longitude of ascending node $\Omega$ & 305$^\circ\!$.5647$\:\pm\:$0$^\circ\!$.0046 & $-$0$^\circ\!$.0589 \\
Orbit inclination $i$ & 103$^\circ\!$.6810$\:\pm\:$0$^\circ\!$.0110
 & +0$^\circ\!$.1374 \\
Perihelion distance $q$ & $\;\:$0.225238$\:\pm\:$0.00011 & +0.001492 \\
Orbital eccentricity $e$ & $\;\:$0.998698$\:\pm\:$0.00010 & +0.001302 \\
\hspace*{2.7cm}osculation & $\;$2280$\:\pm\:$260 & \,\ldots\ldots \\[-0.245cm]
Orbital period (yr)
 $\!\!\left\{ \raisebox{1.7ex}{}\raisebox{-1.7ex}{} \right.$\\[-0.305cm]
\hspace*{2.7cm}original$\:\!^{\rm a}$   & 2110 & \ldots\ldots \\[0.05cm]
\hline\\[-0.26cm]
Orbital arc covered by observations & 1824 Jan.\,2--1824 Mar.\,31
& \ldots\ldots \\
Number of observations employed & 102 & \ldots\ldots \\
Root-mean-square residual & $\pm$5$^{\prime\prime}\!\!.00$ & \ldots\ldots \\
Orbit-quality code$^{\rm b}$ & 2B & \ldots\ldots \\[0.1cm]
\hline\\[-0.25cm]
\multicolumn{3}{l}{\parbox{10cm}{\scriptsize $^{\rm a}$\,With respect to
 the barycenetr of the Solar System.}}\\[-0.08cm]
\multicolumn{3}{l}{\parbox{10cm}{\scriptsize $^{\rm b}$\,Following the
 classification system introduced by Marsden et al.\ (1978).}}
\end{tabular}}
\end{center}
\end{table*}

(15) In terms of the intrinsic brightness, the Lick object was probably
at least 1~mag fainter than C/1847~C1; the temporal gap of 74.4~yr
between the perihelion times suggests that they separated --- with a
relative velocity not exceeding $\sim$1.5~m~s$^{-1}$ and probably lower
than 1~m~s$^{-1}$ --- from their parent during its breakup in a general
proximity of perihelion at its return to the Sun in the course of the
7th millennium BCE.

(16) Campbell's failure to recover the object at sunrise on August~9
could be due to its disintegration near perihelion, an event also
experienced by C/1996~Q1.

(17) The example of the 1921 Lick object shows that members of a
genetically related pair or group of long-period comets could be
separated from one another by many dozens of years and that both the
primary and the companion can survive for thousands of years after
their parent's splitting.  The proposed method of search merits
applications to other instances of unidentified objects with poorly
known motions.

(18) As part of our examination of the true \mbox{orbital} periods of the
promising candidate comets, we recomputed the orbital elements of C/1823~Y1
and established that its true orbital period was $\sim$2100~yr and that the
``definitive'' parabolic orbit by Hnatek (1912) in the Marsden-Williams (2008)
catalog is misleading.\\[-0.1cm]

This research was carried out in part at the Jet Pro\-pulsion Laboratory,
California Institute of Technology, under contract with the National
Aeronautics and Space Administration.\\[-0.1cm]

\begin{center}
APPENDIX\\[0.15cm]
NEW ORBIT DETERMINATION FOR C/1823 Y1
\end{center}
We remarked in Section 5.2 that this comet's orbit determination by
Hnatek (1912), although the most comprehensive on record, suffered
from a defect in that he presented a parabola as his ``definitive''
solution without submitting any compelling evidence to demonstrate
that the orbital period was indeterminate.

In the process of refining the orbit, Hnatek deduced separately the sets
of elements for both the ``most probable parabola'' and the ``most
probable ellipse.''  The eccentricity of this elliptical orbit was
listed by the autor as 0.9995048.  In addition, Hnatek also provided
the sums of squares of residuals from his 8 normal places not only for
the most probable parabola and ellipse, but also for four additional
forced eccentricities between 0.9994 and 1.0002.  Hnatek's paper shows that
the sum of squares of residuals for the most probable ellipse was 0.8
the sum of squares of residuals for the most probable parabola. which
by itself suggests that the two solutions were by no means equivalent.
Yet, immediately following the table of the residual sums, Hnatek
addressed this issue very differently, saying that ``as seen from [the
table], one can vary the orbital period to rather widely stretched
limits without running into conflict with the observations and can
thereby conclude that a parabola can ultimately be adopted as a
definitive shape of the orbit.''  Then he proceeded to optimize the
parabolic solution after discarding the last normal place in both
right ascension and declination because of the residuals that reached
10$^{\prime\prime}$ to 20$^{\prime\prime}$.  This curtailed the orbital
arc covered by the observations by more than three weeks, from 92 days
down to 68 days.

From the entries in the table of the sums of squares of residuals nearest
Hnatek's most probable ellipse we found that the mean error of the
eccentricity was $\pm$0.000120, whereas the deviation of the eccentricity
from unity for this solution was 0.000495, more than a factor of 4 higher.
This result suggests that the choice of the parabolic approximation was 
unjustified.  Since the discarded normal place was based on 8 observations
between 1824 March 17 and 31, while the previous two normal places contained
only 4 observations each, from February 29 through March 3 and from March~4
through March~7, respectively, it probably was a mistake to discard the last
one, with the greatest weight of the three.  We feel that this phase of
Hnatek's work on the orbit of C/1823~Y1 precluded him from capitalizing on
the great potential that his project had offered.

Under the circumstances, our decision to go ahead and undertake the task of
redetermining the orbit of comet C/1823~Y1 from scratch was motivated by two
issues.  One, as this comet was potentially genetically related to the Lick
object, the requirement that its true orbital period be confined to particular
limits was one of the critical conditions that had to be satisfied.  Thus,
the orbit's ellipticity was urgently needed.  And, two, it was desirable to
update the astrometric positions of the comparison stars used by Hnatek with
the more accurate ones from the {\it Hipparcos\/} and {\it Tycho-2\/}
catalogs.\footnote{The search facilities are \mbox{available at the following
websites:} {\tt
http://\,www.rssd.esa.int/index.php?project=HIPPARCOS\&page= hipsearch}
for the\,{\it Hipparcos\/}\,(and~the original\,{\it Tycho\/})\,catalog and {\tt
http:/$\!$/vizier.u-strasbg.fr/viz-bin/VizieR-3?-source=I/259/ tyc2\&-out.add=}
for the {\it Tycho-2\/} catalog.{\vspace{0.1cm}}}

To collect the data for our computations, whose results are summarized in
Tables A-1 and A-2, we employed the comet's observed offsets, in right
ascension and declination, from the comparison stars identified for a majority
of the astrometric observations by Hnatek (1912).\footnote{A sizable minority
of the observations was available only as the comet's apparent positions with
no comparison stars specified; these observations were
ignored.{\vspace{-0.25cm}}}  When, at a given time, a comparison of the
comet's position with a star's position was made only in right ascension
or only in declination, the offset in the other coordinate was determined
by interpolating the neighboring offsets only when measured from the same
comparison star and when involving intervals not exceeding $\sim$1~hr; no
offsets were ever extrapolated.

Once the positions of the relevant comparison stars were identified in the
{\it Hipparcos\/} catalog (in more than 90\% of cases) or the {\it Tycho-2\/}
catalog, a total of 388 positions of the comet were re-reduced and incorporated
into a differential least-squares optimization procedure to compute an initial
set of orbital elements by employing the {\it EXORB7\/} code.

The presence of a dozen observed positions with enormous residuals, exceeding
$\pm$10$^\prime$, suggested that several comparison stars were apparently
misidentified by the observers (or by Hnatek).  After removing these bad data,
we ended up with a total of 376 observations, which served as an input to
another orbit iteration.  Although we still found more than a dozen
observations whose residuals (in at least one coordinate) exceeded
$\pm$200$^{\prime\prime}$, the osculation orbital period already came out
to be close to 2000~yr, with an uncertainty of about $\pm$50\%.  After the
most inferior data were removed, the orbit was iterated again with
360~observations left in the solution.  Next, 32 more positional data
were removed with the residuals exceeding $\pm$100$^{\prime\prime}$
and a fourth iteration was carried out.  We then continued with a fifth
iteration based on 280 observations with the residuals not exceeding
$\pm$50$^{\prime\prime}$; with a sixth iteration based on 178~observations
with the residuals not exceeding $\pm$25$^{\prime\prime}$; and with a seventh
iteration based on 102~observations with the residuals not exceeding
$\pm$12$^{\prime\prime}$; the rms residual of this solution was
$\pm$5$^{\prime\prime}\!$.0.  We tested an eighth iteration based on
93~observations with the residuals not exceeding $\pm$10$^{\prime\prime}$,
but even though the rms residual dropped to $\pm$4$^{\prime\prime}\!$.6,
the errors of the orbital elements remained essentially unchanged from
the solution derived in the seventh iteration, which thus was the final
result of the computations.

The residuals from the 102~employed observations are listed in
Table~A-1, while the elements for the standard 40-day osculation epoch
nearest perihelion are presented in Table~A-2.  This table also shows
the differences between our and Hnatek's results (at the osculation
epoch of Hnatek's choice, 1824 February 15).  The differences are more
than a factor of 10 greater than the errors of our set of elements.
The original reciprocal~semimajor axis~is~equal~to~\mbox{$(1/a)_{\rm
orig} = +0.006077 \pm 0.000471$ (AU)$^{-1}$} and the true orbital period
to \mbox{$P_{\rm orig} = 2110 \pm 260$ yr}.~The comet's orbit is decidedly
elliptical and the true orbital period, derived with a 12\% uncertainty
and compatible with test~7, is much shorter than the orbital period of
Hnatek's most probable ellipse.  The comet definitely did not arrive
from the Oort Cloud.\\[-0.35cm]

\begin{center}
{\footnotesize REFERENCES}
\end{center}
\vspace*{-0.4cm}
\begin{description}
{\footnotesize
\item[\hspace{-0.3cm}]
Abney,\,W.\,de\,W., \& Schuster,\,A. 1884, Phil.\,Trans.\,R.\,Soc.\,London,
 \linebreak {\hspace*{-0.6cm}}175, 253
\\[-0.57cm]
\item[\hspace{-0.3cm}]
A'Hearn, M. F., Millis, R. L., Schleicher, D. G., et al.\ 1995, Icarus,
 {\hspace*{-0.6cm}}118, 223
\\[-0.57cm]
\item[\hspace{-0.3cm}]
A'Hearn, M. F., Schleicher, D. G., Feldman, P. D., et al.\ 1984, AJ,\linebreak
 {\hspace*{-0.6cm}}89, 579
\\[-0.57cm]
\item[\hspace{-0.3cm}]
Ashbrook, J. 1971, Sky Tel., 41, 352
\\[-0.57cm]
\item[\hspace{-0.3cm}]
Baratta, G. A., Catalano, F. A., Leone, F., \& Strazzulla, G. 1989,
 {\hspace*{-0.6cm}}A\&A, 219, 322
\\[-0.57cm]
\item[\hspace{-0.3cm}]
Baum, R. 2007, The Haunted Observatory.  Prometheus Books,
 {\hspace*{-0.6cm}}Amherst, NY
\\[-0.57cm]
\item[\hspace{-0.3cm}]
Bishop, G., \& Hind, J. 1847, MNRAS, 7, 247
\\[-0.57cm]
\item[\hspace{-0.3cm}]
Bond, W. C. 1847, MNRAS, 7, 273
\\[-0.57cm]
\item[\hspace{-0.3cm}]
Bortle, J. E. 1985, Intern. Comet Quart., 7, 7
\\[-0.57cm]
\item[\hspace{-0.3cm}]
Bortle, J. E. 1991, Intern. Comet Quart., 13, 89
\\[-0.57cm]
\item[\hspace{-0.3cm}]
Bortle, J. E. 1997, in Guide to Observing Comets, ed. D. W. E.{\linebreak}
 {\hspace*{-0.6cm}}Green (Cambridge, MA:\ Smithsonian Astrophysical
 Observa-{\linebreak}
 {\hspace*{-0.6cm}}tory), 90
\\[-0.57cm]
\item[\hspace{-0.3cm}]
Campbell, W. W. 1921a, Harv. Coll. Obs. Bull., 757
\\[-0.57cm]
\item[\hspace{-0.3cm}]
Campbell, W. W. 1921b, AN, 214, 69
\\[-0.57cm]
\item[\hspace{-0.3cm}]
Campbell, W. W. 1921c, Nature, 107, 759
\\[-0.57cm]
\item[\hspace{-0.3cm}]
Campbell, W. W. 1921d, PASP, 33, 258
\\[-0.57cm]
%
%
\item[\hspace{-0.3cm}]
Carusi, A., Perozzi, E., Valsecchi, G. B., \& Kres\'ak. L'. 1985,
 in{\linebreak}
{\hspace*{-0.6cm}}Dynamics of Comets: Their Origin and Evolution, IAU Coll.
83,{\linebreak}
{\hspace*{-0.6cm}}ed. A. Carusi \& G. B. Valsecchi (Dordrecht, Netherlands:
Reidel{\linebreak}
{\hspace*{-0.6cm}}Publ. Co.), 319
\\[-0.57cm]
\item[\hspace{-0.3cm}]
Cooper, E. J. 1852, Cometic Orbits. Alex.\,Thom, Dublin, Ireland
\\[-0.57cm]
\item[\hspace{-0.3cm}]
Crovisier, J., Colom, P., G\'erard, E., et al.\ 2002, A\&A, 393, 1053
\\[-0.57cm]
\item[\hspace{-0.3cm}]
de B\`eze, C., \& Comilh, P. 1729, M\'em.\,Acad.\,R.\,Sci.\,Paris, 7, 821
\\[-0.57cm]
\item[\hspace{-0.3cm}]
Emmert, H. C. 1921a, Harv. Coll. Obs. Bull., 759
\\[-0.57cm]
\item[\hspace{-0.3cm}]
Emmert, H. C. 1921b, PASP, 33, 261
\\[-0.57cm]
\item[\hspace{-0.3cm}]
Fellows, S. 1921a, Engl. Mech. \& World Sci., 114, 49
\\[-0.57cm]
\item[\hspace{-0.3cm}]
Fellows, S. 1921b, PASP, 33, 260
\\[-0.57cm]
\item[\hspace{-0.3cm}]
Fellows, S. 1921c, Nature, 108, 69
\\[-0.57cm]
\item[\hspace{-0.3cm}]
Fink, U., \& DiSanti, M. A. 1990, ApJ, 364, 687
\\[-0.57cm]
\item[\hspace{-0.3cm}]
Fulle, M. 1989, A\&A, 218, 283
\\[-0.57cm]
\item[\hspace{-0.3cm}]
Fulle, M., Cremonese, G., Jockers, K., \& Rauer, H. 1992, A\&A,{\linebreak}
 {\hspace*{-0.6cm}}253, 615
\\[-0.57cm]
%
%
\item[\hspace{-0.3cm}]
Fulle, M., Miku\v{z}, H., Nonino, M., \& Bosio, S. 1998, Icarus, 134,{\linebreak}
 {\hspace*{-0.6cm}}235
\\[-0.57cm]
\item[\hspace{-0.3cm}]
Gambart, J. 1825, Conn.\ Tems pour 1828, 273
\\[-0.57cm]
\item[\hspace{-0.3cm}]
Green, D. W. E. 1988, Minor Plan. Circ., 13459
\\[-0.57cm]
\item[\hspace{-0.3cm}]
Green, D. W. E. 2015, Centr. Bur. Electr. Tel., 4084
\\[-0.57cm]
%
%
%
\item[\hspace{-0.3cm}]
Hansen, P. A. 1824, AN, 2, 491
\\[-0.57cm]
\item[\hspace{-0.3cm}]
Harding, K. L. 1824, Berl. Astron. Jahrb. f\"{u}r 1827, 131
\\[-0.57cm]
\item[\hspace{-0.3cm}]
Harker, D. E., Woodward, C. E., Wooden, D. H., et al.\ 1999, AJ,{\linebreak}
 {\hspace*{-0.6cm}}118, 1423
\\[-0.57cm]
\item[\hspace{-0.3cm}]
Hasegawa, I., \& Nakano, S. 2001, PASJ, 53, 931
\\[-0.57cm]
\item[\hspace{-0.3cm}]
Hilton, J. L. 2005, AJ, 129, 2902
\\[-0.57cm]
\item[\hspace{-0.3cm}]
Hind, J. 1847, MNRAS, 7, 256
\\[-0.57cm]
\item[\hspace{-0.3cm}]
Hnatek, A. 1912, Denkschr.\,Akad.\,Wiss.\,Wien,\,Math.-Naturw.\,Kl.,{\linebreak}
{\hspace*{-0.6cm}}87, 1
\\[-0.57cm]
\item[\hspace{-0.3cm}]
Hoffmeister, C. 1921, AN, 214, 69
\\[-0.57cm]
\item[\hspace{-0.3cm}]
Holetschek,\,J.\,1891, Sitzungsber.\,Akad.\,Wiss.\,Wien, Math.-Naturw.{\linebreak}
{\hspace*{-0.6cm}}Cl., 100, 1266
\\[-0.57cm]
\item[\hspace{-0.3cm}]
Holetschek, J. 1892, AN, 129, 323
\\[-0.57cm]
\item[\hspace{-0.3cm}]
Holetschek, J. 1913, Denkschr. Akad. Wiss. Wien, Math.-Naturw.{\linebreak}
{\hspace*{-0.6cm}}Kl., 88, 745
\\[-0.57cm]
\item[\hspace{-0.3cm}]
Hornstein,\,C.\,1854a,\,Sitzungsber.$\:\!$Akad.$\:\!$Wiss.$\:\!$Wien,\,\mbox{Math.-Naturw.}{\linebreak}
{\hspace*{-0.6cm}}Cl., 12, 303
\\[-0.57cm]
\item[\hspace{-0.3cm}]
Hornstein, C. 1854b, AN, 38, 323
\\[-0.57cm]
\item[\hspace{-0.3cm}]
Hornstein, C. 1870, Sitsungsber.\,Akad.\,Wiss.\,Wien,\,Math.-Naturw.{\linebreak}
{\hspace*{-0.6cm}}Cl., Abt.\,2, 62, 244
\\[-0.57cm]
\item[\hspace{-0.3cm}]
Howarth, I. D., \& Bailey, J. 1980, J. Brit. Astron. Assoc., 90, 265
\\[-0.57cm]
\item[\hspace{-0.3cm}]
Hsia, F. C. 2009, Sojournes in a Strange Land:\ Jesuits and Their
 {\hspace*{-0.6cm}}Scientific Missions in Late Imperial China
 (Chicago:\ University
 {\hspace*{-0.6cm}}of Chicago Press)
\\[-0.57cm]
%
%
%
\item[\hspace{-0.3cm}]
Jockers, K., Bonev, T., \& Credner, T. 1999, Ap\&SS, 264, 227
\\[-0.57cm]
%
%
\item[\hspace{-0.3cm}]
Kanda, S. 1922, PASP, 34, 68
\\[-0.57cm]
\item[\hspace{-0.3cm}]
Kawakita,\,H., Furusho,\,R., Fujii,\,M., \& Watanabe,\,J.-i.\,1997,
 PASJ,{\linebreak}
 {\hspace*{-0.6cm}}49, L41
\\[-0.57cm]
\item[\hspace{-0.3cm}]
Kendall, E. O. 1843, AN, 20, 387
\\[-0.57cm]
%
%
\item[\hspace{-0.3cm}]
Kres\'ak, L.$\!\!$'\, 1982, Bull. Astron. Inst. Czech., 33, 150
\\[-0.57cm]
\item[\hspace{-0.3cm}]
Kreutz, H. 1901, Astr. Abh., 1, 1
\\[-0.57cm]
\item[\hspace{-0.3cm}]
Kronk, G. W. 1999, Cometography, Vol. 1, Ancient--1799:~A~Cata-{\linebreak}
 {\hspace*{-0.6cm}}log of Comets (Cambridge,\,UK:\,Cambridge
 University~Press),\,380
\\[-0.57cm]
\item[\hspace{-0.3cm}]
Lara, L. M., Schulz, R., St\"{u}we, J. A., \& Tozzi, G. P. 2001,
 Icarus,{\linebreak}
 {\hspace*{-0.6cm}}150, 124
\\[-0.57cm]
\item[\hspace{-0.3cm}]
M\"{a}kinen, J. T. T., Bertaux, J.-L., Pulkkinen, T. I., et al.\
 2001,{\linebreak}
 {\hspace*{-0.6cm}}A\&A, 368, 292
\\[-0.57cm]
\item[\hspace{-0.3cm}]
Mallama, A., Wang, D., \& Howard, R. A. 2006, Icarus, 182, 10
\\[-0.57cm]
\item[\hspace{-0.3cm}]
Marcus, J. N. 2007, Intern. Comet Quart., 29, 39
\\[-0.57cm]
\item[\hspace{-0.3cm}]
Markwick, E. E. 1921, Engl. Mech. \& World Sci., 114, 88
\\[-0.57cm]
\item[\hspace{-0.3cm}]
Marsden, B. G. 1967, AJ, 72, 1170
\\[-0.57cm]
\item[\hspace{-0.3cm}]
Marsden, B. G. 1989a, AJ, 98, 2306
\\[-0.57cm]
\item[\hspace{-0.3cm}]
Marsden, B. G. 1989b, Minor Plan. Circ., 15379
\\[-0.57cm]
\item[\hspace{-0.3cm}]
Marsden, B. G. 1996, IAU Circ., 6521
\\[-0.57cm]
\item[\hspace{-0.3cm}]
Marsden, B. G., \& Sekanina, Z. 1971, AJ, 76, 1135
\\[-0.57cm]
\item[\hspace{-0.3cm}]
Marsden, B. G., \& Williams, G. V. 2008, Catalogue of Cometary{\linebreak}
{\hspace*{-0.6cm}}Orbits 2008, 17th ed. (Cambridge, MA:\
Smithsonian Astrophysi-{\linebreak}
{\hspace*{-0.6cm}}cal Observatory)
\\[-0.57cm]
\item[\hspace{-0.3cm}]
Marsden, B.\,G., Sekanina, Z., \& Yeomans, D.\,K.\ 1973, AJ,\,78,\,211
\\[-0.57cm]
\item[\hspace{-0.3cm}]
Marsden, B. G., Sekanina, Z., \& Everhart, E. 1978, AJ, 83, 64
\\[-0.57cm]
\item[\hspace{-0.3cm}]
Nelson Day, F. C. 1921a, Nature, 108, 69
\\[-0.57cm]
\item[\hspace{-0.3cm}]
Nelson Day, F. C. 1921b, PASP, 33, 262
\\[-0.57cm]
\item[\hspace{-0.3cm}]
Neslu\v{s}an, L., \& Hajdukov\'a, M. Jr. 2014, A\&A, 566,
A33
\\[-0.57cm]
\item[\hspace{-0.3cm}]
Olbers, H. W. M. 1824a, AN, 2, 456A (Circular to No.\,48)
\\[-0.56cm]
\pagebreak
\item[\hspace{-0.3cm}]
Olbers, H. W. M. 1824b, AN, 3, 5
\\[-0.57cm]
\item[\hspace{-0.3cm}]
\"{O}pik, E. J. 1971, Irish Astron. J., 10, 35
\\[-0.57cm]
\item[\hspace{-0.3cm}]
Pearce, J. A. 1921, JRASC, 15, 364
\\[-0.57cm]
\item[\hspace{-0.3cm}]
Pickering, W. H. 1911, Ann. Harvard Coll. Obs., 61, 163
\\[-0.57cm]
\item[\hspace{-0.3cm}]
Pingr\'e, A. G. 1784, Com\'etographie ou Trait\'e Historique et
 Th\'eo-{\linebreak}
 {\hspace*{-0.6cm}}rique des Com\`etes, Vol. 2 (Paris: L'Imprimerie Royale),
 29, 102
\\[-0.57cm]
\item[\hspace{-0.3cm}]
Pittichov\'a, J., Meech, K. J., Valsecchi, G. B., \& Pittich, E. M.{\linebreak}
{\hspace*{-0.6cm}}2003, BAAS, 35, 1011
\\[-0.57cm]
\item[\hspace{-0.3cm}]
Plummer, W. E. 1892, Observatory, 15, 308
\\[-0.57cm]
\item[\hspace{-0.3cm}]
Porter,\,J.\,G.\,1952,\,Comets and Meteor Streams~(London:~Chapman{\linebreak}
{\hspace*{-0.6cm}}\& Hall)
\\[-0.57cm]
\item[\hspace{-0.3cm}]
Porter, J. G. 1963, in The Moon, Meteorites, and Comets, ed. B.\,G.{\linebreak}
{\hspace*{-0.6cm}}Middlehurst\,\&\,G.\,P.\,Kuiper\,(Chicago:\,University~of~Chicago),\,550
\\[-0.57cm]
\item[\hspace{-0.3cm}]
Rauer, H., \& Jockers, K.~1990, in Asteroids, Comets, Meteors III,{\linebreak}
 {\hspace*{-0.6cm}}ed.\ C.-I. Lagerkvist, H. Rickman, \& B. A. Lindblad
 (Uppsala,{\linebreak}
 {\hspace*{-0.6cm}}Sweden:\ Uppsala Universitet), 417
\\[-0.57cm]
\item[\hspace{-0.3cm}]
Richaud, J. 1729, M\'em. Acad. R. Sci. Paris, 7, 819
\\[-0.57cm]
\item[\hspace{-0.3cm}]
Russell, H. N. 1921, Sci. Amer., 125, 168
\\[-0.57cm]
\item[\hspace{-0.3cm}]
Schmidt, J. F. J. 1847a, AN, 25, 285
\\[-0.57cm]
\item[\hspace{-0.3cm}]
Schmidt, J. F. J. 1847b, AN, 25, 313
\\[-0.57cm]
\item[\hspace{-0.3cm}]
Sekanina, Z. 1976, Center for Astrophysics Preprint (Cambridge,{\linebreak}
{\hspace*{-0.6cm}}MA:\ Smithsonian Astrophysical Observatory), 445
\\[-0.57cm]
\item[\hspace{-0.3cm}]
Sekanina, Z. 1977, Icarus, 30, 374
\\[-0.57cm]
\item[\hspace{-0.3cm}]
Sekanina, Z. 1978, Icarus, 33, 173
\\[-0.57cm]
\item[\hspace{-0.3cm}]
Sekanina, Z. 1982, in Comets, ed. L. L. Wilkening (Tucson, AZ:
{\hspace*{-0.6cm}}University of Arizona), 251
\\[-0.57cm]
\item[\hspace{-0.3cm}]
Sekanina, Z. 2006, in Near Earth Objects, Our Celestial Neighbors:{\linebreak}
{\hspace*{-0.6cm}}Opportunities and Risk, IAU Symp.\ 236, ed.\ G. B.
Valsecchi,{\linebreak}
{\hspace*{-0.6cm}}D. Vokrouhlick\'y, \& A. Milani (Cambridge, UK: Cambridge Uni-
{\hspace*{-0.6cm}}versity), 211
\\[-0.57cm]
\item[\hspace{-0.3cm}]
Sekanina, Z., \& Chodas, P. W. 2007, ApJ, 663, 657
\\[-0.57cm]
\item[\hspace{-0.3cm}]
Sekanina, Z., \& Chodas, P. W. 2012, ApJ, 757, 127 (33pp)
\\[-0.57cm]
\item[\hspace{-0.3cm}]
Sekanina,\,Z., Hanner,\,M.\,S.,{\vspace{-0.03cm}} Jessberger,\,E.\,K., \&
Fomenkova,\,M.\,N.  {\hspace*{-0.6cm}}2001, in Interplanetary Dust, ed.\ E.
Gr\"{u}n, B. {\AA}.  S. Gustafson,
{\hspace*{-0.6cm}}S. Dermott, \& H. Fechtig (Berlin: Springer-Verlag), 95
\\[-0.57cm]
\item[\hspace{-0.3cm}]
Struyck, N. 1740, Inleiding tot de algemeene geographie, benevens
 {\hspace*{-0.6cm}}eenige sterrekundige en andere verhandelingen
 (Introduction to{\linebreak}
 {\hspace*{-0.6cm}}the general geography, with some astronomical and
 other trea-{\linebreak}
 {\hspace*{-0.6cm}}tises) (Amsterdam:\ Tirion)
\\[-0.57cm]
\item[\hspace{-0.3cm}]
Tacchini, P. 1882, Compt. Rend. Acad. Sci. Paris, 95, 896
\\[-0.57cm]
\item[\hspace{-0.3cm}]
Tacchini, P. 1883, Mem. Soc. Spettr. Ital., 11, E1
\\[-0.57cm]
\item[\hspace{-0.3cm}]
Tr\'epied, C. 1882, Compt. Rend. Acad. Sci. Paris, 94, 1636
\\[-0.57cm]
\item[\hspace{-0.3cm}]
Turner, N. J., \& Smith, G. H. 1999, AJ, 118, 3039
\\[-0.57cm]
\item[\hspace{-0.3cm}]
Ud\'{\i}as, A. 2003, Searching the Heavens and the Earth:\,The History{\linebreak}
 {\hspace*{-0.6cm}}of Jesuit Observatories. Astrophysics \& Space Science
 Library,{\linebreak}
 {\hspace*{-0.6cm}}vol.\ 286 (Dordrecht, Netherlands: Kluwer Academic Publishers)
\\[-0.57cm]
\item[\hspace{-0.3cm}]
Vogel, E. 1852a, MNRAS, 12, 206
\\[-0.57cm]
\item[\hspace{-0.3cm}]
Vogel, E. 1852b, AN, 34, 387
\\[-0.57cm]
\item[\hspace{-0.3cm}]
von Biela, W. 1824, AN, 3, 27
\\[-0.57cm]
\item[\hspace{-0.3cm}]
Vsekhsvyatsky, S. K. 1958, Fizicheskie kharakteristiki komet{\linebreak}
{\hspace*{-0.6cm}}(Moscow:\ Gosud.\,izd-vo fiz.-mat.\,lit.); translated:\ 1964,
Physical{\linebreak}
{\hspace*{-0.6cm}}Characteristics of Comets, NASA TT-F-80 
(Jerusalem:~Israel
{\hspace*{-0.6cm}}Program for Scientific Translations)
\\[-0.57cm]
\item[\hspace{-0.3cm}]
Whipple, F. L. 1977, Icarus, 30, 736
\\[-0.57cm]
\item[\hspace{-0.3cm}]
Whipple, F. L. 1978, Moon Plan., 18, 343
\\[-0.57cm]
\item[\hspace{-0.3cm}]
Williams, G. V. 2015, Minor Plan. Circ., 95212
\\[-0.57cm]
\item[\hspace{-0.3cm}]
Wolf, M. 1921a, AN, 214, 69
\\[-0.57cm]
\item[\hspace{-0.3cm}]
Wolf, M. 1921b, AN, 214, 103
\\[-0.57cm]
\item[\hspace{-0.3cm}]
Womack, M., \& Suswal, D. 1996, IAU Circ., 6485
\\[-0.67cm]
\item[\hspace{-0.3cm}]
Wyckoff, S., Heyd, R., Fox, R, \& Smith, A. 2000, BAAS, 32, 1063}
\vspace*{-1.0cm}
\end{description}
\end{document}